\documentclass[aps,prl,longbibliography,twocolumn,showpacs,floatfix]{revtex4-1}
\usepackage{graphicx}
\usepackage[]{natbib}
\usepackage{xspace}
\usepackage{color}
\begin{document}

\title{ 
From the area under the Bessel excursion to anomalous diffusion of cold atoms
 }

\author{E. Barkai}
\author{E. Aghion}
\author{D. A. Kessler}
\affiliation{Department of Physics, Institute of Nanotechnology and Advanced Materials, Bar Ilan University, Ramat-Gan
52900, Israel}

\begin{abstract}
L\'evy flights are random walks in which the probability distribution
of the step sizes 
is fat-tailed. 
  L\'evy spatial diffusion  has been observed
for a collection of  ultra-cold  $^{87}$Rb atoms and  single $^{24}$Mg$^{+}$ ions
in an optical lattice, a system which allows for
a unique degree of control of the dynamics. 
Using the semiclassical theory
of Sisyphus cooling,
 we formulate the problem as a coupled  L\'evy walk,
with strong correlations between the length $\chi$
and duration $\tau$ of the excursions.
Interestingly the problem is related to the
area under the Bessel and Brownian excursions.
 These are overdamped Langevin 
motions 
that start and end at the origin, constrained to remain
positive, in the presence  or absence of an external logarithmic potential,
respectively. 
In the limit of a weak potential, i.e., shallow optical lattices, the
celebrated Airy distribution describing the areal distribution
of the Brownian excursion is found as a limiting case. 
Three distinct phases of the dynamics are investigated:
 normal diffusion, L\'evy diffusion and,
below a certain critical depth of the optical potential,
$x\sim t^{3/2}$ scaling, which is related to Richardson's
diffusion from the field of turbulence.
The main focus of the paper is the analytical
 calculation of the joint probability density function
$\psi(\chi,\tau)$ from a newly developed
 theory of the area under the Bessel excursion. 
The latter describes the spatiotemporal correlations in the problem
and is the microscopic input needed to characterize the spatial diffusion 
of the atomic cloud. 
 A modified Montroll-Weiss equation for the Fourier-Laplace
transform of the density $P(x,t)$ is obtained, which depends on the
statistics of velocity excursions and meanders.
The meander, a random walk in velocity space, 
which starts at the origin and
does not cross it,  describes the last jump event $\chi$ in the sequence.
In the anomalous phases, the statistics of meanders and excursions are
essential for the calculation of the mean square displacement, indicating
that our correction to the Montroll-Weiss equation is crucial,
and pointing to the sensitivity of the transport on a single jump event.
Our work provides general relations between the statistics
of velocity excursions and meanders and that of the diffusivity,
both for normal and
anomalous processes.
\end{abstract}
\maketitle

\section{Introduction}

 The velocity, $v(t)$, of a particle interacting with a heat bath exhibits
stochastic behavior which in many cases is difficult to evaluate.
 The position
of the particle, assumed to start at the origin at time $t=0$,
 is the time integral over 
the fluctuating velocity $x(t) = \int_0 ^t v(t') {\rm d} t'$ and
demands a probabilistic approach to determine its
statistical properties. Luckily the central limit theorem makes it possible,
for  many processes, to predict a Gaussian shape for 
the diffusing packet. 
Then the diffusion constant $K_2$ characterizes the normal motion
through its mean square displacement $\langle x^2 \rangle = 2 K_2 t$.
The remaining goal, for a given process or model, is
to compute $K_2$ and other transport
coefficients. In normal cases this can be done, at least in principle,
via the Green-Kubo
formalism, namely by the calculation of the stationary velocity correlation
function, which gives the diffusivity $K_2=\int_0 ^\infty \langle v(t) v(0)\rangle {\rm d} t$.

 An alternative approach
is investigated in this work and is based on the concept of excursions.
We assume that the random
 process $v(t)$ is recurrent and thus the velocity crosses the zero point,
$v=0$, many times in the observation window $(0,t)$. We divide
the path $x(t)$ into a sum of increments 
\begin{equation}
x(t) = \int_0 ^{t_1} v(t') {\rm d} t' + \int_{t_1} ^{t_2} v(t') {\rm d} t' + .... \int_{t_i} ^{t_{i+1}} v(t') {\rm d} t' + \cdots .
\label{eqInt01}
\end{equation}
Here $\left\{ t_1,t_2 \cdots \right\}$ 
are the points in time of the velocity zero-crossings,
$v(t_i)=0$. In the  interval
$(t_i,t_{i+1})$ the velocity is either strictly
 positive or negative. The velocity
in each interval is thus a stochastic process which starts and ends on the origin
without crossing it in between. Such a random curve is called an
excursion. The random spatial increment 
$\chi_i = \int_{t_i} ^{t_{i+1}} v(t') {\rm d} t'$ is the area under 
the excursion. The position of the particle, according to Eq. (\ref{eqInt01}),
is the sum of the random increments, namely a sum of the signed areas under the
velocity excursions, each having a random duration. 
The goal of this paper is to relate the statistics
 of the areas under these velocity excursions and the corresponding 
random time intervals between zero crossings,  to the problem of
spatial diffusion. This connection is easy to find in the case that 
the increments $\chi_i$ and the
duration of the excursions $\tau_i=t_{i+1} - t_i$
 are mutually uncorrelated, independent and identically distributed
random variables. Over a long measurement time, the number of excursions
is $t / \langle \tau \rangle$ where $\langle \tau \rangle$ is the 
average time for an excursion. Then according to Eq. (\ref{eqInt01})
the mean squared displacement is $\langle x^2 \rangle = \langle \chi^2 \rangle t/ \langle \tau \rangle$, and hence we have
\begin{equation}
K_2 = {\langle \chi^2 \rangle \over 2 \langle \tau \rangle}.
\label{eqIntro03} 
\end{equation}
This equation is reminiscent of the famous Einstein formula, and shows that
diffusion is related to the statistics of excursions. The original work
of Einstein, discussed in many textbooks, is explicitly based on
an underlying random walk picture, and so does not involve the area
under random velocity excursions, nor zero crossings in velocity 
space.  We will arrive at the simple equation (\ref{eqIntro03}) only
at the end of our work, in Sec. \ref{secNormal}, while here it merely
serves as an appetizer to motivate the consideration of velocity
excursions in some detail,  and to 
suggest the usefulness of developing
tools for the calculation 
of $\langle \chi^2 \rangle$ and $\langle \tau \rangle$ (here $\langle \chi \rangle=0$, by symmetry). 
Obviously Eq. (\ref{eqIntro03}) is based on the assumption
that the variance $\langle \chi^2 \rangle$ is finite, and as mentioned,
that correlations are not important. The major effort of this paper
is directed to considering a more challenging case, 
 namely a physically relevant
stochastic process where the variance $\langle \chi^2 \rangle$ diverges, and 
more importantly the process exhibits correlations
between $\chi$ and $\tau$. 
The particular system we  investigate is a model for
diffusion of atoms
in optical lattices, following the experiments \cite{Katori,Wickenbrock,Sagi} 
and the semi-classical theory of Sisyphus cooling \cite{CTphysToday,CT,Zoller}.

 One type of excursion which has been thoroughly investigated 
is the Brownian
excursion \cite{Majumdar1,Majumdar2,Jason}. 
 A Brownian excursion is a conditioned one dimensional Brownian
motion $v(t)$ over the time interval $0<t<\tau$. The motion starts
at $v(0)=\epsilon$ (eventually the  $\epsilon \to 0$ limit is taken) and ends 
at $v(\tau)=\epsilon$ and is constrained
not to cross the origin $v=0$ in the observation time $(0,\tau)$. The area
under this curve is a random variable, whose statistical properties
have been investigated by mathematicians 
\cite{Darling,Louchard,Takacs,Tracy}. 
More recently, this problem was treated with a  
path integral approach  to describe statistics of fluctuating interfaces
\cite{Majumdar1,Majumdar2,Greg,Rambeau} and related
areas until the first passage time
 are used to describe universality of sandpile models 
\cite{Stapleton}. 
 Here, we generalize the
Brownian excursion to a process $v(t)$ described by a Langevin equation,
with
an asymptotically logarithmic potential. We show how the random area under 
this Langevin excursion
determines the dynamics of cold atoms.  
We believe that Langevin excursions, or more generally random excursions,
are useful tools in many areas of statistical physics, hence their
investigation beyond the well studied Brownian excursion,
is worthwhile. 

 If $v(t)$ is a Brownian motion, or as we will proceed to  show for cold atoms in shallow lattices,
the set of points where  $v(t)=0$ is non-trivial,
 as it contains no isolated points
and no interval. This is the case since we are dealing with a continuous
path with power law statistics of the crossing times (see details below).
Mathematician have investigated the statistics of level  crossing,
e.g. zero crossing, 
of continuous paths in great detail. The concept of local time,
introduced by P. L\'evy in the context of
Brownian motion and Ito's excursion theory for continuous paths,
are pillars in this field, see \cite{Chung,Yor} and references within.  
 Here we use a heuristic approach, with a 
 modification of the original process $v(t)$,
 by introducing a 
cutoff $\pm \epsilon$: the starting point of the particle
after each zero hitting, which is taken to zero at the end. 
This makes it possible to use renewal theory, and continuous time random
walks, which are tools well studied in physics literature
of discrete processes. 
In this sense we differ from rigorous mathematical approaches.
Thus we avoid the problem of continuous paths, for
example the infinite
number of zero crossings, by replacing the original path with
an $\epsilon$ modified path for which  the number
of zero crossings is finite (when $\epsilon$ is finite). 
We show that physical quantities characterizing the entire process have
a finite $\epsilon\to 0$ limit.  For example
in Eq. 
(\ref{eqIntro03}), for Langevin dynamics of $v(t)$,
 both $\langle \tau \rangle$ and $\langle \chi^2\rangle$ approach
zero as $\epsilon \to 0$, their ratio $K_2$  approaches a finite limit.

 One might wonder why we wish to
use excursions and their peculiar properties
to evaluate the spreading of atoms or more generally other transport
systems. The answer is that it turns 
into a useful strategy when the friction forces
are non-linear. In particular we will investigate laser cooling,
where within the semi-classical theory, the dimensionless friction 
force is \cite{CT} (see details below) 
\begin{equation}
F(v) = - {v \over 1 + v^2}. 
\label{intro02}
\end{equation}
This friction force, induced by the laser fields, is linear for
small velocities $F (v) \sim - v$ similar to the Stokes friction law
for a massive Brownian particle in water at room temperature.
However, unlike such friction, which increases in magnitude with velocity,
here for large $v$, $F(v) \sim - 1/v\rightarrow 0$. 
Asymptotically, then, the system is frictionless. 
This implies that fast particles have to remain fast for a
long time, which in turn induces 
heavy-tailed populations of fast particles \cite{Renzoni}.
In this case, as we show later, the standard picture of diffusion breaks down. 

 More specifically, the problem of diffusion of cold atoms
under Sisyphus cooling was partially treated
by Marksteiner, et al. \cite{Zoller}. 
While clearly indicating the anomalous nature of the
diffusion process,   
the main tool used was the evaluation of the
stationary velocity correlation function of the process, followed by the use of
the
Green-Kubo formalism for the evaluation of $K_2$. 
They showed 
 that for a certain critical value of the depth of the optical
lattice the value of $K_2$ diverges (see also
\cite{Hodapp}). 
 Katori et al. \cite{Katori}  measured the mean square displacement $\langle x^2 \rangle \sim t^{2 \xi}$
and recorded  the onset of super-diffusion  $\xi>1/2$
beyond a critical depth of the optical lattice  \cite{Katori}. 
Wickenbrock et al. \cite{Wickenbrock}
in the context of driven  optical lattice experiments demonstrated
with Monte Carlo
simulations an upper limit
on the  spreading of the atoms 
 $\langle x^2 \rangle \le \mbox{const} \times t^3$.
Sagi et al. \cite{Sagi}
showed that a packet of spreading
 Rb atoms can be fitted with a L\'evy distribution
instead of a Gaussian distribution found for normal diffusion.
 These findings clearly indicate the breakdown
of the usual strategy of treating normal diffusion and hence
promoted further theoretical investigations. 
The velocity correlation function is
not stationary and hence the Green-Kubo formalism must be replaced 
\cite{DechantPNAS}.
 The moments
exhibit  multifractal behavior \cite{DechantPRL},  there is an enhanced
sensitivity to the initial preparation of the system \cite{Mukamel,Ori},
momentum fluctuations are described by an infinite covariant density 
\cite{KesslerPRL}
and  in certain parameter regimes the
L\'evy central limit theorem applies instead  of the standard Gaussian version 
\cite{Zoller,KesBarPRL}. 
In this regime of shallow optical lattices the power of the analysis of
the area under Langevin  excursions
becomes essential as we will show here.  
We note that the relation of laser cooling with L\'evy statistics
is not limited to the case of Sisyphus cooling considered in this manuscript. 
Sub-recoil laser cooling, a  
 setup different from ours, also leads to fundamental relations
between statistical physics of rare events, and laser-atom physics 
\cite{Bardou,Levybook}.  
For a recent mini-review on the departure from Boltzmann-Gibbs statistical
mechanics for cold atoms in optical lattices, see \cite{LutzNature}. 

\subsection{Scope and organization of paper}

The current work significantly
extends the investigation of 
the properties of the spatial distribution of atoms in Sisyphus cooling begun in Ref. \cite{KesBarPRL}.  There we uncovered three phases
of the motion which are controlled
by the depth $U_0$ of the optical lattice: a Gaussian phase $x \sim t^{1/2}$,
a L\'evy phase $x \sim t^\xi$ and $1/2< \xi <3/2$, and a Richardson phase
$x \sim t^{3/2}$ (see details below). Within the intermediate phase, the
density of particles, in the central region of the packet,
 is described by a symmetric L\'evy distribution, similar
to the fitting procedure used in a recent Weizmann Institute experiment \cite{Sagi}.
 However, this cannot be the
whole story. 
 As is well known the variance of the L\'evy
distribution diverges, which implies that $\langle x^2 \rangle=\infty$,
which is unphysical. Indeed, as mentioned,
 Katori et al. experimentally determine
a finite mean square displacement $\langle x^2 \rangle \sim t^{2 \xi}$
(see also \cite{Wickenbrock}). 
 We showed related  {\em numerical} evidence that the 
L\'evy distribution is cut off at distances of the order $x \sim t^{3/2}$.
  This breakdown of
L\'evy statistics arises due to the importance of correlations between jump
lengths $\chi$ and jump duration $\tau$, which are neglected in the derivation
of the L\'evy distribution. Our purpose here is to investigate these correlations
in detail. As we proceed to show, the correlations are given by the statistical
properties of the Langevin excursions discussed above, for all except the last
interval. Because the velocity is not constrained to be zero at the measurement
time $t$, this last interval is not described by an excursion
but rather  by a Langevin meander, where the random
walk $v(t)$  begins at the origin and does not return for the entire
duration of the walk.
The properties of the excursions and meanders enter 
in a modified Montroll-Weiss \cite{Montroll} 
equation for the Fourier-Laplace transform of the density
$P(x,t)$ which we derive. We then use this equation to calculate a
quantity which is sensitive to the correlations, namely the mean square displacement. The mean square displacement exhibits anomalous diffusion and is sensitive
to the last jump event, i.e., to the statistics of the meander. 
Thus, our treatment modifies both the celebrated Montroll-Weiss equation,
to include  the last jump in the sequence (i.e., the meander)
 and the existing
theory of areas under 
Brownian meanders and excursions to include the dissipative
 friction
force, which is responsible to the cooling of the atoms.
Our analysis  illuminates the  rich physical behavior and
provides the needed set of mathematical tools beyond the 
decoupling approximation 
used to obtain the L\'evy distribution in our previous work.
For completeness, we present a detailed decoupled and  coupled analysis,
the latter being the main focus of the current work.

 The paper is organized as follows. We start with a brief survey
of the semi-classical theory of Sisyphus cooling \cite{CT,Zoller},
and show the connection of the dynamics to L\'evy walks following 
Marksteiner et al. 
\cite{Zoller}. The importance of the correlations between
$\tau$ and $\chi$ is emphasized, a theme which as mentioned
 has not received
its deserved attention. In Sec. 
\ref{SecSca}, a simple scaling theory is presented which yields
the exponents describing the dynamics of the atomic packet. The main
 calculation
of the distribution of the area under the Bessel excursion is found in
Sec. 
\ref{SecArea}, the calculation of the area under the Bessel meander
is given in an Appendix. A new 
coupled continuous time random walk theory in Sec.
\ref{Secctrw} provides the connection between the statistics
of excursions and meanders, and the evolution of the density profile. Asymptotic
behaviors of the Fourier-Laplace transform of the joint probability 
density function (PDF)
of jump lengths and waiting times are investigated in Sec. 
\ref{SecAsy}. These in turn give us the asymptotic behaviors
of the  atomic density packet, the mean square displacement,
 and the different phases of the dynamics which are
investigated in  Secs.
\ref{SecLev} -
\ref{secNormal}.
Derivation of the  distribution of
the  time interval straddling time
$t$ for the Bessel process is carried out in Appendix F,
which allows as to connect between our heuristic renewal
 approach and more rigorous treatments \cite{Chung,Bertoin}
 and further discuss the
nontrivial fractal  set of zero crossings.

\section{Semi-classical description of cold atoms- Langevin dynamics}
 We briefly
present the semi-classical picture for the dynamics of the atoms.
The trajectory of a single particle with mass $m$
 is $x(t) = \int_0 ^t p(t) {\rm d} t/m$ where
$p(t)$ is its momentum.
Within the standard picture \cite{CT,Zoller,Grynberg}
of Sisyphus cooling, two competing
mechanisms describe the dynamics.
The cooling force
$F(p) = - \overline{\alpha} p /[1 + (p/p_c)^2 ]$
acts to restore the momentum to the minimum energy state $p=0$.
Momentum diffusion is governed by  a
diffusion coefficient which is momentum dependent,
$D(p) = D_1 + D_2 / [1 + (p/p_c)^2]$.  The latter
describes momentum fluctuations which
lead to heating due to random emission events which stochastically
jolt the atom.
We use dimensionless units, time $t \to
t \overline{\alpha}$, momentum $p \to p/p_c$,
distance 
$x \to x m \overline{\alpha}  / p_c$, and introduce the dimensionless
momentum diffusion
constant  $D=D_1/ (p_c)^2 \overline{\alpha}$.
For simplicity,
we set $D_2=0$  since
 it does not modify the asymptotic $|p|\to \infty$
 behavior of the
diffusive heating term, nor that of the force
 and therefore  does not modify our main conclusions.

\begin{figure}\begin{center}
\includegraphics[width=0.41\textwidth]{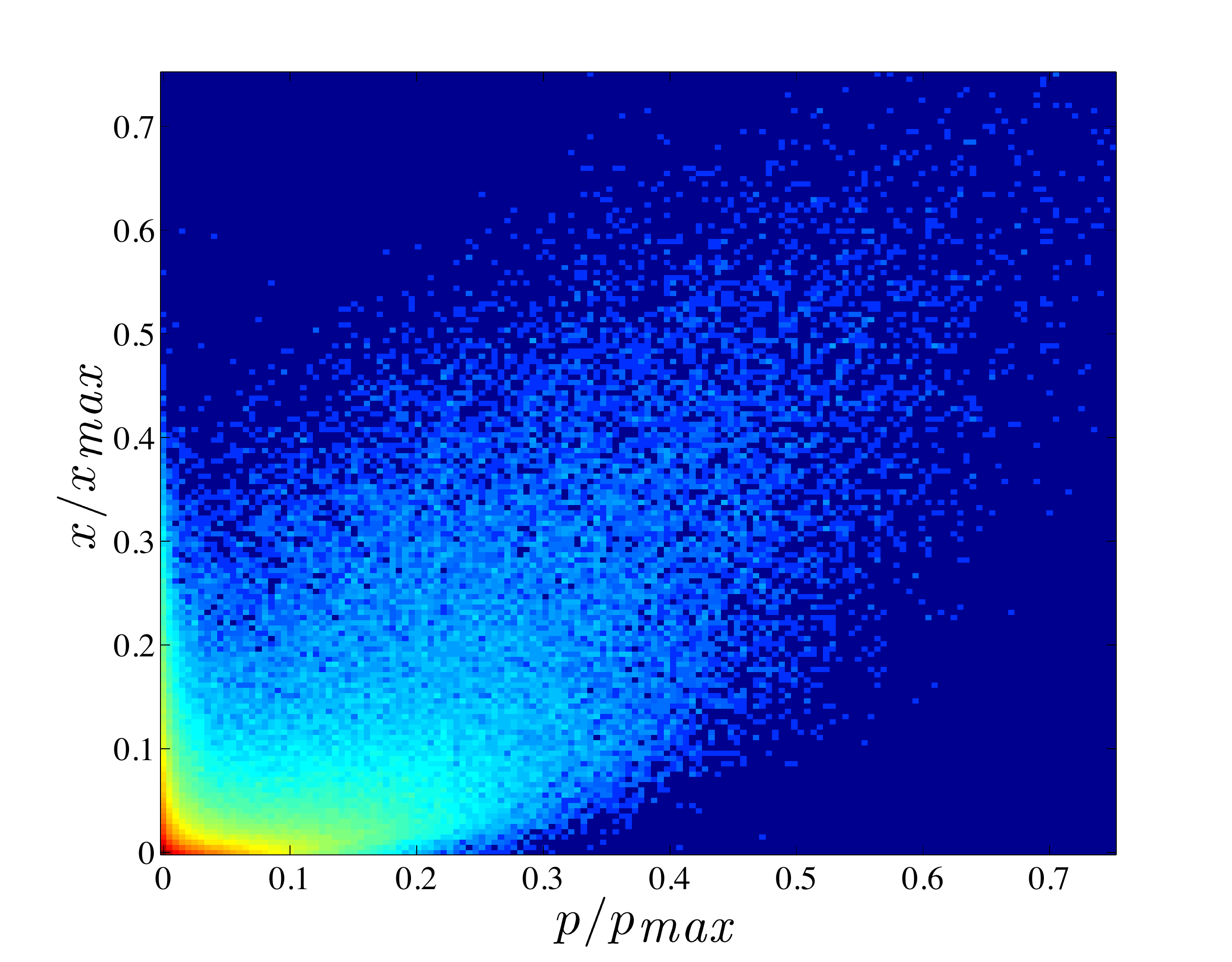}
\end{center}
\caption{
The joint density of $x$ and $p$ for $D=2/3$, $10^6$ particles and $t=10^6$.  
 This scatter plot of phase space shows correlation between $x$ and $p$ and
rare events with large fluctuations both for $x$ and $p$.
} 
\label{fig2d}
\end{figure}

The Langevin equations
\begin{equation}
{{\rm d} p \over {\rm d} t } = F(p) + \sqrt{ 2 D} \xi(t), \ \ \ \ \ \
{{\rm d} x \over {\rm d} t } = p
\label{eq05}
\end{equation}
describe the dynamics in phase space.
Here the noise term is Gaussian, has zero mean
and is white, $\langle \xi(t) \xi(t') \rangle = \delta(t- t')$.
The now dimensionless cooling force is
\begin{equation}
F(p) = - { p \over 1 + p^2} .
\label{eq03}
\end{equation}
The force $F(p)$  
is a
very peculiar non-linear friction force. 
We note that friction forces which decrease with increasing velocity
or momentum like $1/p$ 
 are found also for some nanoscale devices, e.g., an atomic tip on a surface
\cite{Urbakh}.  

One goal of the paper is to find the spatial distribution of particles
governed by Eq. (\ref{eq05}). 
The stochastic Eq. (\ref{eq05})
gives the trajectories of the standard
Kramers picture  for   the semi-classical dynamics in an optical lattice
which in turn was derived from microscopic considerations \cite{CT,Zoller}.
Denoting  the joint PDF of $(x,p)$ by $W(x,p,t)$, the Kramers equation
reads
\begin{equation}
{\partial W \over \partial t} + p {\partial W \over \partial x} = \left[
D {\partial^2 \over \partial p^2 }  - {\partial \over \partial p } F(p) \right]
W.
\label{eqKramers}
\end{equation}
From the semiclassical treatment of the interaction of the atoms with
the counter-propagating laser beams, we have
\begin{equation}
D= c E_R/ U_0,
\label{eqDc}
\end{equation}
where $U_0$ is the depth of the optical potential,  $E_R$
the recoil energy and the dimensionless parameter
$c$ \cite{remark2}
depends on the atomic transition involved \cite{CT,Zoller,Lutz}.
$U_0$ is a control parameter; hence different
values of $D$ are attainable in experiment, and exploration of
different phases of the dynamics are within reach \cite{Katori,Wickenbrock,Sagi,Renzoni}. 
Eq. (\ref{eqDc}) is rather intuitive since deep optical lattices,
i.e. large $U_0$, implies small $D$ while large recoil energy
leads to a correspondingly large value of $D$. 

The behavior of the distribution for momentum  only
(when $x$ is integrated out from the Kramer's Eq., yielding the Fokker-Planck Eq.) is much simpler, and has been presented in previous work~\cite{KesslerPRL}.
The equilibrium properties are governed by the effective
potential in momentum space,
\begin{equation}
V(p) = - \int_0 ^p F(p) {\rm d} p= (1/2) \ln(1 + p^2)
\label{eqvp}
\end{equation}
which for large $p$ is  logarithmic, $V(p) \sim \ln(p)$.
This large $p$ behavior of $V(p)$
is responsible
for several unusual equilibrium and non-equilibrium
properties of the momentum distribution
\cite{Katori,Renzoni,KesslerPRL,Dechantprl,Mukamel}.
 The  equilibrium momentum  distribution function is given by~\cite{Renzoni}
\begin{equation}
W_{{\rm eq}} (p) = \frac{1}{\cal{Z}} e^{-V(p)/D} =  \frac{1}{\cal{Z}}\left( 1 + p^2 \right)^{ - 1/2 D}.
\label{eqequil}
\end{equation}
Here 
\begin{equation}
{\cal Z}=\int_{-\infty} ^\infty \exp[- V(p)/ D ] {\rm d} p = 
\sqrt{\pi}\Gamma\left({1-D\over 2 D} \right)/\Gamma\left({1 \over 2 D }\right)
\label{eqZ}
\end{equation}
is the normalizing partition function. Eq. (\ref{eqequil}) is
Student's $t$ distribution,
also sometimes called a Tsallis distribution \cite{Renzoni}. 
Actually the problem is related coincidentally to Tsallis statistics,
since as mentioned the equilibrium PDF is proportional
to a Boltzmann-like factor,
$W_{{\rm eq}} (p) \propto \exp[- V(p)/D]$,
so $D$ acts like a temperature. 
More importantly, the power-law tail of the equilibrium PDF,
for sufficiently large $D$,  implies a large population
of fast particles,
which in turn will spread in space faster than what one would expect
using naive 
Gaussian central limit theorem 
arguments. For example if $1/D < 3$ the ensemble averaged kinetic energy
in equilibrium diverges, since $\langle p^2 \rangle_{{\rm eq}} = \infty$ while
when $1/D<1$ the partition function diverges and a steady-state
 equilibrium is never reached. A dramatic increase of
the energy of atoms  when the optical lattice parameter $U_0$ approaches 
a critical
value was found experimentally in \cite{Katori} and a
power-law  momentum distribution was measured in \cite{Renzoni}.
Of course, the kinetic energy of a physical system cannot be infinite, and
the momentum distribution must be treated as a time dependent
object within the infinite covariant density approach
\cite{KesslerPRL}. While much is known
about the momentum distribution,  both experimentally and theoretically,
 the experiments \cite{Katori,Wickenbrock,Sagi}
demand a theory for the spatial spreading. 
We note that diffusion in logarithmic potentials 
has a vast number of applications 
\cite{Mukamel,Ori,KesslerPRL,Farago,Bray,Lo,Micciche,Martin} and refs. therein,
 e.g.  Manning condensation \cite{Manning}, 
and unusual mathematical properties,
including ergodicity breaking
\cite{Dechantprl}. In general,   
logarithmic potentials play a special role in statistical
physics \cite{Dyson,Beenakker,Campisi}. 

\section{Mapping the problem to a L\'evy Walk process}

 In principle one may attempt to directly solve the Kramers equation 
(\ref{eqKramers})
to find the joint
PDF of the random variables $(x,p)$ at a given time $t$. 
In Fig. \ref{fig2d} we plot a histogram of the phase space obtained
numerically. We see a complicated structure: roughly along
the diagonal, clear correlations between $x$ and $p$ are visible;
on the other hand, along the  $x$ and $p$ axis, decoupling
between momentum and position is evident, together with broad (i.e. non-Gaussian)
 distributions of $x$ and $p$. At least to the naked eye no simple 
scaling structure is found in $x-p$ space, and hence we shall turn
to 
different, microscopic, variables  which do exhibit simple 
scaling.  
This leads to an analysis centered on 
the mapping
of the Langevin dynamics to a L\'evy walk scheme \cite{Zoller,Klafter} 
and the statistics of areas under random excursions. 

\begin{figure}\begin{center}
\includegraphics[width=0.41\textwidth]{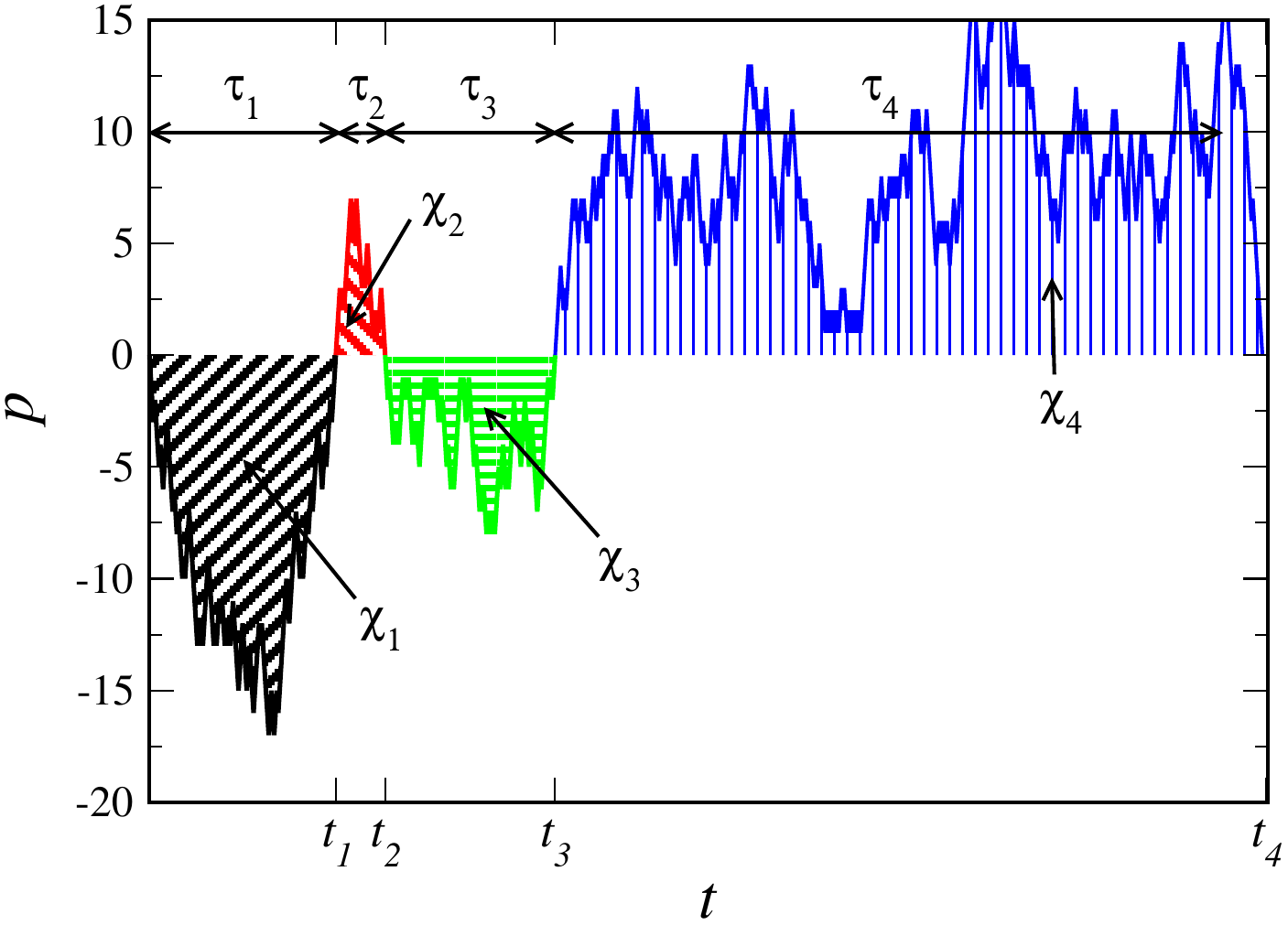}
\end{center}
\caption{
Schematic presentation of momentum of the particle versus time.
The times between consecutive
zero crossings are called the waiting times $\tau$ and the shaded
area under each excursion are the random flight displacements $\chi$.
The $\tau$'s and the $\chi$'s are correlated, since 
a long waiting time typically  implies a large displacement.
}
\label{fig1}
\end{figure}

Starting at the origin, $p=0$, the particle along its stochastic path in
momentum space crosses $p=0$ many times (see Fig. \ref{fig1}). 
In other words the random
walk in momentum space is recurrent; this being the case 
even when $D\to \infty$ since then the process in momentum space 
is one of pure diffusion (i.e. the force is negligible)
and from Polya's theorem we
know that such  one dimensional walks are recurrent. 
The cooling force being attractive clearly maintains this property. 

 Let $\tau>0$ be the random time between one crossing event and the
next and let $-\infty <\chi<\infty$ be the random displacement
for the corresponding $\tau$. As shown schematically in Fig.
\ref{fig1} the process starting with zero momentum
is defined by a sequence of jump durations 
$ \{ \tau_1 , \tau_2, \cdots \} $
with corresponding displacements 
$ \{ \chi_1,\chi_2,\cdots \}.$
These random waiting times and displacements generate a L\'evy
walk \cite{Zoller}, as we explain below. 
Here $\chi_1= \int_0 ^{\tau_1} p(t') {\rm d} t'$,
$\chi_2= \int_{\tau_1}  ^{\tau_1+ \tau_2}  p(t') {\rm d} t'$ etc. 
Let the points on the time axis
$\{ t_1,t_2, \cdots\}$ denote times $t_n>0$ where the particle crossed the 
origin of momentum $p=0$ (see Fig. \ref{fig1}). These times  are
related to the waiting times: $t_k= \sum_{i=1} ^k \tau_i$.
We see that
 $\chi_k = \int_{t_{k-1}} ^{t_{k}} p(t') {\rm d} t'$ is the area under
the random momentum curve constrained in such a way that
$p(t')$ in the time interval $(t_{k-1},t_{k})$ does not cross the origin,
while it started and ended there. 
The total displacement, namely the position of the particle
at time $t$, is a sum of the individual displacements 
$ x= \sum_{i=1} ^n \chi_i + \chi^*$. Here $n$ is the random number of
crossings of zero momentum in the time interval $(0,t)$ and
$\chi^*$ is the displacement made in the last interval $(t_n,t)$. 
By definition, in the time interval
 $(t_n,t)$ no zero crossing was recorded. The time 
$\tau^*=  t-t_n$ is sometimes called the backward recurrence time \cite{GL}.
The measurement time is clearly $t = \sum_{i=1} ^n \tau_i + \tau^*$.
In standard transport problems the effect of the last displacement 
 $\chi^*$  on the position
of particle $x$ is negligible, and similarly one usually assumes
$t\simeq t_n$ when $t$ is large. However, for 
anomalous diffusion of cold atoms,  where the distributions of displacements
and jump durations are wide, these last events cannot be ignored. 

 One  goal is to find the long time
 behavior of $P(x,t)$, the normalized
 PDF of the  spatial position
of a particle that  started
at the origin $x=0,\ p=0$ at $t=0$.
 It is physically clear that in the long time limit
our results will not be changed if instead we consider narrow initial
conditions, for example Gaussian PDFs of initial position and momentum.
Initial conditions with  power-law tails will likely lead to very
different behaviors \cite{Mukamel,Ori}. 
Once we find $P(x,t)$ we have the semi-classical
approximation for the spatial density of particles. The latter can
be compared with the Weizmann Sisyphus cooling experiments \cite{Sagi} provided
that collisions among the atoms are negligible. 

The
L\'evy walk process under investigation is a renewal process \cite{GL}
in the sense that once the particle crosses the momentum origin the 
process is renewed (since the Langevin dynamics is Markovian). 
This is crucial in our treatment, and it implies that the waiting
times $\tau_i$
 are statistically independent identically distributed random variables
as are the $\chi_i$. However, as we soon discuss, 
the pairs $\{\tau_i,\chi_i\}$ are correlated. 

 Since the underlying Langevin dynamics is continuous, we need a refined 
definition of the L\'evy walk process. Both the $\tau_i$'s
 and the $\chi_i$'s are infinitesimal;
 however the  number of renewal events, $n$, diverges 
for any finite measurement time $t$, 
in such a way that the total displacement
$x$ is finite. In this sense the L\'evy walk process under investigation is
different from  previous works where the number of renewals, 
for finite measurement  time is finite. 
One way to treat the problem is to discretize the dynamics,
as is necessary in any event to perform a computer simulation, and then $\chi_i$ and $\tau_i$ are
of course finite. In our analytical treatment, following Marksteiner, et al.
\cite{Zoller},  we consider the
first passage time problem
 for a particle starting with momentum 
$p_i$ and  reaching $p_f<p_i$ for the first time at $\tau$. 
We take $p_f=0$ and eventually take $p_i =\epsilon \to 0$.  The L\'evy walk scheme
is hence summarized
with the following steps:
\begin{itemize}
\item[1.] Choose with probability $1/2$ either
$+p_i$ or $-p_i$.
\item[2.] Follow the Langevin dynamics until the particle
reaches $p_f=0$.
\item[3.] Record the random displacement $\chi$ and random duration
$\tau$ during this excursion.
\item[4.] Go to 1.
\end{itemize} 
This loop is terminated at time $t$, the final displacement
$\chi^*$ calculated, and as mentioned the
 total displacement is $x= \sum_{i=1} ^n \chi_i + \chi* $. 
 In the first step we have probability
$1/2$ to start with either $+ p_i$ or $-p_i$ since the cooling force is
antisymmetric and so vanishes at $p=0$. The advantage of
presenting the problem as a set of recurrent random walks through the
$\chi$'s and $\tau$'s instead of the direct Langevin picture stems 
from the fact that we can treat analytically the former L\'evy walk 
picture. 

 We denote the joint PDF
of the pair $\{\chi,\tau\}$ of a 
single excursion by $\psi(\chi,\tau)$.  
The theoretical development starts from the analysis of $\psi(\chi,\tau)$ 
and then from this we use the L\'evy walk scheme to relate
this single excursion information  to the properties of the entire walk, and in particular,
 $P(x,t)$ for large $t$. 

\begin{figure}\begin{center}
\includegraphics[width=0.41\textwidth]{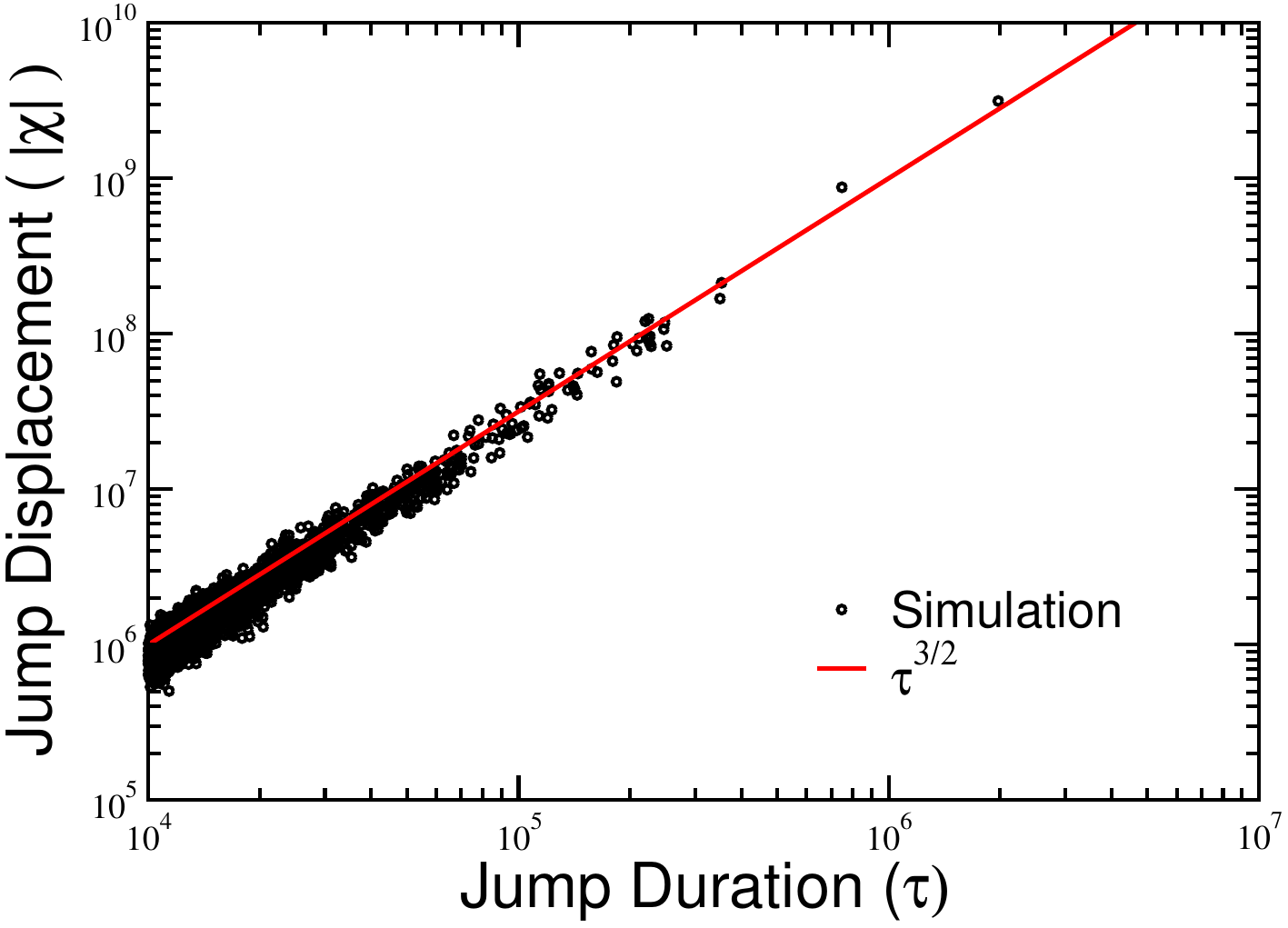}
\end{center}
\caption{
We plot the flight distance $|\chi|$ versus
the jump duration $\tau$ to demonstrate the strong correlations between
these two random variables. Here $D=2/3$ and the red line has a slope $3/2$
reflecting the $\chi\sim \tau^{3/2}$ scaling discussed in the text.
}
\label{fig2}
\end{figure}

 There exists a strong correlation between the excursion duration
$\tau$ and the displacement length $\chi$ which is encoded in
$\psi(\chi,\tau)$.  Let the PDFs of $\chi$ and  $\tau$ be denoted
$q(\chi)$, $g(\tau)$, respectively.
 Weak correlations would imply the
decoupled scheme $\psi(\chi,\tau) \simeq q(\chi) g(\tau)$ 
which is far easier to analyze; however as we shall see this decoupling is not 
generally true and leads to
wrong results for at least some observables of interest. Properties
of $q(\chi)$ and $g(\tau)$ were investigated in \cite{Zoller,Lutz} and are also studied
below. The problem becomes more interesting and challenging
due to these correlations. As we show,  these are responsible 
for the finiteness of the moments of $P(x,t)$,
in particular the mean square displacement $\langle x^2 \rangle$,
and for the existence of a rapidly decaying tail of $P(x,t)$.  
This in turn is related to
 the L\'evy flight versus L\'evy walk dilemma \cite{Klafter},
to multifractality \cite{DechantPRL}, 
and to the physical meaning of  the fractional  diffusion
 equation \cite{Review}  used
as a fitting tool for the Weizmann experiment \cite{Sagi} 
(see Eq. 
(\ref{eqDis}) and discussion there).
As we show below, beyond a critical value of $D=1$ the correlations 
can never be ignored and 
govern the behavior of the  entire packet, not only the large $x$ tails
of $P(x,t)$. 
Physically, the correlations are obvious, since long durations of flight $\tau$
involve large momentum $p$, which in turn induce large 
displacements $\chi$.  As an example of these correlations,
we plot in Fig. \ref{fig2} the displacement $|\chi|$ versus the corresponding
$\tau$ obtained from computer simulation. The figure
clearly demonstrates the correlations 
and  it also shows a 
$|\chi| \propto \tau^{3/2}$ scaling
which we now turn to investigate.  
Notice should be paid to how much simpler the $\chi-\tau$ distribution
is, compared to the $x-p$ distribution in Fig. 
\ref{fig2d}. Simulations presented in Fig. \ref{fig2} were performed on a 
discrete lattice in momentum space,
starting on  the first lattice point, see
Appendix  
\ref{SecAppDSim} for details.

\section{Scaling theory- relation between exponents}
\label{SecSca}

 As shown by Marksteiner, et al.~\cite{Zoller}, and Lutz
\cite{Lutz}  the PDFs
of the excursion durations and displacements satisfy the asymptotic
laws
\begin{equation}
g \left( \tau \right) \sim g_{*} \tau^{- (3/2 + \gamma)}, \ \qquad
q\left( \chi \right) \sim q_{*} | \chi |^{ - ( 4/3 + \beta)}.
\label{eqSca01}
\end{equation}
with
\begin{equation}
\gamma= {1 \over 2 D}, \ \  \ \ \ \ \ \beta= {1 \over 3 D} .
\label{eqSca02}
\end{equation}
In Appendices A and B we  study these PDFs
using  backward Fokker-Planck equations. 
 In particular we find the amplitudes $g^{*}$ and $q^{*}$,
and the relevant PDFs moments.
What is most crucial are
the exponents $\beta$ and $\gamma$.
When  $D \to \infty$ we get $\gamma=\beta=0$ and Eq. (\ref{eqSca01})
gives familiar limits. In the ``high temperature" limit 
of large $D$, the cooling force
is negligible and then the Langevin equation reduces
to Brownian motion in
momentum space. Then $g(\tau) \propto \tau^{-3/2}$, which is 
the well-known asymptotic behavior of the 
 PDF of first passage times for unbounded one dimensional
Brownian motion \cite{Redner,Gardiner}. Less well
known is  $\lim_{D \to \infty} q(\chi) \propto |\chi|^{-4/3}$ which describes
the distribution of the area under  a Brownian motion
until its first passage time
(see \cite{Kearney,MajRev} who give this PDF explicitly). 
Notice that the power-law behavior Eq. (\ref{eqSca01}) yields a diverging
second moment of the displacement $\chi$ for $D> 1/5$ which in 
turn gives rise to anomalous statistics for $x$.

 The correlations between $\chi$ and $\tau$ are now related to the
asymptotic behaviors of their PDF's, Eq. (\ref{eqSca01}).
We rewrite the joint PDF
\begin{equation}
\psi(\chi,\tau) = g(\tau) p(\chi|\tau)
\label{eqSca03}
\end{equation} 
where $p(\chi|\tau)$ is the conditional PDF
to find jump length $\chi$ for a given jump
duration $\tau$. We introduce a scaling ansatz which is expected
to be valid at large $\tau$:
\begin{equation}
p(\chi|\tau) \sim \tau^{-\eta} B \left( { \chi \over \tau^\eta} \right).
\label{eqSca04}
\end{equation} 
Since $g(\tau) = \int_{-\infty} ^\infty \psi(\chi,\tau) {\rm d} \chi$, 
we have the normalization condition $\int_{-\infty}^\infty B(z) {\rm d} z =1$. 
By definition 
\begin{equation}
 q(\chi) = \int_0 ^\infty \psi(\chi,\tau) \,{\rm d} \tau\propto 
 \int_0 ^\infty {\rm d} \tau\, \tau^{ -3/2- \gamma - \eta} B\left( {\chi \over \tau^\eta} \right)  \ 
\label{eqSca05}
\end{equation} 
Changing variables to $z=\chi/\tau^\eta$, we get
\begin{equation}
q\left( \chi \right) \sim \mbox{const} \,\chi^{ - ( 1 + { 1 \over 2 \eta} + {\gamma \over \eta} ) } .
\label{eqSca06}
\end{equation} 
Comparing with Eq.
(\ref{eqSca01}) we get a simple equation for the unknown exponent $\eta$
which is
$1+ 1/(2 \eta) + \gamma/\eta= 4/3 + \beta$, so using Eq. (\ref{eqSca02}) 
we find $\eta =3/2$.
This is precisely the scaling behavior $\chi \sim \tau^{3/2}$ we observe
in our simulation, Fig. \ref{fig2}. 
Hence the natural scaling solution of the problem is
\begin{equation}
p(\chi|\tau) \sim {1 \over \sqrt{D} \tau^{3/2}}  B\left( { \chi \over\sqrt{D}  \tau^{3/2}} \right).
\label{eqSca07}
\end{equation}
 In hindsight this result is related to Brownian scaling. 
For a particle free of the cooling force, its momentum will exhibit 
Brownian scaling  $p \sim  (D \tau)^{1/2}$ and hence the excursions
which are integrals of the velocity scale as $\chi\sim \sqrt{D} \tau^{3/2}$.
The crucial  point is that this simple Brownian scaling is maintained 
even in the presence of the cooling force for all $D$. This
is due to the marginal nature of the weak  $1/p$ friction force for large $p$ which leaves the Brownian scaling intact,
but changes the scaling function.   While the
$3/2$ scaling is simple and $D$ independent, the shape of $B(\cdot)$
is sensitive to this control parameter. In the next section
we investigate $B(\cdot)$ and for this we must
go beyond simple scaling arguments.

\section{Area Under the Bessel Excursion}
\label{SecArea}

\begin{figure*}\begin{center}
\includegraphics[width=0.90\textwidth]{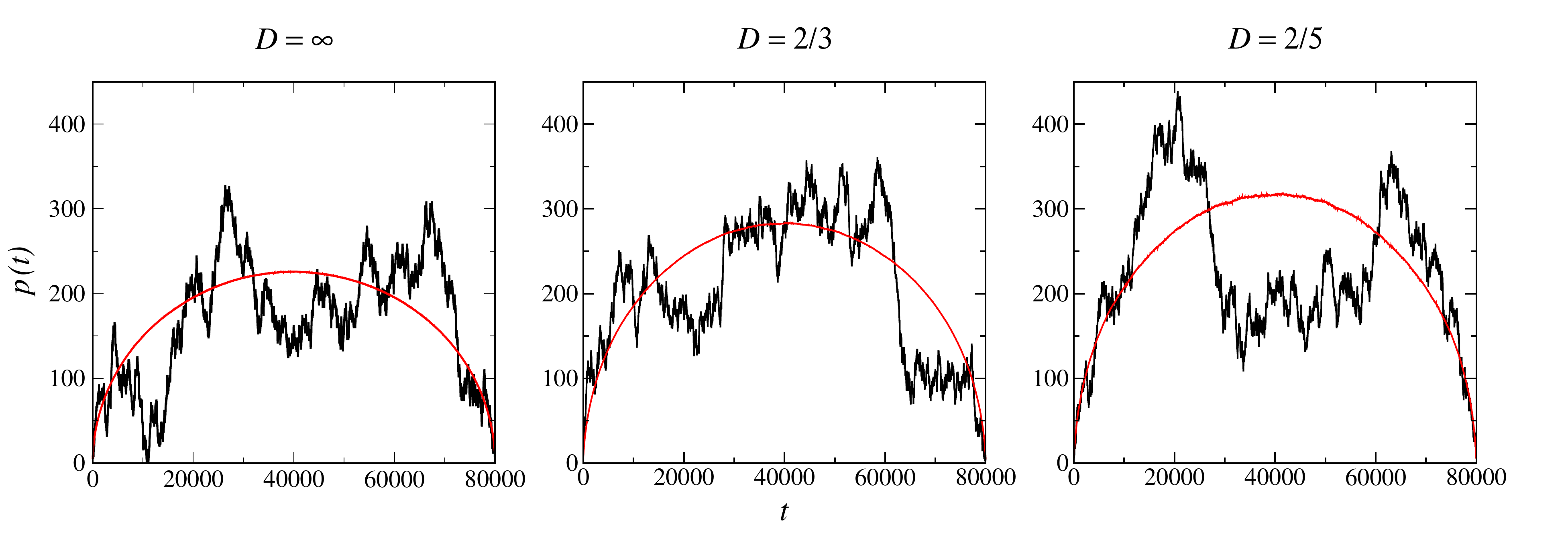}
\end{center}
\caption{
We show three typical Bessel excursions for
$D=\infty,2/3,2/5$ (left to right panel) and the mean (red curve).  The 
random excursions 
are  constrained Langevin paths, with a $-1/p$ force,  that do not cross the
origin in the observation time $t$ while start and end on $\epsilon \to 0$.
For simulations we used the regularized force field, 
which only alters the dynamics when $p\simeq 1$, 
and is negligible in the the long time limit (see Appendix D). 
When $D\to \infty$ we get a Brownian excursion. We see that as $D$ is decreased
the excursions are further pushed from the origin $p=0$, since small
$D$ implies effectively large forces, hence to avoid the zero crossing the 
particles must drift further away from origin. 
Thus the attractive force, repels particles, which
at first might sound counter intuitive and seems to be an 
unexplored property of excursions. 
}
\label{figPath}
\end{figure*}


 A natural generalization of the Brownian excursion is a Langevin
excursion. Such a stochastic curve is the path $p(t')$, 
given by the Langevin equation,  in the time
interval $0\le t' \le \tau$, such that it starts and ends at $p_i=p_f=\epsilon$, 
but is constrained to remain positive in between. Here, $p_i=p(0)$ is 
the initial  and $p_f=p(\tau)$ is the final location in momentum space.
For our application,
the  path is considered
in the limit $\epsilon \to 0$. Since the path never crosses the origin,
the area under such a curve is $\chi=\int_0 ^\tau p(t') {\rm d} t'$, 
and hence the PDF of $\chi$ for fixed $\tau$ yields the sought after
conditional PDF, $p(\chi|\tau)$. Obviously, $\chi$ is the integral over
the constrained path $p(t')$; hence by definition it is the area under
the excursion. 
The meander will describe the last jump $\chi^*$ since at the measurement
time the particles velocity is generally non-zero. 
 For now we will discuss only excursions, and find $p(\chi|\tau)$, and
return to the meander later.  

 Here we focus our attention on a specific excursion we call the Bessel
excursion, corresponding to the case $F(p) = -1/p$:
\begin{equation}
{{\rm d} p \over {\rm d} t } = -1/p + \sqrt{ 2 D} \xi(t), \ \ \ \ \ \
\label{eqBesselLang}
\end{equation}
 so that the
effective potential is the non-regularized logarithm, $V(p) = \ln(p)$ and
$p>0$.  
Since the scaling approach is valid for long times,
 where excursions are long, the typical momentum $p$ is large and the details of the 
force field close to the origin are negligible for the purpose
of the calculation of the scaling function $B(\cdot)$. We will check this
assumption with numerical simulations, 
which of course use a regularized form of the force law.
Some sample paths of Bessel excursion are presented in
Fig. \ref{figPath}.
 The name Bessel excursion stems from the fact that
Langevin dynamics in the non-regularized friction force field $1/p$ 
corresponds to a well-known stochastic
process called the Bessel
process \cite{Schehr,Martin}. 
More information
on the regularized and non-regularized processes is given in Appendixes
A,B.  

 From symmetry, $p(\chi |\tau) = p(-\chi|\tau)$, and so we may restrict
our discussion to $p_i=\epsilon > 0$ and $\chi>0$.  
Later we will use this symmetry which implies that for
 $-\infty<\chi<\infty$
 we have $p(\chi|\tau) = p(|\chi|,\tau)/2$ where
$p(|\chi|,\tau)$ is the PDF of positive excursions, 
normalized according to $\int_0 ^\infty p(|\chi|,\tau) {\rm d} |\chi| = 1$. 

Let $G_\tau(\chi,p|p_i)$ be the joint PDF
of the random variables $\chi$ and $p$.  Since $\chi= \int_0 ^\tau p(t') {\rm d} t'$, we have 
$\chi=0$ at time $\tau=0$ so initially $G_{\tau=0}(\chi,p |p_i)= \delta(p-p_i)\delta(\chi)$.
Later we will take $p$ to be the final momentum $p_f$, which similarly to $p_i$
will be set to 
the value $\epsilon\to 0$ (for the sake of notational brevity we omit
the subscript in $p_f$). 
The calculation
of $p(\chi|\tau)$ follows three steps. 
For the Brownian case $F(p)=0$, this method was successfully applied
by Majumdar and Comtet \cite{Majumdar2}. 

\begin{itemize}
\item[(i)]
  We find the Laplace
$\chi \to  s D $ transform of  $G_\tau(\chi,p|p_i)$ 
\begin{equation}
\int_0 ^\infty e^{- s D \chi} G_\tau\left(\chi,p|p_i\right) {\rm d} \chi  \equiv \widehat{G}_\tau\left(s,p|p_i\right).
\label{eqBessel01}
\end{equation}
The reason why we multiply $s$ with $D$ in the Laplace transform will become clear soon. 
Since $\chi$ is a functional of the path $p(t')$ we will use the Feynman-Kac
(FK)
formalism to find $\widehat{G}_\tau(s,p|p_i)$ (see details below). 
The constraint that the path $p(t')$ is always positive enters as an absorbing
boundary condition at $p=0$ \cite{Martin}.
\item[(ii)]
The second step is to consider 
$p_i=p=\epsilon \to 0$
 and obtain
the Laplace transform
\begin{equation}
\hat{p}(s|\tau) = \int_0^\infty p(\chi|\tau)e^{ - s D  \chi} {\rm d \chi} =
\lim_{\epsilon \to 0} {\widehat{G}_\tau (s,\epsilon|\epsilon) \over \widehat{G}_\tau(s,\epsilon|\epsilon)|_{s=0} }. 
\label{eqBessel02}
\end{equation}
The denominator in the above equation ensures the normalization,
since we must  have $\hat{p}(s|\tau)|_{s=0}= 1$. 
\item[(iii)]
Finally, we invert Eq. (\ref{eqBessel02}) back to 
$\chi$ space 
using the inverse Laplace transform,
 $s D \to \chi$, 
which yields  
 $p(\chi|\tau)$. From there we can immediately find also the
scaling function $B(\cdot)$, 
Eq. (\ref{eqSca04}).
\end{itemize}
 We now implement these steps to solve the problem.

  The FK formalism \cite{MajRev}  treats
functionals of Brownian motion. Here we  use a  modified
version of the FK equation
to treat over-damped Langevin paths \cite{Carmi}.  
Let $A = \int_0 ^\tau U[p(t')]{\rm d} t'$
be a functional of the Langevin path and assume $U(\cdot)> 0$ so
that we are treating positive functionals. Here $G_\tau(A,p|p_i)$ is 
the joint PDF of $A$ and $p$ and $\widehat{G}_\tau(s,p|p_i)$ is the
corresponding Laplace $A \to s D$ transform. The generalized
 FK equation reads
\begin{equation}
{\partial \widehat{G}_\tau (s,p|p_i) \over \partial \tau} = \left. \left[ \hat{L}_{{\rm fp}} - s D U(p) \right] \widehat{G}_\tau (s,p|p_i)\right.
\label{eqBessel03} 
\end{equation}
Here $\hat{L}_{{\rm fp}} = D (\partial_p)^2 - \partial_p F(p)$ 
 is the Fokker-Planck operator. When $F(p)=0$,
Eq. (\ref{eqBessel03}) is the celebrated FK equation which is an imaginary
time Schr\"odinger equation and $- s D U(p)$ is the potential of the
corresponding quantum problem. The constraint on positive excursions,
namely $p>0$, gives the boundary condition
$\widehat{G}_\tau(s,0|p_i)=0$. In the quantum language
this is an infinite potential barrier 
for $p<0$. This formalism can be used in principle to
obtain the area under Langevin excursions for all forms of $F(p)$.

 For the Bessel excursion under investigation here,
 the functional  $U[p(t)]=p(t)$ is linear since
 $\chi=\int_0 ^\tau p(t') {\rm d} t'$. 
Quantum mechanically, this gives a linear potential  and
hence the connection to the Airy
function found for $F(p)=0$ in \cite{Majumdar1,Majumdar2}. 
With the force field $F(p) = - 1/p$ we have
$\hat{L}_{{\rm fp} } = D (\partial_p)^2+ \partial_p p^{-1}$
and hence we find, using
Eqs. (\ref{eqBesselLang},\ref{eqBessel01},\ref{eqBessel03}) 
(the former gives the $s D$ term)
\begin{equation}
{\partial \widehat{G}_{\tau} (s,p|p_i)  \over \partial \tau} =
\left( D {\partial^2 \over \partial p^2} + {\partial \over \partial p} { 1 \over p} - s D p \right) \widehat{G}_\tau(s,p|p_i) .
\label{eqBessel04} 
\end{equation}
 The solution of this equation is found using the separation ansatz
\begin{equation}
\widehat{G}_{\tau} \left( s, p |p_i \right) = \sum_{k=0} ^\infty a_k \psi_k(p) e^{ - D E_k \tau}, 
\label{eqBessel05} 
\end{equation}
which yields the time independent equation
\begin{equation}
{\partial^2 \over \partial p^2}  \psi_k + \frac{1}{D}{\partial\over \partial p}  {\psi_k\over p} -  s p \psi_k= - E_k \psi_k.
\label{eqBessel06} 
\end{equation}
We now switch to the more familiar one dimensional Schr\"odinger equation
via the similarity transformation $\phi_k = p^{1/(2D)} \psi_k$ which gives
\begin{equation}
- {\partial^2 \over \partial p^2}  \phi_k +\left[ {1 \over 2D} \left( {1 \over 2 D} + 1 \right) p^{-2} + s p\right] \phi_k = E_k \phi_k .
\label{eqBessel07} 
\end{equation}
This Schr\"odinger equation has a binding potential, which yields
discrete eigenvalues, with an effective repulsive potential with a $p^{-2}$
 divergence for $p \to 0$ and  a binding linear potential for  large $p$ provided $s\neq 0$. 
Eq. (\ref{eqBessel07})
 describes a three dimensional non-relativistic
quantum particle in a linear
potential \cite{Ruijgrok,Ateser}, 
the $p^{-2}$ part  corresponding to an angular momentum term.  

As usual $\phi_k(p)$ yields a complete  orthonormal 
basis $\int_0 ^\infty \phi_k(p) \phi_m(p) {\rm d} p = \delta_{km}$ 
and the formal solution of the problem is
\begin{equation}
\widehat{G}_\tau \left( s , p | p_i \right) = \sum_k  \left(\frac{p_i}{p}\right)^{1/(2 D)} 
\phi_k (p_i)\phi_k(p)  e^{- DE_k \tau}.
\label{eqBessel08} 
\end{equation}
We can scale out the Laplace variable $s$ from the time independent
 problem, defining
$\phi_k(p) =  s^{1/6}f_k (s^{1/3} p ) $ and $E_k = s^{2/3} \lambda_k $.
Using Eq. (\ref{eqBessel07}) we find
\begin{equation}
- {\partial^2 \over \partial \tilde{p}^2}  f_k +\left[ {1 \over 2D} \left( {1 \over 2 D} + 1 \right) \tilde{p}^{-2} +  \tilde{p}\right] f_k = \lambda_k f_k, 
\label{eqBessel09} 
\end{equation}
where $\tilde{p} = s^{1/3} p$ . Notice that when $D \to \infty$ the second term
vanishes and we get the Airy equation. For this technical
 reason we 
have chosen in Eq. 
(\ref{eqBessel01})
the Laplace variable as $s D$ not $s$. Also, it should be noted that the $f_k$ are orthonormal in the variable $\tilde{p}$.

It follows from Eq. (\ref{eqBessel09}) and the absorbing boundary condition that
\begin{equation}
f_k(\tilde{p}) \sim \tilde{d}_k \tilde{p}^{1+1/(2D)}, \qquad \tilde{p}\to 0
\label{eq28er}
\end{equation} 
with $\tilde{d}_k$ a $k$-dependent coefficient.
The $\tilde{d}_k$ will soon be seen to be important and they are evaluated 
from solutions of Eq. 
(\ref{eqBessel09}) 
 with the normalization 
condition $\int_0 ^\infty [f_k(\tilde{p})]^2 {\rm d} \tilde{p}=1$.
For the initial and final conditions under investigation,
$p_i=p=\epsilon$, 
we have
\begin{equation}
\widehat{G}_\tau \left( s, \epsilon|\epsilon \right) \sim s^{\nu + 2/3} \epsilon^{1 + 2 \alpha} \sum_k \left( \tilde{d}_k \right)^2 e^{ - D \lambda_k s^{2/3} \tau} 
\label{eqBessel14}
\end{equation}
where the exponents $\nu$ and $\alpha$ will turn out to be useful
\begin{equation}
\nu = \frac{1+D}{3D},\qquad\qquad \alpha = {1 +D \over 2 D } .
\label{eqBessel15}
\end{equation}
Note that classification of boundaries  for a 
non-regularized Bessel process was carried out in  \cite{Martin}
and discussed here  in Sec. 
\ref{SecDis}.
For an absorbing boundary condition 
both the sign of the probability current and usual
condition of the vanishing
of the probability on the absorbing point must be taken into consideration
\cite{Martin}.

\begin{figure}\begin{center}
\includegraphics[width=0.41\textwidth]{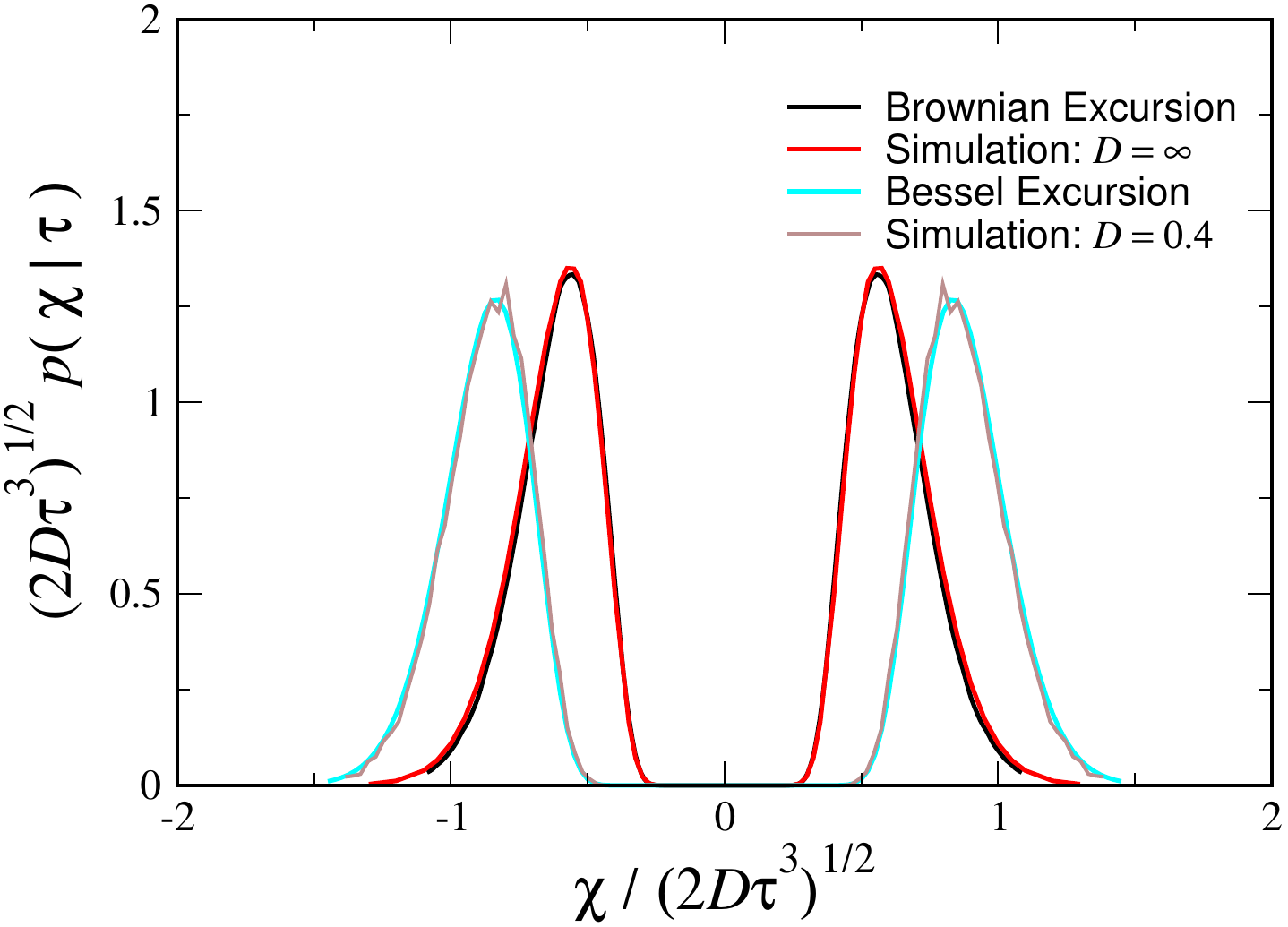}
\end{center}
\caption{
The conditional probability $p(\chi|\tau)$ is 
described by the Airy distribution when $D \to \infty$ otherwise
by the distribution derived here based on the area under the Bessel process
theory,
 Eq.  
(\ref{eqHG}) .
They describe
areas under constrained Brownian (or Bessel) excursions
 when the cooling force $F(p)=0$ (or $F(p) \ne 0$)
respectively.  
 }
\label{fig2a}
\end{figure}

According to Eq. 
(\ref{eqBessel02}),
we need 
$\widehat{G}_\tau(s=0,\epsilon|\epsilon)$, 
 in order to normalize the solution. 
This $s=0$ propagator can be found exactly.
The eigenvalue problem now reads
\begin{equation}
- \frac{\partial^2}{\partial p^2} \phi_k^0  + \left(\frac{1}{2D}\right)\left(\frac{1}{2D}+1\right) \frac{\phi_k^0}{p^2} 
=  E_k^0 \phi_k^0.
\label{eqBessel15a}
\end{equation}
The superscript $0$ indicates the $s=0$ case. Since $s=0$
 the linear field
in Eq. 
(\ref{eqBessel07}) is now absent so the
``particle" is not bounded and hence one finds
a continuous spectrum $E^0 _k =  k^2$.
The wave functions consistent with the boundary condition are 
\begin{equation}
\phi_k ^0 (p) = B_k \sqrt{p}  J_\alpha(k p)
\label{eqBessel16}
\end{equation}
where $J_\alpha(\cdot)$ is the Bessel function of the first kind.
 The second solution with $J_{-\alpha}(k p)$ is unphysical due to the
boundary condition \cite{Martin}. The
 normalization condition is 
\begin{equation}
1= (B_k)^2 \int_0 ^L p  \left[ J_\alpha(k p ) \right]^2\,
{\rm d} p 
\label{eqBessel17}
\end{equation}
where $L\to \infty$ is the ``box" size which eventually will drop 
out of the calculation. Solving the integral in Eq. (\ref{eqBessel17}) 
gives $B_k = \sqrt{ \pi k /L}$. For initial and final conditions
$p_f=p_i = \epsilon$ the $s=0$ propagator then reads
\begin{equation}
\widehat{G}_\tau \left( s=0 , \epsilon|\epsilon\right) = {\pi \epsilon \over L}
\sum_{k=0} ^\infty k \left[ J_\alpha\left( k \epsilon \right)\right]^2 e^{ - D k^2 \tau}.
\label{eqBessel17a}
\end{equation}
Using the small $z$ expansion of the Bessel function $J_\alpha (z) \sim (z/2)^\alpha/ \Gamma(1+\alpha)$ one finds
\begin{equation}
\widehat{G}_\tau\left( s=0,\epsilon|\epsilon\right) \sim { \pi \epsilon^{1 + 2 \alpha} \over L  4^\alpha \Gamma^2 \left( 1 + \alpha\right) } \sum_k k^{2 \alpha + 1} e^{ - D k^2 \tau } .
\label{eqBessel18}
\end{equation}
 Notice that replacing $J_\alpha(\cdot)$ with $J_{-\alpha}(\cdot)$ leads to $G_\tau(s=0,\epsilon|\epsilon)\sim \epsilon^{1- 2 \alpha}$ which diverges since $\alpha>1/2$, as mentioned this unphysical solution is rejected due to the absorbing boundary conditions \cite{Martin}. 
The usual density of states calculation follows the quantization rule of a particle
in a box extending between $0$ and $L$ which gives $k L = n \pi$ with integer
$n$. Hence $\sum_k \cdots  \to \int_0 ^\infty {\rm d}k {L \over \pi} \cdots$
and one finds, after a change of variable to $x=k^2$,
\begin{eqnarray}
\widehat{G}_\tau(s=0,\epsilon|\epsilon) &=&  
{\epsilon^{2 \alpha +1} \over 2^{2 \alpha} \left[ \Gamma\left( 1 + \alpha\right) \right]^2 } \int_0 ^\infty {{\rm d} x \over 2} x^{\alpha} e^{- Dx \tau} \nonumber\\
&=& {\epsilon^{2 \alpha +1} \over 2^{2 \alpha} \Gamma(1 + \alpha)} \left(D \tau \right)^{-(1+\alpha)}. 
\label{eqBessel19}
\end{eqnarray}
Inserting Eqs. (\ref{eqBessel14},\ref{eqBessel19}) in Eq. 
(\ref{eqBessel02})
we find our first main result
\begin{eqnarray}
\hat{p}\left( s | \tau\right) &=& 
2^{ 2 \alpha + 1} \Gamma\left( 1 + \alpha\right) \left[ s \left( D \tau\right)^{3/2} \right]^{\nu + 2/3}\times\nonumber\\
&\ & \qquad \sum_{k=0} ^\infty \left( \tilde{d}_k \right)^2
e^{ - D\lambda_k s^{2/3} \tau } .
\label{eqBessel20}
\end{eqnarray}
In the limit $D \to \infty$ we find
$|\tilde{d}_k|\to 1$ and the $\lambda_k$'s are the energy 
eigenvalues of the Airy equation, related to the zeros of the Airy function, and, up to a rescaling of $s$, we get the result obtained by Darling \cite{Darling}
 and Louchard \cite{Louchard} 
for the area under the Brownian excursion,
namely the Laplace transform of the Airy distribution \cite{Majumdar2}.
 We can show that $\hat{p}(s=0|\tau)=1$ as it should.  

It is easy to tabulate the eigenvalues $\lambda_k$ and the coefficients
 $\tilde{d}_k$ using  Eq. 
(\ref{eqBessel09}) and standard numerically exact
techniques.
 In Table \ref{table1}  we tabulate the first few slopes $\tilde{d}_k$ and eigenvalues
for $D=2/5$. 
For large $k$, i.e. large energy, the $1/p^2$ part of the
 potential is irrelevant and the eigenvalues $\lambda_k$ converge 
to the eigenvalues of the Airy problem considered previously. 
Similarly, we expect $\lim_{k \to \infty} |d_k|=1$ for any finite $D$,
since in the Airy problem limit, i.e.,  $D\to \infty$ case 
\cite{Majumdar1,Majumdar2}, $|\tilde{d}_k|$ is unity for all $k$.

To complete the calculation we need to perform an
inverse Laplace transform, namely  invert from $s D$ back
to $\chi>0$ (and multiply the PDF by $1/2$ if interested in both
positive and negative excursions). 
In Eq. (\ref{eqBessel20}) we have terms with the structure
\begin{equation}
s^{\nu + 2/3} \exp(- c_k s^{2/3})= \sum_{n=0} ^\infty {(-c_k)^n s^{\nu + 2( n+1) / 3} \over n!} . 
\label{eqBessel21}
\end{equation}
Each term on the right hand side can be inverse transformed separately
since the inverse Laplace transform of $(D s)^\gamma$ is  $\chi^{-\gamma -1}/\Gamma(-\gamma)$.
After term by term transformation
we sum the infinite series
(summation over $n$) using  Maple. Thus we arrive
at our first destination, the conditional
PDF for $\chi>0$ 
is found in terms of generalized hypergeometric functions
$$ $$
\begin{widetext}
\begin{eqnarray} p(\chi|\tau) &=& -\frac{\Gamma(1+\alpha)}{2\pi \chi}\left(\frac{4D^{1/3}\tau}{(\chi)^{2/3}}\right)^{\alpha+1}\sum_k [\tilde{d}_k]^{2}\left[\Gamma\left(\frac{5}{3}+\nu\right)\sin\left(\pi\frac{2+3\nu}{3}\right){}_2F_2\left(\frac{4}{3}+\frac{\nu}{2},\frac{5}{6}+
\frac{\nu}{2};\frac{1}{3},\frac{2}{3};-\frac{4D\lambda_k^3\tau^3}{27 \chi^2}\right)
\right.\nonumber\\
&\ &\qquad\qquad\qquad {} - \frac{D^{1/3}\lambda_k\tau}{(\chi)^{2/3}} \Gamma\left(
\frac{7}{3}+\nu\right)\sin\left(\pi\frac{4+3\nu}{3}\right){}_2F_2\left(\frac{7}{6}+\frac{\nu}{2},\frac{5}{3}+\frac{\nu}{2};\frac{2}{3},\frac{4}{3};-\frac{4D\lambda
_k^3\tau^3}{27 \chi^2}\right) \nonumber\\
&\ &\qquad\qquad\qquad \left. {} + \frac{1}{2}\left(\frac{D^{1/3}\lambda_k \tau}{
\chi^{2/3}}\right)^2 \Gamma\left(3+\nu\right)\sin\left(\pi\nu\right){}_2F_2\left(2
+\frac{\nu}{2},\frac{3}{2}+\frac{\nu}{2};\frac{4}{3},\frac{5}{3};-\frac{4D\lambda_k^3\tau^3}{27 \chi^2}\right)\right],
\label{eqHG}
\end{eqnarray}
\end{widetext}
where the summation is over the eigenvalues. In Fig. \ref{fig2a} we plot
the solution for $D=2/5$ and $D \to \infty$ corresponding the the Brownian
case.  
Notice the $\chi \sim \sqrt{D}  \tau^{3/2}$ scaling which proves the
scaling hypothesis Eq. 
(\ref{eqSca07}).
The sum in Eq. (\ref{eqHG}) converges quickly as long as $\chi$ is not too large, and so we can use it to construct a plot of $p(\chi|\tau)$.
 For example, for Fig. 
\ref{fig2a}
only $k=0,...4$ is needed to obtain excellent convergence. 
We stress that Eq. (\ref{eqHG}) gives an explicit representation
of the scaling function for $0<\chi<\infty$, and hence by symmetry for all $\chi$. The scaling variable
is 
\begin{equation}
v_{3/2} = {\chi \over  \sqrt{D} \tau^{3/2}}
\label{eqScdef}
\end{equation}
 and the scaling function
 $B(v_{3/2})$  in Eq. 
(\ref{eqSca07}) can be read directly  off of  Eq.
(\ref{eqHG}). 

\begin{table}[ht]
\centering
\begin{tabular}{|c|c|c|}
\hline
$k$ & $\tilde{d}_k$ & $\lambda_k$ \\
\hline
$0$ & $0.41584$ & $3.5930$ \\
\hline
1 & -0.56973 & 5.0753 \\
\hline
2 & 0.68170 & 6.3754 \\
\hline
3 & -0.77287 & 7.5582 \\
\hline
4 & 0.85117 & 8.6567 \\
\hline
\end{tabular}
\caption{ Table of the first five 
 eigenvalues and coefficients $\tilde{d}_k$, for $D=2/5$.}
\label{table1}
\end{table}

\subsection{The Bessel Meander}

As noted, the last zero crossing of the momentum process
$p(t)$  takes place at a random time
$t_n$, and hence at time $t$ the particle is unlikely to be on 
the origin of momentum. Since the particle, by definition, did not
cross the origin in $(t_n,t)$ its momentum remains either
positive or negative in the backward recurrence time
interval $\tau^{*}= t-t_n$  (with equal probability). This means that in
this time interval the motion is described by a meander, not an excursion.
The area under the  Brownian meander was investigated previously
 \cite{Majumdar2,Jason,Perman}. For our purposes
we need to investigate the Bessel meander. This is  
a Langevin path described by Eq.  
(\ref{eqBesselLang})
 constrained to remain positive, which starts at
the origin 
but is free to end with $p>0$. More specifically, the diffusive scaling of
$\chi^*\sim (\tau^*)^{3/2}$  still holds. Then as for the pairs $(\chi,\tau)$,
we have the conditional PDF
\begin{equation}
p_M(\chi^* | \tau^*) \sim D^{-1/2}  (\tau^*)^{-3/2} B_M \left( { \chi^* \over \sqrt{D} (\tau^*)^{3/2}} \right)
\label{eqMeander01}
\end{equation}
Here the subscript $M$ stands for a meander. The scaling function $B_M(.)$
is different from $B(.)$ though clearly both are symmetric, with mean equal
to $0$ (since positive and negative meanders and excursions are
equally probable).  
The calculation of $B_M(.)$ runs parallel to that for the excursion and
is presented in Appendix E.
 Soon 
we will demonstrate the importance of
the meander  for specific observables of interest. 
Having  explicit information on  $B(.)$ and $g(\tau)$, Eq. 
(\ref{eqSca01}) (see Appendix A for details) and the scaling form 
of $B_M(.)$ (see Appendix E), we can now
investigate the packet $P(x,t)$.

\section{Montroll-Weiss Equation for Fourier-Laplace transform of $P(x,t)$}
\label{Secctrw}

We now use tools developed in the random walk community 
\cite{Wong,KBS,Blumen,Carry}
to relate the
 joint PDF of a single excursion
Eqs. 
(\ref{eqSca03},\ref{eqSca04}), 
\begin{equation}
\psi({ \chi}, \tau)=g(\tau) D^{-1/2} \tau^{-3/2}B\left[ { \chi}/(D^{1/2} \tau^{3/2})\right]
\label{eqpsiBg}
\end{equation}
 to the probability density $P(x,t)$ for the entire walk. 
We find a modified  Montroll-Weiss \cite{Montroll,Bouchaud,Review} type 
of equation for the Fourier-Laplace transform of $P(x,t)$
\begin{equation}
\hat{P}(k,u) = \int_{-\infty} ^\infty {\rm d} x e^{ i k x} \int_0 ^\infty {\rm d} \tau  e^{- u t} P(x,t)
\label{eqMW01}
\end{equation} 
in terms of the Fourier-Laplace transform of $\psi(\chi,\tau)$
which will be denoted $\hat{\psi}(k,u)$. One  modification
is that we include here the correct treatment of the last jump.
Usually, the continuous time random walk (CTRW) model \cite{Review},
 has as an input a single joint PDF of jump lengths
and times, while in our case we have essentially two such functions
describing the excursions (i.e. $B(.)$) and the meander (i.e. $B_M(.)$). 
  Generally, Montroll-Weiss equations
are the starting point for derivation of fractional diffusion equations 
and the asymptotic behaviors of the underlying random walks \cite{Review}.
The original work of Montroll and Weiss \cite{Montroll}
 assumed there were no correlations
between step size  $\chi$ and and the waiting time
 $\tau$, corresponding to a situation called a decoupled
CTRW. The diffusion of atoms in optical
lattices corresponds to a coupled spatial-temporal  random walk theory
first considered by Scher and Lax \cite{Lax} (see \cite{Marcin,Bao,Akimoto}
 for recent
developments).

Define $\eta_s(x,t) {\rm d} t {\rm d} x$ as the probability that the
particle crossed the  momentum state $p=0$ for the $s$th time in
the time interval $(t,t+{\rm d} t)$ and that the particle's position
was in the interval  $(x,x + {\rm d} x)$.
This probability is related to the probability of the previous crossing 
according to
\begin{eqnarray}
\eta_s (x,t) &=& \int_{-\infty}^\infty  {\rm d} { \chi} \int_0 ^t {\rm d} \tau\, 
\eta_{s-1} \left( x - { \chi} , t - \tau \right) \times \nonumber\\
&\ &\qquad{1 \over \sqrt{D}  \tau^{3/2}} B \left( {\chi \over \sqrt{D}  \tau^{3/2} }\right) g \left( \tau \right),
\label{eq08}
\end{eqnarray} 
where we have used  Eq. (\ref{eqpsiBg}).
We change variables according to ${ \chi} = v_{3/2} D^{1/2}  \tau^{3/2}$
and obtain 
\begin{eqnarray}
\eta_s(x,t) &=& \int_{-\infty} ^\infty\!\! {\rm d} v_{3/2} \int_0 ^\infty\!\! {\rm d} \tau\, \eta_{s-1} \left( x - v_{3/2} \sqrt{D\tau^3}  , t- \tau\right) \times\nonumber\\
&\ &\qquad\qquad B\left(v_{3/2} \right)
g\left( \tau \right) . 
\label{eqMW09}
\end{eqnarray}
The process is now described by a sequence of waiting times
$\tau_1, \tau_2, \cdots$ and the corresponding generalized velocities
$v_{3/2}(1), v_{3/2}(2) , \cdots$.
The displacement in the $s$th
  interval is:  
\begin{equation}
{ \chi}_s = v_{3/2}(s) \sqrt{D} (\tau_s)^{3/2}.
\label{eqMW09a}
\end{equation}
The advantage of this representation of the problem  
in terms of the pair of  microscopic stochastic variables 
$\tau,v_{3/2}$ 
(instead of the correlated pair $\tau, { \chi}$) 
is clear from Eq. (\ref{eqMW09}): we may treat $v_{3/2}$ and $\tau$ as
independent random variables whose corresponding PDFs are
$g(\tau)$ and $B(v_{3/2})$ respectively. Here  $\tau>0$ and $-\infty<v_{3/2} <\infty$.  
The initial condition $x=0$ at time $t=0$
implies $\eta_0(x,t) = \delta(x) \delta(t)$.
Then $P(x,t)$, the probability of finding the particle in
$(x,x+ {\rm d} x)$ at time $t$, is found according to
\begin{equation} 
\begin{array}{c}
P(x,t) =\\
\sum_{s=0} ^\infty \int_{-\infty} ^\infty  {\rm d} v_{3/2} \int_0 ^t {\rm d} \tau^* \, \eta_s \left( x - v_{3/2} \sqrt{D(\tau^*)^3} , t - \tau^* \right)\\
\times  B_M\left(v_{3/2} \right) W \left( \tau^* \right).
\end{array}
\label{eqMW10}
\end{equation}
%
%
Here the 
survival probability $W(\tau^*)\equiv 1 - \int_0 ^\tau g(\tau^*) {\rm d} \tau^*$
enters  since
the last jump event took place at $t - \tau^*$ and in the time period
$(t-\tau^*,t)$ the particle did not cross the momentum origin. 
For the same reason, we have in  Eq. (\ref{eqMW10}) 
$B_M(.)$, not $B(.)$, since the last time interval in the sequence is
a meander and not an excursion.  
 The summation in Eq. (\ref{eqMW10}) is
a sum of the number  $s$ of  returns to the momentum
origin $p=0$.  
 We note that
in the analysis  the particle
is always moving unlike the ``wait and then jump" approach used
in the original CTRW model. 

As usual~\cite{Review,KBS},
we consider the problem in Laplace-Fourier space where
 $t \to u$ and $x \to k$. Using the convolution theorem
and Eq. (\ref{eqMW09})
we find 
\begin{equation}
\widehat{\eta}_s\left( k , u \right) =\widehat{ \eta}_{s-1} \left( k , u \right) \widehat{{\cal L}} \left[ \widehat{B} \left( k D^{1/2} \tau^{3/2} \right) g \left( \tau \right) \right]. 
\label{eqMW11}
\end{equation}
Here $\widehat{\eta}_s(k,u)$ is the Fourier-Laplace transform of $\eta_s(x,t)$, $\widehat{B}$ the Fourier transform of $B(v_{3/2})$,
\begin{equation}
\widehat{B} \left( k \sqrt{D\tau^3}  \right) = \int_{-\infty} ^\infty \exp\left( i k v_{3/2}\sqrt{D\tau^3}  \right) B (v_{3/2} ) {\rm d} v_{3/2},
\label{eqMW12}
\end{equation}
 and $\widehat{{\cal L}}$
is the Laplace
transform operator $\widehat{{\cal L}}[h(\tau)] = \int_0 ^\infty\exp( - u \tau) h(\tau) {\rm d} \tau$. 
By definition, the Fourier-Laplace transform of $\psi({ \chi}, \tau)$
is
$\hat{\psi}(k,u) = \widehat{{\cal L}}\left[ \widehat{B} \left( k\sqrt{D\tau^3}  \right) g(\tau) \right]$ and  hence 
\begin{equation}
\widehat{\eta}_s (k,u) = \widehat{\psi}(k,u) \widehat{\eta}_{s-1} (k,u) .
\label{eqMW13}
\end{equation} 
This implies that
\begin{equation} 
\widehat{\eta}_s(k,u) = \left[ \widehat{\psi}(k,u)\right]^s
\label{eqMW13a}
\end{equation} 
reflecting the renewal property
of the underlying random walk. 
Summing
the Fourier-Laplace transform of Eq. (\ref{eqMW10}), applying again the 
convolution theorem for Fourier and Laplace transforms and using Eq.
(\ref{eqMW13a}),  we find
a Montroll-Weiss type of equation,
the Fourier-Laplace transform of $P(x,t)$: 
\begin{equation}
\widehat{P}(k,u) = {\widehat{\Psi}_M (k,u)   \over 1 - \widehat{\psi}\left({k,u} \right)}.
\label{eqMW14}
\end{equation} 
Here
$\widehat{\Psi}_M (k,u)$ is the Fourier-Laplace
transform of $D^{-1/2} \tau^{-3/2} B_M (\chi/\sqrt{D} \tau^{3/2}) W(\tau)$.
Eq. (\ref{eqMW14}) relates statistics of velocity excursions and
meanders  to the
Fourier-Laplace transform of the particle density. 
The approach is not limited to the specific problem under investigation,
namely the $\chi\sim \tau^{3/2}$ scaling
 is not a necessary  condition for the validity
of Eq. (\ref{eqMW14}). In the general case one must revert to $\psi(\chi,\tau)=
g(\tau) p(\chi|\tau)$ instead of the scaling form captured by $B(.)$.  
Such an approach might be useful for other systems where the friction 
is non-linear. 

 At this stage,
Eq. (\ref{eqMW14})  still depends
on $\epsilon$ since the first passage times $g(\tau)$ from $p=\epsilon$
to the momentum origin $p=0$ is $\epsilon$ dependent (see Appendix A
and next Sec. for details). 
In fact as $\epsilon\to 0$ the number of renewals (i.e. zero crossings) 
tends to infinity, while in usual CTRWs,
the number of renewals (or jumps)
is finite for finite observation time $t$. 
In the next sections we will show how the long time results become
independent of $\epsilon$ in the limit of $\epsilon \to 0$.  

\section{Asymptotic behavior of $\widehat{\psi}(k,u)$}
\label{SecAsy}

 The Fourier-Laplace transform of the joint distribution of 
jump times and lengths $\psi(\chi,\tau)$ is
\begin{equation}
\widehat{\psi}\left(k,u\right) =  \int_{-\infty} ^\infty {\rm d} \chi \int_0 ^\infty {\rm d} \tau e^{ - u \tau + i k \chi} g(\tau) {1 \over \sqrt{D\tau^3}   } B\left( 
{\chi \over \sqrt{D\tau^3} }\right).
\label{eqAs01}
\end{equation}
In this section we investigate 
the small $k$ and small $u$ behaviors of $\hat{\psi}(k,u)$ which in the next
sections will turn out to be important in the
determination of the long time behavior of $P(x,t)$.
The small $k$ ($u$) limit corresponds to large distance (time)
as is well known \cite{Review}.

\subsection{$k=0$ and $u$ small} 

 Using the normalization condition
$\int_{-\infty} ^\infty B(v_{3/2}) {\rm d} v_{3/2}=1$ it is easy to verify
that 
\begin{equation}
\psi(k,u)|_{k=0} = \widehat{g}(u), 
\label{eqAs02}
\end{equation}
the Laplace transform of
the waiting time PDF.  This waiting time PDF is investigated analytically
in Appendix A. 
The small $u$ behavior of $\widehat{g}(u)$ differs
depending on the value of $D$. According to Eq. 
(\ref{eqSca01}), if $D<1$ the average  waiting
 time $\langle \tau \rangle$ is finite  and so
\begin{equation}
\widehat{g}(u) \sim 1 - \langle \tau \rangle u\ \ \ \mbox{when} \ \ \  D<1.
\label{eqAs03}
\end{equation} 
Here $\langle \tau \rangle = \int_0 ^\infty \tau g (\tau) {\rm d} \tau$
is given in terms of a well known formula for the first passage time \cite{Gardiner}.
The average waiting time for a particle starting with momentum $p_i=\epsilon$
to reach  $p=0$ for the first time is
\begin{equation}
\langle \tau \rangle = {1 \over D} \int_0 ^\epsilon {\rm d} y e^{ V(y) / D} 
\int_y ^\infty e^{ - V(z) /D} {\rm d} z,
\label{eqAs04}
\end{equation}
where $V(p)$ is the effective momentum
potential, Eq.  
(\ref{eqvp}).
Eq. (\ref{eqAs04}) reflects the absorbing boundary
condition at the origin and a reflecting boundary at infinity.
Notice that 
$\lim_{\epsilon \to 0} \langle \tau \rangle = 0 $ as it should 
and the leading order  Taylor expansion yields
\begin{equation}
\langle \tau \rangle \sim  { { \cal Z} \over 2 D} \epsilon. 
\label{eqAs05}
\end{equation} 
Here ${\cal Z}$ is the normalizing partition function, Eq. (\ref{eqZ}).
Since  ${\cal Z}$ diverges when $D \to 1$ from below,
we see that average waiting time
diverges in that limit.  

For $D>1$, the mean waiting time is infinite, and so    Eq. 
(\ref{eqAs03}) does not hold and instead, as we show in Appendix A,
\begin{equation}
\hat{g} (u) \sim 1 - g^{*} |\Gamma(-\alpha)| u^{\alpha} ; \qquad 1/2<\alpha<1,
\label{eqAs06aa}
\end{equation}
with
\begin{equation}
g^{*} = { 2 \alpha \over \left( 2 \sqrt{D} \right)^{2 \alpha} \Gamma(\alpha) }
\epsilon. 
\label{eqAs07}
\end{equation}
For large $\tau$, this yields the power-law behavior in Eq. 
(\ref{eqSca01}), $g(\tau) \sim g^{*} \tau^{-(1+\alpha)}$, which in fact
describes the tail also for $\alpha>1$.
The prefactor $g^{*}$ vanishes as $\epsilon \to 0$
and importantly does not depend on the full shape of the effective
potential $V(p)$, but rather only on the value of the dimensionless parameter
$D$.  
The case $D=1$ contains logarithmic corrections and 
 will not be discussed here. 
Using
Eqs. 
(\ref{eqAs06aa},
\ref{eqAs07})
we find  non-trivial distributions
of  the time interval straddling time $t$, the backward
and forward recurrence times,  
which were previously treated by mathematicians without the $\epsilon$
trick. This connection between renewal theory and statistics
of zero crossing of the Bessel process is  discussed in Appendix F.

\begin{widetext} 
\subsection{The $u=0$, small $k$ limit}

 Similarly to Eq.  
(\ref{eqAs02}), by definition
\begin{equation}
\widehat{\psi}(k,u)\Big|_{u=0} = \widetilde{q}(k),
\label{eqAs06}
\end{equation}
where $\widetilde{q}(k)$ is the Fourier transform of the symmetric 
jump length distribution $q(\chi)$. 
Alternatively we can use Eq. (\ref{eqAs01}) with $u=0$, and then, upon
 changing variables
according to $\chi/(\sqrt{D} \tau^{3/2})= v_{3/2}$ and using
the normalization of the scaling function $B(v_{3/2})$, we find
\begin{equation}
\widehat{\psi}(k,u)\Big|_{u=0} = 1 - \int_0 ^\infty {\rm d} \tau \int_{-\infty} ^\infty
{\rm d} v_{3/2} \left(1 - e^{i k v_{3/2} \sqrt{D\tau^3}  }  \right) g(\tau) B(v_{3/2}).
\label{eqAs07aa}
\end{equation}
Using $\widehat{B}(k)$,
the Fourier transform of $B(v_{3/2})$, Eq. (\ref{eqMW12}), we can 
rewrite Eq. (\ref{eqAs07aa}),
\begin{equation}
\widehat{\psi}(k,u)\Big|_{u=0} = 1 - \int_0 ^\infty {\rm d} \tau  \left[ 1 - \widehat{B} \left(k \sqrt{D} \tau^{3/2}\right) \right] g\left( \tau \right). 
\label{eqAs09}
\end{equation}
This expression is the starting point for a small $k$ expansion
carried out in Appendix C which gives
for $\nu<2$,  ($D > 1/5$), 
\begin{equation}
\widehat{\psi}(k,u)\Big|_{u=0} \sim 1 - g^{*} {2 \over 3} \langle \left|v_{3/2}\right|^\nu \rangle
\left(- \Gamma(-\nu)\right) \cos \left( {\nu \pi \over 2} \right) \left|\sqrt{D}k\right|^\nu.
\label{eqAs10}
\end{equation}
\end{widetext}
The non-analytical character of this expansion is responsible for the anomalous diffusion
of the atom's position. 
The first term on the right hand side is the normalization condition,
the second, $\left|\sqrt{D} k\right|^\nu$, is consistent with the fat-tailed PDF
of jump lengths $q(\chi) \propto \left|\chi\right|^{-(1 + \nu)}$, Eq.
(\ref{eqSca01}),
 namely an
excursion length whose variance diverges since $\nu<2$. 
As expected, this behavior is in full agreement with Eq.   
(\ref{eqSca01}) since $4/3 + \beta= 1 + \nu$. 
In Eq. (\ref{eqAs10}), there appears the non-integer moment
of the scaling function $B(.)$,
\begin{equation}
\langle |v_{3/2}|^\nu \rangle = \int_{-\infty} ^\infty |v_{3/2}|^\nu B(v_{3/2}) {\rm d} v_{3/2}.
\label{eqAs11} 
\end{equation} 
Given the formidable structure of the scaling function
$B(v_{3/2})$, we do  not describe here
\cite{Kesslerfut} the direct method 
to obtain
non-integer
moments like Eq. (\ref{eqAs11}). Instead, we present here a method
which gives  $\langle \left|v_{3/2}\right|^\nu \rangle$ indirectly.  
In a future publication \cite{Kesslerfut}
we will discuss this and other moments of $B(v_{3/2})$.
 
We can use Eq. 
 (\ref{eqAs06}) together with $q(\chi) \sim q_{*} |\chi|^{-(1 + \nu)}$
to find in Fourier space
\begin{equation}
\tilde{q}(k) \sim 1 - { \pi q_{*} |k|^\nu \over \sin\left({ \pi \nu \over 2} \right) \Gamma(1 + \nu) } 
\label{eqAs14} 
\end{equation} 
for $\nu<2$. 
In Appendix B, we investigate the area under a Bessel excursion
regularized at the origin using a backward Fokker-Planck equation
\cite{Zoller}  which gives the amplitude of jump
lengths 
\begin{equation}
q_{*} \sim \epsilon  {\nu \over  3^{2 \nu -1} } { \left( 3 \nu -1 \right)^\nu \over 2  \Gamma(\nu)}.
\label{eqSca06a}
\end{equation}
Comparing with Eq. (\ref{eqAs10}), we arrive at the following simple relation
\begin{equation}
\langle \left|v_{3/2}\right|^\nu \rangle = \lim_{\epsilon \to 0} {3 q_* \over D^{\nu/2}  g_*}. 
\label{eqAs15}
\end{equation}
Thus we may use
Eqs. (\ref{eqAs07},\ref{eqSca06a})
to find the desired  $\nu$th moment of the scaling function
$B(.)$, 
\begin{equation}
\langle \left|v_{3/2}\right|^\nu \rangle = { 2^{2 \alpha -1} \over 3^{2 \nu -1}} {\Gamma(\alpha) \over  \Gamma(\nu)}. 
\label{eqAs16}
\end{equation}
In the limit $D\to \infty$, we  have $\nu=1/3$ and then
\begin{equation}
\lim_{D \to \infty} \langle |v_{3/2}|^\nu \rangle = 
{\sqrt{\pi} 3^{4/3} \over 2 \Gamma(1/3)} \simeq 1.431..
\label{eqAs17}
\end{equation}
We see that while the amplitudes $g^{*}$ and $q^{*}$ vanish as the convenient
theoretical tool $\epsilon \to 0$,
$\langle |v_{3/2}|^\nu\rangle$ is independent of it in the limit,
indicating the usefulness
of this 
variable. 

\subsection{The second moment $\langle (v_{3/2})^2 \rangle$}

\begin{figure}\begin{center}
\includegraphics[width=0.41\textwidth]{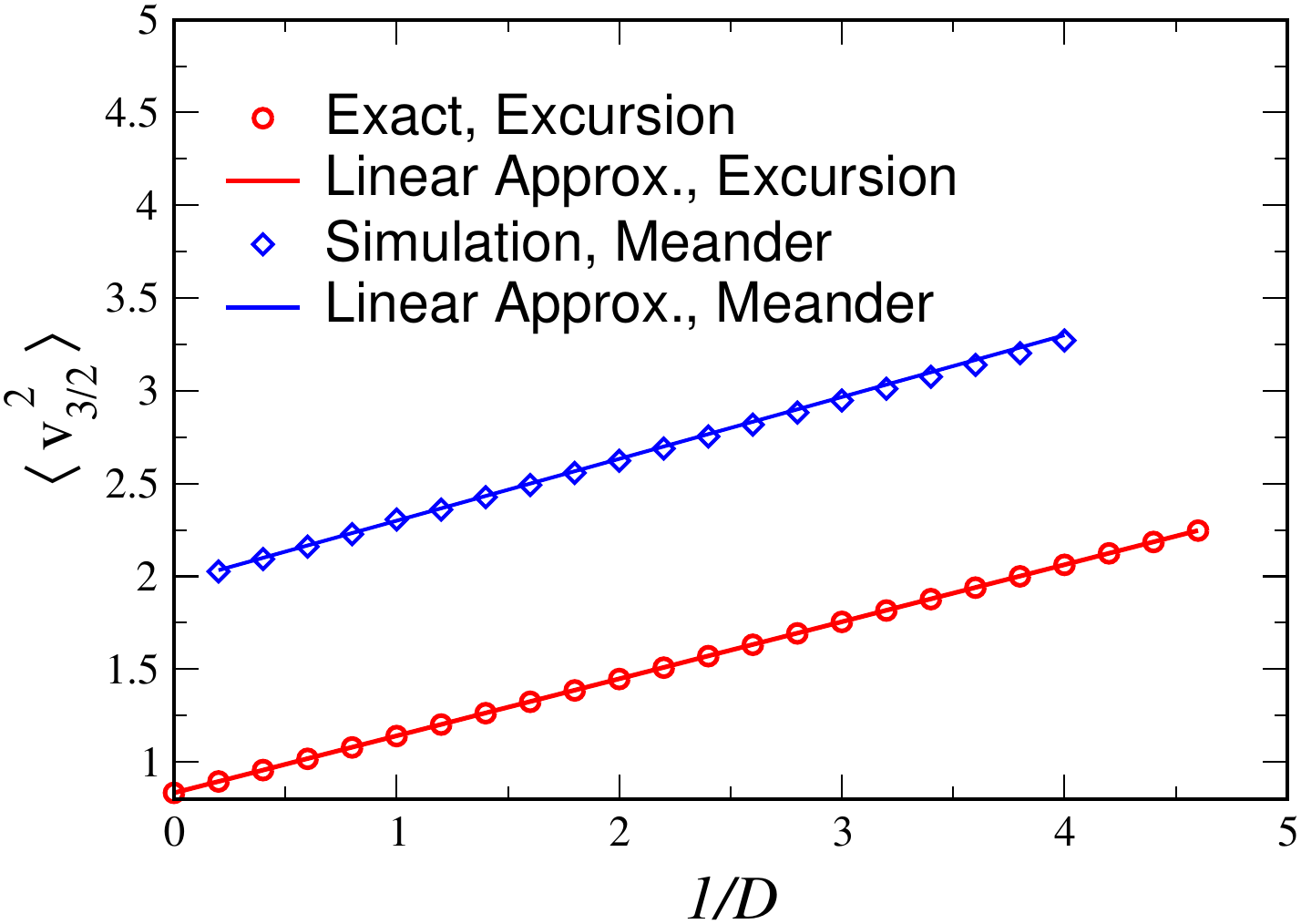}
\end{center}
\caption{
The variance of the area under the  Bessel excursion and meander
$\langle (v_{3/2})^2\rangle$ and $\langle (v_{3/2})^2 \rangle_M$
versus $1/D$. Exact solution for the excursion
\cite{Kesslerfut}
is well approximated by 
Eq. (\ref{eqSecMom}) and simulation for the meander with
Eq. (\ref{eqerez18}) (averages over $2\times 10^5$ particles
and $\tau=10^5$). 
}
\label{figSec}
\end{figure}

As we show below the second moment $\langle (v_{3/2})^2 \rangle$ determines
the mean square displacement of the particles.
The mean vanishes since in the underlying
random walk positive and negative excursions are equally likely.
As for the $\nu$th moment, $\langle |v_{3/2}|^\nu \rangle$, the extraction
of integer moments from the exact solution, is not straightforward.
In the regime $1/5 < D<\infty$, an excellent approximation is
\begin{equation}
\langle (v_{3/2})^2 \rangle \simeq { 5 \over 6} + {83 \over 270 D}. 
\label{eqSecMom}
\end{equation}
Similarly, for the meander  we find 
\begin{equation}
\left\langle (v_{3/2})^{2}\right\rangle _{M}\simeq {1 \over 3 D} + {59 \over 30}.
\label{eqerez18}
\end{equation}
This and Eq. (\ref{eqSecMom}) agrees with known results in the Brownian
limit $D\to \infty$ \cite{Majumdar2}  
(units and notations used in \cite{Majumdar2} are not those used
by us).
In Fig. \ref{figSec} we show that
 $\langle (v_{3/2})^2 \rangle$  
and  $\langle (v_{3/2})^2 \rangle_M$, 
nicely match  their linear approximations 
\cite{Kesslerfut}. 

\section{The L\'evy phase}
\label{SecLev}

 We now explain why L\'evy statistics, and hence the generalized
central limit theorem, describes the central part of the diffusion profile
$P(x,t)$ at long times for $1/5<D<1$.  
The L\'evy profile is cut off in the tails of the distribution,
due to the correlations between jump length and time investigated here.
We focus on the L\'evy phase first, because it  has been reported on
experimentally \cite{Sagi}. 

 The key idea is that in the regime $D<1$, $\langle \tau \rangle$ is 
finite 
and hence 
the number of jumps $n$ scales with  $t/\langle \tau \rangle$  
when $t$ is large
\cite{GL}.
At the same time 
the jump lengths PDF still does not have a variance (since $1/5<D$)
which means that the usual Gaussian
central limit theorem does not hold. Instead, due
to the power-law distribution $q(\chi) \propto |\chi|^{-(1+\nu)}$, the process
belongs to the domain of attraction of a L\'evy stable law.
 As long as $x$ is not too large,
 the correlations are not important. 
However when $x \propto t^{3/2}$ the
simple L\'evy picture breaks down, since clearly we cannot perform a jump
larger than the order of $t^{3/2}$. Thus for $1/5<D<1$, 
and $t^{-3/2}\ll k \ll 1$,  we can approximate
\begin{equation}
\widehat{\psi}(k,u) \simeq 1 - u \langle \tau \rangle + c_0 q_{*} |k|^\nu ...
\label{eqLevy01} 
\end{equation}
where we have used Eqs. 
(\ref{eqAs03},
\ref{eqAs14}). Here from Eq. (\ref{eqAs14})  $c_0=\pi/[ \sin(\pi \nu/2) \Gamma(1+\nu)]$. Eq. (\ref{eqLevy01}) corresponds to a decoupling scheme,
 $\widehat{\psi}(k,u) \simeq \hat{g}(u) \tilde{q}(k)$, 
 which
according to   
 arguments in
\cite{Blumen}
is exact in the long time limit in the regime under investigation.
Notice that $1/5<D<1$ gives $2/3<\nu<2$.

Using the Montroll-Weiss type Eq. (\ref{eqMW14}) and
$\tilde{\Psi}_M (k,u)  \sim \langle \tau \rangle$,
which is easy to prove, we find the Fourier-Laplace
representation of the solution
\begin{equation}
\widehat{P}(k,u) \sim { \langle \tau \rangle \over u \langle \tau \rangle + c_0 q_{*} |k|^{\nu} }.
\label{eqLevy02} 
\end{equation}
As mentioned, both $\langle \tau \rangle$ and $q_{*}$ vanish as $\epsilon$
approaches  zero. Rearranging,
we have
\begin{equation}
\widehat{P}(k,u) \sim { 1 \over u + K_\nu |k|^\nu},
\label{eqLevy03} 
\end{equation}
where 
\begin{equation}
K_\nu = \lim_{\epsilon \to 0} { c_0 q_{*} \over \langle \tau \rangle}
\label{eqLevy04m1} 
\end{equation} 
and from Eqs. 
(\ref{eqAs05},
\ref{eqSca06a}),
\cite{KesBarPRL}
\begin{equation}
K_\nu = \lim_{\epsilon \to 0} { c_0 q_{*} \over \langle \tau \rangle}
= {1 \over {\cal Z} } { \pi \over \sin\left( { \pi \nu \over 2} \right)}
{ \left( 3 \nu - 1\right)^{\nu-1} \over 3^{ 2 \nu -1} |\Gamma\left(\nu\right)|^2}. 
\label{eqLevy04} 
\end{equation} 
$K_\nu$ is called the anomalous diffusion coefficient. When returning
to physical units,  
we get
\begin{equation}
\tilde{K}_{\nu} = 
{p_c ^\nu \over  m^\nu \overline{\alpha}^{\nu -1} } K_\nu
\label{eqKesF}
\end{equation}
which has units $\mbox{cm}^\nu/\mbox{sec}$. 
An equivalent expression is 
$K_\nu = c_0 \langle |v_{3/2}|^{\nu} \rangle D^{\nu/2}  \lim_{\epsilon \to 0} g^{*} / 3 \langle \tau \rangle$. 
$\widehat{P}(k,u)$  as given in Eq. (\ref{eqLevy03}), is in fact precisely
the symmetric L\'evy PDF in Laplace-Fourier space, whose $(x,t)$ presentation
(see Eq. (B17) of \cite{Bouchaud}) is
\begin{equation}
P(x,t) \sim {1 \over \left( K_\nu t \right)^{1/\nu}} L_{\nu,0} \left[ 
{ x \over \left(K_\nu t \right)^{1/\nu} } \right] 
\label{eqLevy05} 
\end{equation} 
for $2/3 < \nu < 2$. 
The properties of the L\'evy function $L_{\nu,0}(.)$ are well known. 
The Fourier transform of this solution is $\exp(- K_\nu t |k|^\nu)$
for $2/3<\nu<2$ which can serve as the working definition of the solution,
via  the inverse Fourier transform. 

\begin{figure}\begin{center}
\includegraphics[width=0.41\textwidth]{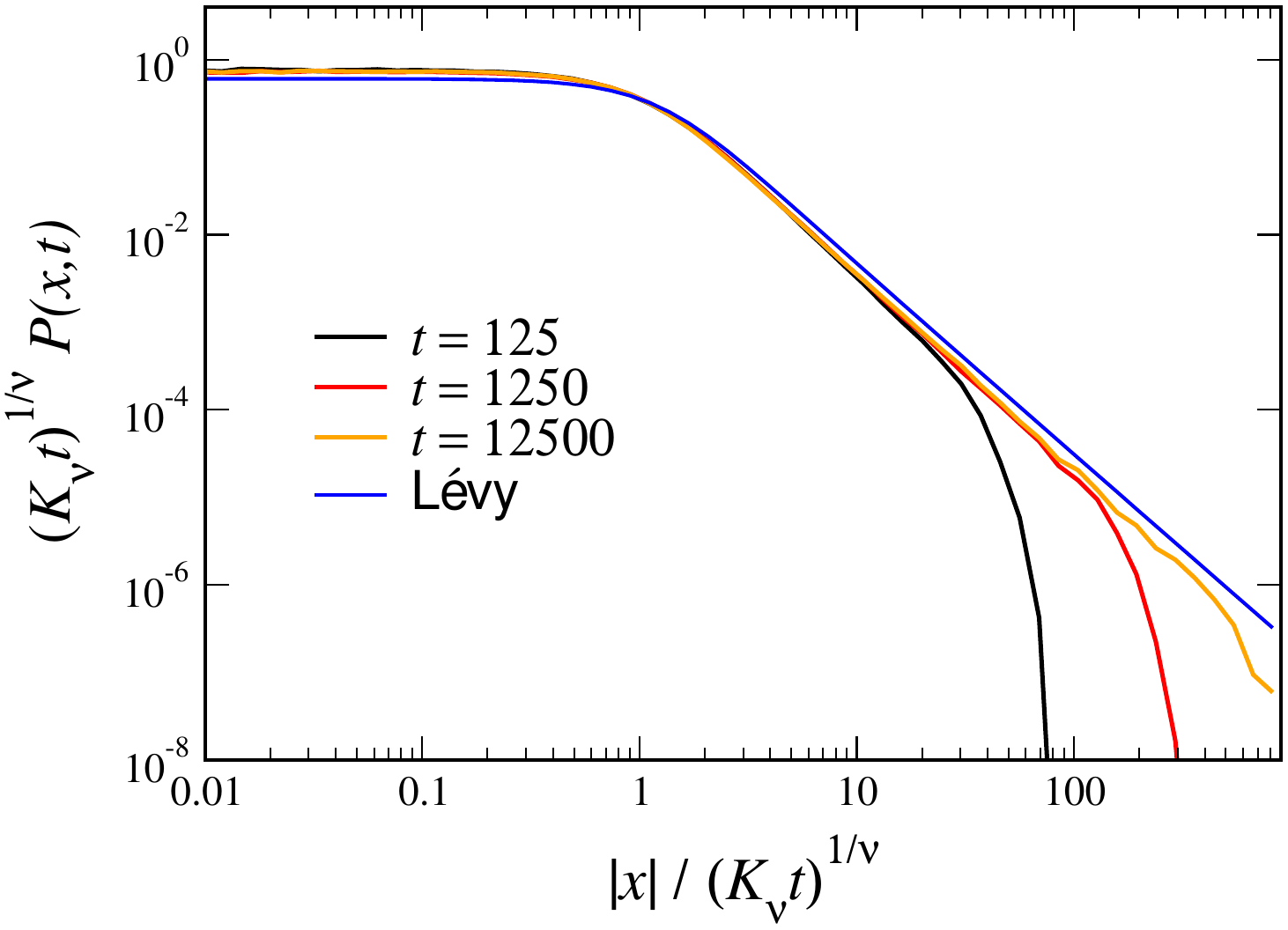}
\end{center}
\caption{
$(K_\nu t)^{1/\nu} P(x,t)$ versus $x/(K_\nu t)^{1/\nu}$ for $\nu=7/6$  $(D=2/5)$. The theory:
L\'evy PDF Eq. 
(\ref{eqLevy05}) 
 with $K_\nu$ Eq. 
(\ref{eqLevy04}), 
perfectly matches 
simulations without fitting. 
Notice the cutoff for large $x$ which is due to the coupling between jump
lengths and waiting times, making $x>t^{3/2}$ unphysical. 
}
\label{fig3}
\end{figure}

 Fig. \ref{fig3}  shows excellent agreement between
simulations and the theory. It 
also illustrates the cut off on L\'evy statistics
which is found at distances $x \propto t^{3/2}$. Beyond this length scale,
the density falls off rapidly. This, as noted above,
is the result of the correlation between $\chi$ and $\tau$, as there 
is essentially no weight associated with
 paths whose displacement is greater than the order of
$t^{3/2}$. 

\begin{figure}\begin{center}
\includegraphics[width=0.41\textwidth]{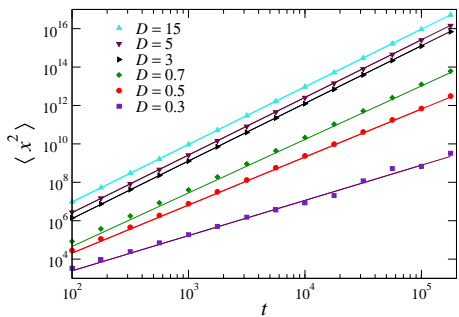}
\end{center}
\caption{
 Simulations and theory for the mean square displacement versus time 
nicely match without fitting. 
The $D>1$  Obukhov phase  yields  $\langle x^2 \rangle \propto t^{3}$
scaling,
hence for $D=15,5,3$ the curves are parallel.
In the L\'evy phase $1/5<D<1$,
 diffusion is anomalous with an exponent which
is $D$ and hence $U_0$ dependent.
 For $D=0.3$, we see relatively large fluctuations, possibly due
to the nearby transition to the Gaussian phase. 
Here we used a Langevin simulation, with a time step $dt=0.1$,
and  averaged over $10^5$ particles.
Theoretical lines are based on Eqs. 
(\ref{eqMSD07},
\ref{eqMSD09}) and 
the excursion and meander areal variance 
Eqs. (\ref{eqSecMom},\ref{eqerez18}).
}
\label{figMSDvsT}
\end{figure}

This cutoff ensures the finiteness of the mean square displacement,
see Fig. \ref{figMSDvsT}.
Using the power-law tail of the L\'evy PDF $L_{\nu} (x) \propto x^{-(1+\nu)}$
and the time scaling of the  cutoff we get: 
\begin{equation}
\langle x^2 \rangle \simeq \int^{t^{3/2}} t^{-(1/\nu)} (x/t^{1/\nu})^{-(1 + \nu)}
x^2 {\rm d} x \propto t^{4- 3 \nu/2}
\label{EQhand}
\end{equation}
for $2/3<\nu < 2$ ($1/5<D<1$). If we were to rely only 
on the L\'evy PDF, Eq. (\ref{eqLevy05}),
we would get $\langle x^2 \rangle = \infty$. Thus the L\'evy PDF solution 
must 
be treated with care, realizing its limitations in the statistical description
of the moments of the distribution and its tails. As we soon show, 
if $\nu>2$ we get
normal diffusion $\langle x^2 \rangle \propto t$ while for $\nu<2/3$ we have 
$\langle x^2 \rangle \propto t^{3}$. This last behavior is due to the correlations,
which restrict jumps longer than  $t^{3/2}$. So, we have three phases
of motion \cite{KesBarPRL}:
\begin{equation}
\langle x^2 \rangle \propto 
\left\{
\begin{array}{c c c}
t \ \ \  & 2<\nu & \ \  \mbox{Normal} \\
t^{ 4 - 3 \nu/2} \ \ \  & 2/3 < \nu < 2 & \ \ \mbox{L\'evy}  \\
t^{ 3} \ \ \  & 1/3< \nu < 2/3 & \ \ \mbox{Obukhov-Richardson}.
\end{array}
\right. 
\label{eqexponents}
\end{equation}
These simple scaling arguments for the mean square displacement
can be derived from a more rigorous first-principle approach, to
which we will turn in Sec. \ref{SecMSD}.  

\subsection{ The diffusion exponent}

Wickenbrock, et al. \cite{Wickenbrock}
 investigated the additional  effect of a high frequency (HF) oscillating
force $F_{\rm HF} = A_{\rm HF} \sin\left(\omega_{\rm HF} t +\phi_0\right)$
 on the dynamics of the atoms, where the frequency $\omega_{\rm HF}$
 is much larger than other frequencies in the system. 
According to \cite{Wickenbrock}
 in the limit of a strong drive,
the depth of the optical lattice potential is renormalized $U_0 \rightarrow
U_0 |J_0 (2 k r)|$ where $J_0(.)$ is the Bessel function of the first kind,
$r=A_{\rm HF}/(m\omega_{HF} ^2)$ and $k$ is the
laser field wave vector. This elegant set-up  allows the control
of the transport via the renormalization
of the optical depth $U_0$ and according to Eq. (\ref{eqDc})
 the control of the dimensionless parameter $D$. 
For example, in the vicinity of the zeroes
of the Bessel function $J_0$ we clearly  find an effective  shallow lattice,
which according to the theory corresponds to the Richardson phase.   
In the experiment \cite{Wickenbrock}
resonances in the transport are observed close to the
zeros of Bessel functions, namely an enhanced spreading of the atoms.
However, as pointed out in \cite{Wickenbrock} many 
 super-diffusing atoms are lost, which leads to an underestimate of the diffusion
exponent. In \cite{Wickenbrock}
the diffusion exponent was found  using
Monte-Carlo simulation (see Fig. 1 there). To demonstrate
the predictive power of our theory, at least for exponents,
 we compare  between
theory and the  numerics \cite{Wickenbrock}. As mentioned in  our
summary
we  postpone a comparison  of experiments
to theory until the losses become insignificant.

\begin{figure}\begin{center}
\includegraphics[width=0.41\textwidth]{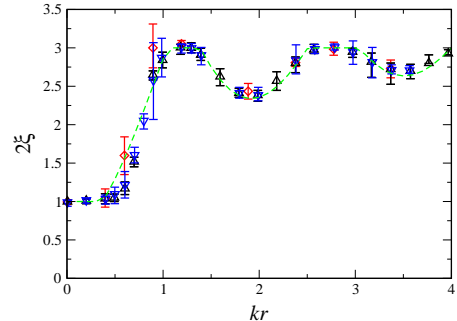}\end{center}
\caption{
The anomalous diffusion exponent $\langle x^2 \rangle \sim t^{2 \xi}$
 versus the high frequency control parameter $kr$ defined in the text. 
The triangles and diamonds
are simulation results from Fig. 1 in  \cite{Wickenbrock}  with $U_0=200 E_r$
(with varying photon scattering rate, see details in \cite{Wickenbrock}).
 The curve is 
the theory Eqs. (\ref{eqDc}, 
 \ref{eqexponents}) with a renormalized $U_0$ as explained in the text 
with $c=34.7$. The transition between normal $2 \xi=1$,  L\'evy
$ 1<2 \xi <3$ and Richardson $2 \xi=3$ behaviors  is clearly visible,
and as shown by Wickenbrock et al. controlled by the high frequency
field. }
\label{figRenzoni}
\end{figure}

 From Eq. (\ref{eqexponents}) and
within the L\'evy phase we find superdiffusion and
$\langle x^2 \rangle \propto t^{2 \xi}$ with
$2 \xi= 4 - 3 \nu/2= 7/2 -1/(2D)$ for $1/5 < D<1$ so $1<2 \xi <3$.
For $D>1$ we find  $\langle x^2 \rangle \propto t^3$  so $2 \xi=3$
and normal diffusion with $2 \xi=1$
when $D<1/5$. With a renormalized Eq. (\ref{eqDc})  $D= c E_R / [ U_0 |J_0( 2 kr)|]|$ 
we  present in Fig. \ref{figRenzoni} the mean square displacement exponent
 versus the control parameter
$kr$,  
with nice  agreement between theory and simulations.

\section{The mean square displacement}
\label{SecMSD}

Here we present the calculation of the mean square displacement of the atoms
using
the Montroll-Weiss equation. 
Our aim, as declared in the Introduction, is to unravel the 
quantitative connections between the transport and the statistics
of excursions, for example the relation between the mean square displacement
and moments of the area under the Bessel excursion and meander.  
Such relations are expected to be general beyond the model under investigation. 
A different
strategy for the  calculation 
can be based on a super-aging 
velocity-velocity correlation function  approach
\cite{Dechantprl,DechantPNAS}.

To derive Eq. (\ref{eqLevy05}), we assumed in Eq. 
(\ref{eqLevy03}) that $u$ and $K_\nu |k|^\nu$ are of the same order
of magnitude. We now consider a different  small $k,u$ limit of $\hat{P}(k,u)$.
We first expand the numerator and denominator of $\hat{P}(k,u)$ in the 
small parameter $k$ to second order. We leave $u$ fixed and
find 
\begin{equation}
\hat{P}(k,u) \sim {1 \over u} { 1 - {1 \over 2} D \langle (v_{3/2})^2 \rangle_M \hat{f}_1(u) k^2  \over
1 + {1 \over 2} D \langle (v_{3/2})^2 \rangle \hat{f}_2 (u) k^2 }. 
\label{eqMSD01}
\end{equation}
This second order expansion contains the second order
moment of the scaling function $\langle (v_{3/2} )^2 \rangle =
\int_{-\infty} ^\infty (v_{3/2} )^2 B(v_{3/2}) {\rm d} v_{3/2}$
and similarly for $\langle (v_{3/2})^2 \rangle_M$.
Thus the numerator (denominator) in (\ref{eqMSD01}) contains a term 
describing the meander (excursions) contribution, respectively. 
From symmetry $\langle v_{3/2} \rangle =0$ hence the expansion does
not contain linear terms. Here 
\begin{equation}
\hat{f}_1(u) = -{1 \over \hat{W}(u)} { {\rm d}^3 \over {\rm d} u^3} \hat{W}(u) 
\label{eqMSD02}
\end{equation}
where $\hat{W}(u) = [1 - \hat{g}(u)]/u$ is the Laplace transform of the
survival probability $W(\tau)$ and
\begin{equation}
\hat{f}_2 (u) = - { {{\rm d}^3 \over {\rm d}u^3 } \hat{g}(u) \over 1 - \hat{g} (u) } . 
\label{eqMSD03}
\end{equation}
The third order derivative with respect to $u$
is clearly related to the $\chi \propto \tau^{3/2}$ scaling we have
found and to the second order expansion. While the
$k^2$ expansion in Eq. (\ref{eqMSD01}) works fine for small
$k$ and finite $u$, when $u\to 0$ we get divergences.
For example when $\alpha<1$ we have 
$\hat{g}(u) \sim 1 - G u^\alpha+\cdots$ and
hence the third order derivative of $\hat{g}(u)$ diverges as $u \to 0$.
In fact it is easy to see that
$\hat{f}_2(u)$ will diverge when $u\to 0$ when $g(\tau) \propto \tau^{-(1 + \alpha)}$ and $\alpha<3$. Thus  $\alpha=3$ marks a transition from
anomalous diffusion to normal which is consistent with
what we found in the previous section since
$\alpha = 3$ gives $D=1/5$ and hence $\nu=2$. Of course, the $k^2$ behavior
in Eq. (\ref{eqMSD03}) is very different from the non-analytical
$|k|^\nu$ found in Eq.
(\ref{eqAs10}). 
 This indicates that the order
of taking the limits $k \to 0$ and $u \to 0$ is non-commuting \cite{KBS}. 

The Laplace transform of the mean square displacement of the
atoms is given by
\begin{equation}
\langle \hat{x}^2 (u) \rangle = - {{\rm d}^2 \hat{P}(k,u) \over  {\rm d} k^2}\Big|_{k=0} . 
\label{eqMSD04}
\end{equation}
Hence for particles starting on the origin one can easily see
\begin{equation}
\langle \hat{x}^2 (u) \rangle = \lim_{\epsilon \to 0} {D \over u}  \left[
 \langle (v_{3/2})^2 \rangle_M  \hat{f}_1 (u) +  
 \langle (v_{3/2})^2 \rangle  \hat{f}_2 (u) \right]  . 
\label{eqMSD05}
\end{equation}
The second moment is finite due to the observed fast decay of the scaling 
function $B(v_{3/2})$ for $v_{3/2} \gg 1$ ensuring that
$\langle (v_{3/2})^2 \rangle$
and $\langle (v_{3/2})^2 \rangle_M$ are both finite.

\subsection{Obukhov-Richardson diffusion} 

\begin{figure}\begin{center}
\includegraphics[width=0.41\textwidth]{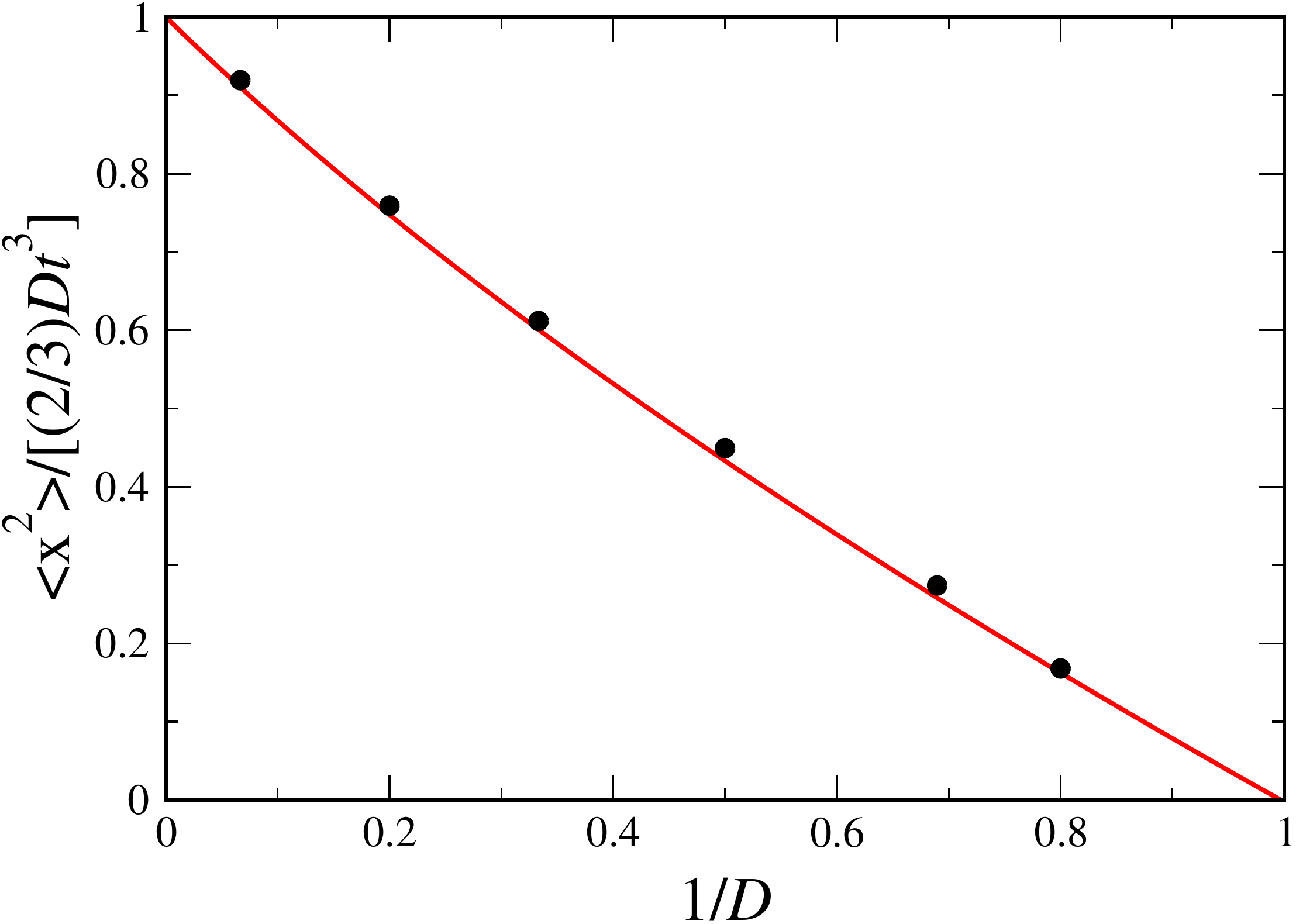}
\end{center}
\caption{
Scaled mean square displacement for the Richardson-Obukhov phase.
Simulations and theory nicely match without fitting. Close to the transition
to the L\'evy phase, i.e., $D \rightarrow 1$, 
 the finite time simulations  slowly converge
to the asymptotic limit, as might be expected. 
In the  simulations we used $t=10^5$ and averaged over $10^5$ particles.}
\label{figMSDOR}
\end{figure}

When $\alpha<1$ ($D>1$), we have $\hat{g}(u) \sim 1 - G u^\alpha + \cdots$ where 
$G=g^{*}|\Gamma(-\alpha)|$. 
Using Eqs. (\ref{eqMSD02},\ref{eqMSD03}),  $\hat{f}_1(u) \sim c_1 u^{-3}$ and
$\hat{f}_2 \sim c_2 u^{-3}$ for small $u$ with 
\begin{eqnarray}
c_1 &=& (1-\alpha)(2-\alpha)(3-\alpha) \nonumber \\ 
c_2 &=& \alpha(1-\alpha)(2-\alpha).
\label{eqc1c2}
\end{eqnarray}
 The small $k,u$ 
expansion of $\hat{P}(k,u)$, Eq. 
(\ref{eqMSD01}), is 
\begin{equation}
\hat{P}(k,u) \sim { 1 \over u} { 1 - {1 \over 2} D \langle (v_{3/2})^2 \rangle_M c_1 u^{-3} k^2 \over 1 + {1 \over 2} D \langle (v_{3/2})^2 \rangle c_2  u^{-3} k^2} 
\label{eqMSD06}
\end{equation}
which is $g^{*}$ and $\epsilon$  independent.
The mean-square displacement in the small $u$ limit is
\begin{equation} 
\langle \hat{x}^2 (u) \rangle \sim D \left[c_1 \langle (v_{3/2})^2 \rangle_M  + c_2\langle (v_{3/2})^2 \rangle\right]  u^{-4} .
\label{eqMSD07u}
\end{equation}
Converting to the time domain 
\begin{equation} 
\langle x^2 (t) \rangle \sim  \left[c_1 \langle (v_{3/2})^2 \rangle_M +
c_2 \langle (v_{3/2})^2 \rangle\right]  {D t^3 \over 6}.
\label{eqMSD07}
\end{equation}
The scaling $x^2 \propto t^3$ in this $\alpha<1$ regime
 is similar to Richardson's
observations concerning
the relative diffusion of a pair of particles in turbulence
(see discussion below). We see that in this regime, both the meander
and the excursion contribute to the computation 
of the mean square displacement. 
The theory agrees with the finite time simulations 
presented in Fig. \ref{figMSDOR}, where we see that
$\langle x^2 \rangle/t^3$ approaches zero as $D\rightarrow 1$
 in the long time limit.

 We can verify Eq. (\ref{eqMSD07}) in the limit of large $D$,
where the friction force is negligible. In that case, 
Eq. (\ref{eq05}) with $F(p)=0$ easily gives $\langle x^2 \rangle = 2 D t^3 /3$.
On the other hand we have for Brownian excursions and
meander $\langle (v_{3/2})^2\rangle= 5/6$ and $\langle (v_{3/2})^2 \rangle_M = 59/30$, Eqs. 
(\ref{eqSecMom},\ref{eqerez18})
 \cite{Majumdar2}. Using $\lim_{D\to \infty} \alpha = 1/2$, we have
in this limit $c_1= 15/8$ and $c_2=3/8$. Plugging these numbers in Eq.  
(\ref{eqMSD07}) we get $\langle x^2 \rangle = 2 D t^3 /3$, as it should.
This simple demonstration, implies that it is essential to treat the
last jump event properly (as a meander) and previous CTRW approaches
relying on a unique jump length distribution, lead to wrong conclusions
(e.g. replacing $\langle (v_{3/2})^2 \rangle_M$ with $\langle (v_{3/2})^2 \rangle$ is wrong). Furthermore, in this limit the contribution of the meander
is numerically larger then the contribution of the excursions,
 even though the number
of excursions is large.  Notice however that for the calculation of the L\'evy
density $P(x,t)$,
Eq. (\ref{eqLevy05}), 
 the statistics of the meander did not enter. Hence
depending on the observable of interest, and the value of $D$ the meander
may be either a relevant part of the theory or not. 


\subsection{Super-diffusion} 

\begin{figure}\begin{center}
\includegraphics[width=0.41\textwidth]{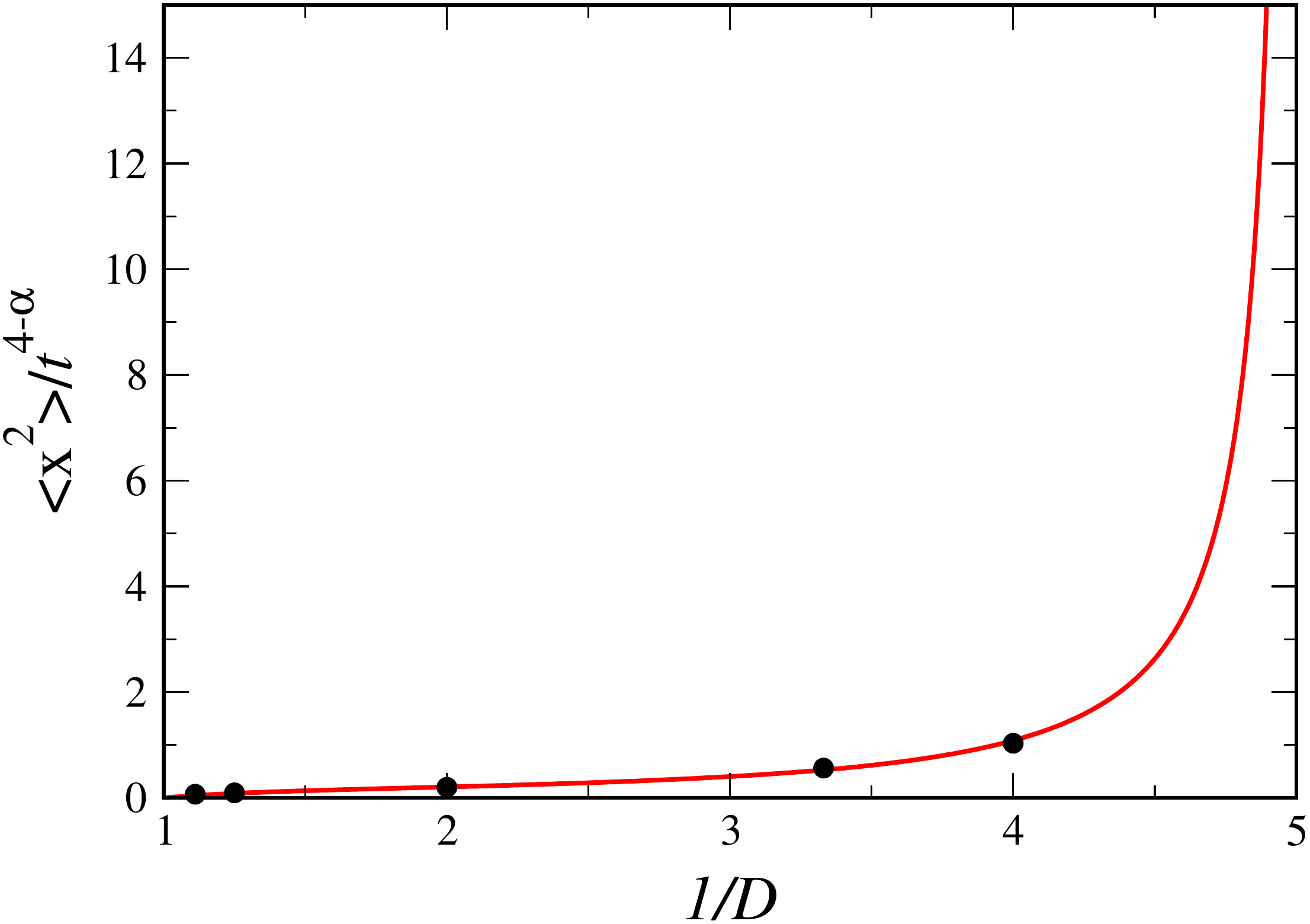}
\end{center}
\caption{
Scaled mean square displacement in the L\'evy phase.
 In the  simulations we used $t=10^5$ and averaged over $10^5$ particles,
to reduce fluctuations in the vicinity of the transition to the Gaussian phase
we averaged over trajectory times in the range $10^3-10^5$ (last $2$ points).}
\label{figMSDSD}
\end{figure}

 For $1<\alpha<2$ ($1/3<D<1$) we use the expansion 
\begin{equation}
\hat{g}(u) = 1 - u \langle \tau \rangle + G u^\alpha + \cdots. 
\label{eqgu}
\end{equation}
Here the first moment of the waiting time is
finite while the second moment diverges. Crucially
 both $\langle \tau \rangle$ and
$G=|g^{*}\Gamma(-\alpha)|$ vanish as $\epsilon \to 0$. For this parameter regime,
 we have found L\'evy behavior for 
 the central part of $P(x,t)$. 
Using Eq. (\ref{eqMSD05}) we find
\begin{equation}
\langle \hat{x}^2 (u) \rangle \sim D \lim_{\epsilon \to 0} \left[
|c_1| \langle (v_{3/2})^2 \rangle_M + |c_2| \langle (v_{3/2})^2 \rangle \right]  { G \over \langle \tau \rangle} u^{\alpha - 5}.
\label{eqMSD08}
\end{equation}
Using Eqs.
(\ref{eqAs05},\ref{eqAs07})
\begin{equation}
 \lim_{\epsilon \to 0} {G\over \langle \tau \rangle}  =  {2 D \over {\cal Z}} { 2 \alpha \over (2 \sqrt{D})^{2\alpha} } { | \Gamma(-\alpha) | \over \Gamma(\alpha)}.
\label{eqMSD09a}
\end{equation}
The inversion of Eq. (\ref{eqMSD08}) to the time 
domain and inserting 
$D= (2 \alpha -1)^{-1}$ 
gives
$$ \langle x^2 (t) \rangle \sim  $$
\begin{equation}
  {\alpha 4^{1-\alpha}  
   | \Gamma(-\alpha) | 
\left[ |c_1| \langle (v_{3/2})^2 \rangle_M + 
|c_2| \langle (v_{3/2})^2 \rangle \right] 
\over {\cal Z} (2 \alpha -1)^{2-\alpha}
\Gamma(5-\alpha) \Gamma(\alpha)}
t^{4-\alpha}.
\label{eqMSD09}
\end{equation}
The same result is valid for $2<\alpha<3$. 
This behavior depends on the normalizing partition
function ${\cal Z}$, namely 
the shape of the potential $V(p)$ in the vicinity of
the momentum origin becomes important, unlike the Richardson-Obukhov phase,
where only the large $p$ behavior of $V(p)$ is important in the long time 
limit 
(i.e., the  case $\alpha<1$, Eq. 
(\ref{eqMSD07})).
Notice that $|c_1|$ in Eq. (\ref{eqMSD09}) tends to zero when $\alpha \to 3$ from
below. This means that the meander becomes irrelevant when 
we approach the normal diffusion phase $x^2 \sim t$. 
Fig. \ref{figMSDSD} compares simulation and theory and demonstrates that
$\langle x^2 \rangle/t^{4-\alpha}$ diverges as the transition to the
Gaussian phase $D<1/5$ is approached. 

\subsection{Breakdown of scaling assumption- Normal Diffusion}

 Our starting point was the scaling hypothesis $\chi^2 \sim \tau^3$,
Eq. 
(\ref{eqSca04}) and Fig. \ref{fig2}. 
Indeed we have shown that for large $\chi$ and $\tau$
the scaling function $B(.)$ describes the conditional probability 
density $p(\chi|\tau)$ for all values of $D$. However this does not
imply that the scaling solution $B(.)$ is always relevant for the calculation
of the particle density $P(x,t)$. 
 Roughly speaking, so far we have assumed that 
large jumps $\chi$ and long waiting
times $\tau$ dominate the underlying process.
No one guarantees, however,  that this large 
$\chi$ limit (long jumps) is important while small $\chi$ (small jumps)
 can be neglected.
Indeed when we switch over to the normal diffusion phase $D<1/5$,
 the small scale aspects
of the process become important (meaning the regularization of the potential
$V(p)$ when $p \to 0$  becomes crucial). This is similar to
the L\'evy versus  Gaussian central limit theorem arguments,
where the former is controlled
by the tails of the distribution while the latter by the variance. 
Let us see this breakdown of scaling in more detail.

\begin{figure}\begin{center}
\includegraphics[width=0.41\textwidth]{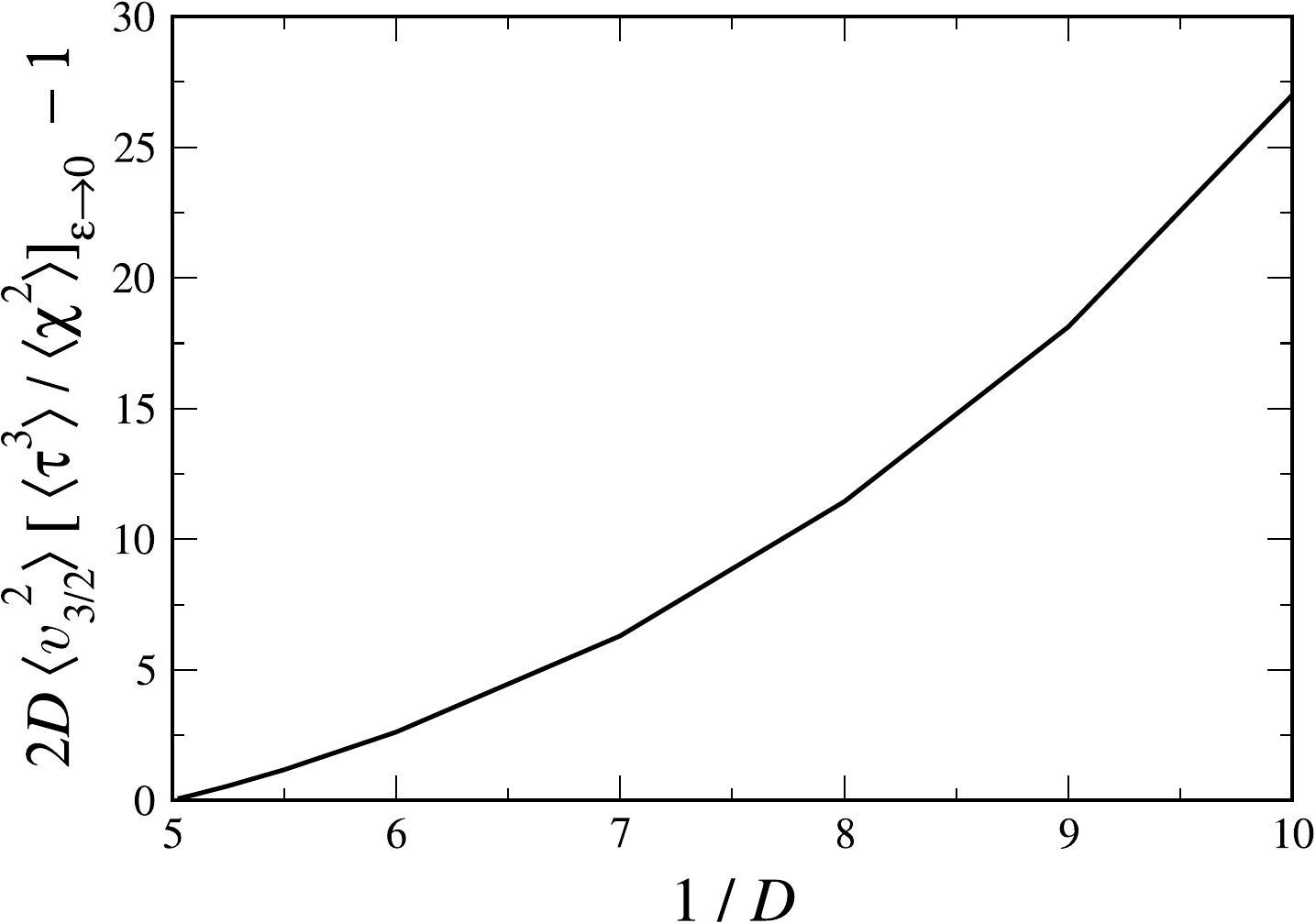}\end{center}
\caption{
We show $2 D \lim_{\epsilon \to 0}
\langle (v_{3/2})^2 \rangle / \langle \tau^3 \rangle -1 $ versus $1/D$ which
is zero if scaling holds, and $D<1/5$. 
We see that scaling holds only 
when $D \to 1/5$ which marks the transition from 
normal to L\'evy type of diffusion. For $D>1/5$
scaling holds since  then $\langle \chi^2\rangle$ and 
$\langle \tau^3 \rangle$ diverge. 
}
\label{figdev}
\end{figure}

When $\alpha > 3$ we have the expansion $\hat{g} (u) \sim 1 - u \langle \tau \rangle + u^2 \langle \tau^2 \rangle /2 - u^3 \langle \tau^3 \rangle / 6 + \cdots$
where the first three integer moments of the waiting time PDF
are finite (see Appendix A). 
Then the function
$\hat{f}_1(u)$ is negligible when $u \to 0$ while $\hat{f}_2(u) \sim \langle \tau^3 \rangle / u \langle \tau \rangle$. We find
\begin{equation}
\langle x^2 \rangle \sim D \langle (v_{3/2})^2\rangle \lim_{\epsilon \to 0} 
{ \langle \tau^3 \rangle \over \langle \tau \rangle} t.
\label{eqMSD10}
\end{equation}
 Thus $\alpha=3$ (or $D=1/5$)  marks the transition
between the anomalous super-diffusive phase and the normal diffusion phase.
Furthermore, in this case the meander is of no importance. 

However, Eq. (\ref{eqMSD10}) is correct  only if our assumptions
about  scaling are valid. 
By definition
\begin{equation}
\langle \chi^2 \rangle =\int_0 ^\infty \int_{-\infty} ^\infty \psi(\chi,\tau) 
\chi^2 {\rm d} \chi {\rm d} \tau ,
\label{eqMSD10a}
\end{equation}
and if the scaling
hypothesis holds 
\begin{equation}
\langle \chi^2 \rangle =\int_0 ^\infty \int_{-\infty} ^\infty g(\tau) { B \left( {\chi/ \sqrt{D} \tau^{3/2} } \right) \over \sqrt{D} \tau^{3/2}} 
\chi^2 {\rm d} \chi {\rm d} \tau ,
\label{eqMSD10c}
\end{equation}
which gives
\begin{equation}
\langle \chi^2 \rangle = D \langle (v_{3/2})^2 \rangle \langle \tau^3 \rangle.
\label{eqMSD10b}
\end{equation}
Inserting 
Eq. (\ref{eqMSD10b}) in 
(\ref{eqMSD10}) 
we get the expected result, discussed in the introduction
\begin{equation}
\langle x^2 \rangle \sim \lim_{\epsilon\to  0} {\langle \chi^2 \rangle \over \langle \tau \rangle} t .
\label{eqMSD10ba}
\end{equation}
This simple result is the correct one, 
even though the scaling assumption is not valid in
this regime. Sometimes we reach truthful
conclusions, even though the
assumptions on the way are invalid. To see the breakdown of scaling we
plot in Fig. \ref{figdev}  $Y_{dev}= \lim_{\epsilon \to 0} \langle \chi^2 \rangle / [ D \langle (v_{3/2})^2 \rangle \langle \tau^3 \rangle] - 1$ 
versus $1/D$ which should be zero in the normal phase 
if the scaling hypothesis holds
(where $\langle \chi^2 \rangle$
and $\langle \tau^3 \rangle$ are finite). 
The calculations of $\langle \chi^2 \rangle$ and $\langle
\tau^3\rangle$ are given in Appendix A and B. 
 We see that at the transition point, $D=1/5$,  $Y_{dev}=0$,
but otherwise $Y_{dev} \neq 0$. Hence the scaling hypothesis does not work
in the normal phase $D<1/5$. This means we need another approach for normal
diffusion, which luckily is easy to handle. For $D>1/5$ long jumps
dominate, since $\langle \chi^2 \rangle$ diverges, and then our
scaling theory
works fine.

\begin{figure}\begin{center}
\includegraphics[width=0.41\textwidth]{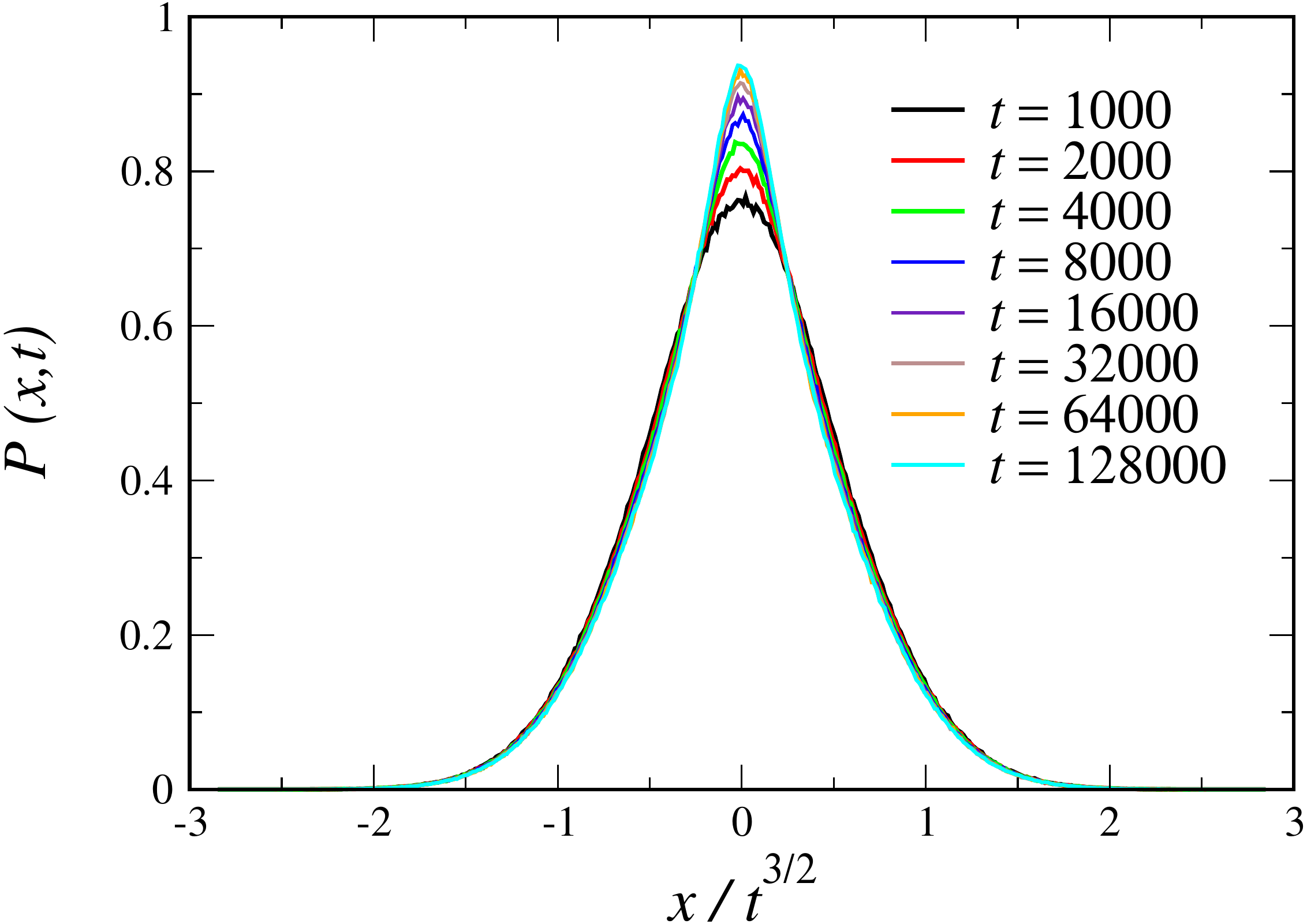}\end{center}
\caption{ 
$P(x,t)$ versus $x/t^{3/2}$ obtained from numerical simulation for $D=20/3$.
 Increasing  time the solution  converges to a scaling function. 
 Such a $x \sim t^{3/2}$ Richardson-Obukhov scaling is found for $D>1$.
}
\label{figObu}
\end{figure}

\section{The Richardson-Obukhov phase }
\label{SecRO}

In Sec. \ref{SecLev} we discussed the L\'evy phase which is
found for $1/5<D<1$. 
 When the average jump duration, $\langle \tau \rangle$, diverges, i.e. for $D>1$, the dynamics of $P(x,t)$ enters a new phase. Since the index of the L\'evy
PDF $\nu$ approaches $2/3$ as $D$ approaches $1$, $x$ scales like $t^{3/2}$
in the limit. Due to the correlations between $\chi$ and $\tau$, $x$
cannot grow faster than this, so in this regime, $P(x,t)\sim
t^{-3/2} h(x/t^{3/2})$. An example for this behavior is presented
in Fig. 
\ref{figObu}.
 This scaling is that of free diffusion, namely
momentum scales like $p\sim t^{1/2}$ and hence the time integral over the
momentum scales like $x \sim t^{3/2}$. Indeed in the absence of
the logarithmic potential, namely in the limit $D \gg  1$ 
the Langevin equations (\ref{eq05}) give
\begin{equation}
P(x,t)  \sim  \sqrt{{ 3 \over 4 \pi D t^{3} }} \exp\left( - { 3 x^2 \over 4 D t^3} \right). 
\label{eqRich}
\end{equation}
This limit describes the Obukhov model for a tracer particle in
turbulent flow, where the velocity follows simple Brownian motion
\cite{Obukhov,Friedrich}. These scalings are related to Kolmogorov's theory of
$1941$ [see Eq. (3) in Ref. (\cite{Friedrich})] and to Richardson's diffusion,
$\langle x^2 \rangle \sim t^3$ \cite{Rich}.
Eq. (\ref{eqRich}) is valid when the optical potential depth is small
since $D \to \infty$ when $U_0 \to 0$. This limit should be
taken with care, as the observation time must
be made large before considering the limit of weak potential. In the
opposite scenario, $U_0 \to 0$ before $t \to \infty$, we expect ballistic 
motion $|x| \sim t$, since then the optical lattice has not had time to make 
itself felt \cite{Sagi}. 
Physically, the atoms in this phase are undergoing a random walk in 
momentum space, due to random emission events, which in turn give
the $x \sim t^{3/2}$ scaling.  For shallow lattices,
 the Sisyphus cooling mechanism
breaks down, in the sense that transitions from maximum to minimum of the 
potential field created by the laser fields, are not preferred over the
inverse
transitions. Thus the deterministic dissipation is not effective,
and we are left with Brownian scaling in momentum space, $p \sim t^{1/2}$. 

\section{The Normal Phase}
\label{secNormal}

 When the  variance of the jump length is finite, namely $\nu>2$ $(D<1/5)$,
 we get normal
diffusion. Here the  variance of jump lengths is finite and hence the
scale free dynamics breaks down. 
The breakdown of scaling means that instead of 
using the scaling function
$B(.)$, e.g. in Eq. 
(\ref{eqpsiBg}) for 
$\psi({ \chi}, \tau)$ we must use  the joint PDF
$\psi({ \chi}, \tau) = g(\tau) p(\chi|\tau)$ Eq. 
(\ref{eqSca03}) and in principle not limit ourselves to large
$\tau$. However, luckily there is no need for a new calculation.  

 We focus on the central part of of the density $P(x,t)$ where
central limit theorem arguments hold. 
 In this normal case  the spatio-temporal distribution of
jump times and jump lengths effectively decouples, similar to the L\'evy
phase.
Since  the variance of jumps size and the averaged time for jumps are
finite, many small jumps contribute to the total displacement, and hence
in the long time limit, 
we expect Gaussian behavior with no correlations between jump lengths and
waiting  times,
i.e., the decoupling approximation is expected to work. 
  More precisely,  the average waiting
time is finite so $\hat{g}(u) \sim 1 -  \langle \tau \rangle u + \cdots$ and
the variance of jump lengths is also finite so the Fourier transform
of $q(\chi)$ has the following small $k$ expansion
$\tilde{q}(k) = 1 - \langle \chi^2 \rangle k^2 / 2 + \cdots$  
where 
$\langle \chi^2 \rangle= \int_{-\infty} ^\infty \chi^2 q(\chi) {\rm d} \chi$ 
is the variance of the jump lengths. 
This variance is investigated 
in Appendix B
using a backward Fokker-Planck equation.
 In the small $k,u$  limit 
$\hat{\psi} (k,u ) \sim 1 - u \langle \tau \rangle + k^2 \langle \chi^2 \rangle / 2+ \cdots $
and the Montroll-Weiss equation
(\ref{eqMW14}) is
\begin{equation}
\hat{P}(k,u) \sim  { 1 \over u + 
K_2  k^2 } .
\label{eqGa01}
\end{equation} 
This is the expected Gaussian behavior for the position probability density
\begin{equation}
P(x,t) \sim 
{1 \over  \sqrt{ 4 \pi K_2 t} }
\exp\left( - { x^2 \over 4 K_2 t} \right)
\label{eqGGAA}
\end{equation}
 and
\begin{equation}
K_2 = \lim_{\epsilon \to 0} { \langle \chi^2 \rangle \over 2 \langle \tau \rangle }  
\label{eqGa02}
\end{equation} 
is the diffusion constant, namely 
\begin{equation}
\langle x^2 (t) \rangle \sim 2 K_2 t. 
\label{eqGa03}
\end{equation} 
Eq. (\ref{eqGa02}) 
relates the statistics of the excursions, i.e. the variance of
the area under the excursion $\chi$
 and its average duration $\langle \tau \rangle$ to the diffusion
constant (here $\langle \chi \rangle=0$). 
As noted in the introduction
the equation has the structure of the famous Einstein relation,
relating the variance of jump size and the time between jumps
to the diffusion.

While $\langle \tau\rangle$,  Eq. (\ref{eqAs05}),
and $\langle \chi^2 \rangle$ 
Eq. (\ref{eqssvar08}),
approach zero when $\epsilon \to 0$,
their ratio remains finite and gives
\begin{equation}
K_2 = {1 \over D {\cal Z} } \int_{-\infty} ^\infty {\rm d} p e^{ V(p) /D} \left[ \int_p ^\infty {\rm d} p' e^{- V(p') /D} p' \right]^2  .
\label{eqGa04}
\end{equation}
This equation was derived previously using different approaches
\cite{Hodapp,Dechantprl}.  
The diffusion constant $K_2$
 diverges as $D \to 1/5$ from below
indicating the transition to the super-diffusive phase. A sharp increase in
the diffusion constant $K_2$ as the intensity of the laser reaches
a critical value was demonstrated experimentally \cite{Hodapp}. 
Eq. (\ref{eqGa04}) can be derived using the Green-Kubo formalism \cite{Zoller},
so in the normal phase, the  analysis of the statistics of excursions is
an alternative to usual methods. It seems that in the
L\'evy phase, 
the analysis of statistics of excursions is vital. 
Specifically the usual Green-Kubo formalism breaks down
since $K_2$ is infinite, and the calculation of the
anomalous diffusion coefficient $K_{\nu}$
cannot be based on a computation of a stationary velocity correlation
function.

\section{Discussion}
\label{SecDis}

 Starting with  the Langevin description of the semi-classical motion of
atoms undergoing a Sisyphus cooling process, we mapped the problem
onto a random walk scheme using excursions as a tool. 
 Thus our work combines ideas from the theory of stochastic processes with
cold atoms physics. We now summarize the key ingredients of our
results and its predictions from the point of view of these two communities.

\subsection{Fractional Diffusion Equation and CTRW theory revisited} 

The first ingredient of the theory was the calculation
of the coupled joint distribution $\psi(\chi,\tau)$. 
Rather generally, coupled waiting time - jump length distributions
are the microscopical ingredient for coupled 
continuous time random walks 
\cite{Review,Wong,KBS,Blumen,Carry,Bouchaud,Lax}. These distributions, however 
are difficult to
obtain from first principle calculations and usually
treated in a simplified manner. For example, 
postulating $\psi(\chi,\tau)= g(\tau) \delta(|\chi| - \tau)/2$
is common in stochastic theories.  
Hence we provided an important  pillar for the foundation of this widely
applied approach. 

 Our work gives the relation between velocity excursions,
and random walk theory and hence diffusion phenomenon.
 Since zero crossing in velocity space is obviously
a very general feature of physical paths of random processes,
 we expect the areal distributions of 
 excursions and meanders will play an important
role in other systems, at  least as a tool for the calculation
of  transport and diffusion (e.g., in principle our approach
can treat also the case of a constant applied force, 
where a mean net flow is induced).
 The celebrated Montroll-Weiss equation needed
two important modifications. First, since the underlying process
is continuous we regularized the process, such that the excursions
and meander start at $\epsilon\to 0$. 
 Secondly, the statistics of the last
jump event, i.e., the meander, must be treated with care. In contrast,
usually it is assumed that only $\psi(\chi,\tau)$ is needed
for a microscopical description of a continuous
time random walk model. This and the correlations
between $\chi$ and $\tau$ make the problem challenging and
usual approaches to diffusion fail.  With our approach, 
the behavior of the packet
is  mapped onto a problem of the calculation of areas 
under Bessel excursions
and meanders. 
For example we derived
the  relation between the mean square displacement of the atoms and the areas
under both the Bessel excursion and meander 
Eqs. 
(\ref{eqMSD05},
\ref{eqMSD07},
\ref{eqMSD09}).
  A decoupled scheme which neglects the correlations
$\psi(\chi,\tau) \simeq g(\tau) q (\chi)$ gives a diverging result for $\nu<2$,
which is unphysical.


In the regime $1/5 < D <1$, which we have called the L\'evy 
regime,  the correlations between jump lengths and waiting times
point to  the
limitations  of the fractional diffusion equation, a popular
framework based on fractional calculus \cite{Review}.
The fractional diffusion
equation was previously investigated in the context of random
walk theory, and it describes a L\'evy flight processes \cite{Review}.
 Here we have provided a microscopic justification for
it. The fractional diffusion equation \cite{Review,Saichev,MBK},
\begin{equation}
{\partial^\beta \over \partial t^\beta} P(x,t) = K_\nu \nabla^\nu P(x,t)
\label{eqDis}
\end{equation}
 was the phenomenological
starting point for the description of the experiments
in the work of Sagi et al. \cite{Sagi}.
Our work (see also \cite{KesBarPRL}) provides the exponent $\nu$,
Eqs. 
(\ref{eqDc},
\ref{eqBessel15}),
in terms of the recoil energy and lattice depth, $\beta =1$ and
the anomalous diffusion coefficient, $K_\nu$ from Eq. 
(\ref{eqLevy04}). 
Indeed, the experiment \cite{Sagi} found the value 
$\beta=1$, so the time derivative
on the left hand side is 
a first order time derivative. The fractional
space derivative  $\nabla^\nu$ is a Weyl-Reitz fractional
derivative \cite{Review}.
To see the connection between our results and the fractional equation
we re-express Eq. (\ref{eqLevy03}) as 
\begin{equation}
 u  
\widehat{P}(k,u) - 1 = -  K_\nu |k|^\nu 
\widehat{P}(k,u) 
\label{eqDCCCv}
\end{equation}
which is the Fourier-Laplace transform of Eq. (\ref{eqDis}). 
 Here we recall \cite{Review} that the Fourier space
representation of, $\nabla^\nu$, is $-|k|^\nu$ and that 
$u \widehat{P}(k,u) - 1$ is the representation of the time derivative
in Laplace-Fourier space, of a $\delta$ function initial condition
centered on the origin. We see that $P(x,t)$ for $2/3<\nu<2$ satisfies
the fractional diffusion equation. 
 In other words, for initial
conditions starting on the origin the solution
of Eq. (\ref{eqDis})  is the L\'evy PDF Eq.
(\ref{eqLevy05}).

However the use of the fractional diffusion equation must be performed
with care. 
It 
predicts a seemingly
 unphysical behavior: 
the mean square displacement is infinite. 
Indeed as mentioned in the introduction,  the mean square displacement was
shown experimentally to exhibit superdiffusion by Katori et al.  \cite{Katori}
and in simulations in \cite{Wickenbrock}. 
 In fact the mentioned coupling implies that the fractional
equation is valid only in a scaling region,
$|x|<t^{3/2}$ and thus the L\'evy distributions 
obtained analytically here and used phenomenologically in
the  Weizmann experiment \cite{Sagi} describing
 the center part of the spreading distribution  of the
particles, but the tails exhibit a cutoff for $x > t^{3/2}$.
In other words the L\'evy distribution describing the center part of the
packet for $1/5 < D < 1$ does not contain information on the
correlations and to experimentally investigate correlations in this regime
one must probe the tails of the packet. 

 The renewal approach is the basis of the 
coupled  CTRW theory developed here.
Renewal theory  turned out to be predictive in the sense that
while the set of zero crossings is nontrivial in the continuum limit,
we could avoid this mathematical obstacle by introducing modified
paths with an $\pm \epsilon$ starting point after each zero hitting.
 Physical observables, like $\nu$,
$P(x,t)$, $K_\nu$ and $\langle x^2 \rangle$ do not depend
on $\epsilon$ in the limit, as expected.
 Technically, this is due to a cancellation of $\epsilon$, for example
both 
$\langle \tau \rangle$, Eq. 
(\ref{eqAs05}), and
$q^{*}$,
Eq. (\ref{eqSca06a}),
 depend linearly on $\epsilon$ hence
their ratio is $\epsilon$ independent and the anomalous  diffusion coefficient
$K_{\nu}$, 
Eq. (\ref{eqLevy04}), 
becomes insensitive to $\epsilon$. The same is true for the normal
diffusion constant $K_2$  Eq.  
(\ref{eqGa02}) and other physical observables Eqs.
(\ref{eqAs17},\ref{eqMSD07},\ref{eqMSD09a},\ref{eqMSD09}).
A closer reading of Appendix A and B  will reveal that this cancellation is
not exactly trivial. In our work we use the {\em regularized}
 Bessel process
for the calculation of the PDFs of the first passage time, $g(\tau)$,
and jump lengths, $q(\chi)$.
Hence for the calculations  the  details
of the potential $V(p)$  for small $p$ are important (but not for 
$B(v_{3/2})$!). If we use
instead the PDFs 
$g_B(\tau)$ and $q_B(\chi)$ (see Appendix A and B)
of  the non-regularized process, i.e. a logarithmic potential
for any $p$,  some (but not all)  of the $\epsilon$  cancelation will not take place
(see Eq. (\ref{eqAs15}) where both $q^{*}\propto \epsilon$ and $g^{*}\propto \epsilon$ for the regularized process which allows for the cancelation). 
For example
 both for the  regularized and the non-regularized
processes $g(\tau) \sim g^{*} \tau^{- (1 + \alpha)}$
(so the exponent $\alpha$ is identical in both cases),
however  $g^{*} \propto \epsilon$ for the regularized
process, while $g^{*} \propto \epsilon^{2 \alpha}$ for 
the non-regularized case
Eq. (\ref{eqApA06}). 
The cancelation of $\epsilon$ is further discussed in Appendix
F for three types of random durations.
That Appendix  shows that while the first passage time PDF
$g(\tau)$  depends on $\epsilon$ (see Appendix A),  
the time interval straddling time $t$ and the backward
 and forward recurrence times are insensitive to this parameter. 
Thus one upshot of Appendix F is a further  demonstration that this 
intuitive
$\epsilon$ renewal trick actually works.
A rigorous mathematical treatment of the problem will lead
to a stronger foundation
 of our results. 

 Regularization of the process is crucially important in the L\'evy
and Gaussian phase $D<1$ where observables depend on the details
of the potential $V(p)$.  This is related to work of
Martin et al. \cite{Martin}
on the  classification
of boundaries  for the non-regularized
 Bessel process. They showed that, using our notation, for
$D<1$, $p=0$ is an exit boundary while for $D>1$ it is
 a regular boundary ($D<0$ corresponds to an entrance boundary condition,
which is not relevant to our work). For an exit boundary starting on 
$p\simeq 0$  
it is impossible to reach a finite momentum state $p$ so clearly
we cannot afford such a boundary in our physical problem. For that reason,
we need to consider the regularized first passage time problem at least 
 when $D<1$,
and then the boundary is regular. A regular boundary on $p=0$ means
the the diffusion process can enter and leave from the boundary. Therefore, 
for $D>1$ our final results do not depend on the shape of $V(p)$,
besides its asymptotic limit,  namely one may replace
the regularized Bessel process with the non-regularized one.
To see this, notice that $\langle x^2 \rangle$ for $D>1$ depends
on $\alpha<1$ but not on small scale properties of  $V(p)$
 and since $\alpha$ is the first passage
exponent for both the regularized and non-regularized processes, it does not
really matter which process is the starting point of the calculation. 
Further evidence for such a behavior is in Appendix F, where statistics
of durations for $D>1$ are shown to depend on $\alpha<1$ but not on the
shape of $V(p)$. 
Thus the $D=1$ marks transition from Richardson to L\'evy behavior,
the diverging of the partition function $Z$ namely the
equilibrium velocity distribution turns non-normalizable, and
according to Martin et al. the boundary of the Bessel
process switches from regular to exit.  
 
 Additional theoretical work is needed on the leakage of particles, and more
generally evaporation (see more details below). 
  Assuming that energetic particles are
those which get evaporated, 
and that this takes place through the boundary of the system, 
one would be interested in the first passage time properties
of particles, from the center of the system to one of its boundaries.
Previous theoretical
work  on first passage times for L\'evy walks and flights
might give a useful first  insight, 
\cite{Buldyrev,Zoia1,Katzav,Zoia,Korabel} 
however it is left
for future work to compare these simplified models
with the microscopic semiclassical picture
of the underlying dynamics. Evaporation is 
an important ingredient of cooling (beyond the Sisyphus cooling)
since it gets rid of the 
very energetic particles; hence this line of research has practical 
applications.     
Theoretically, understanding the boundary conditions for fractional
diffusion equations, needed for the calculation of statistics
of  first passage times,
is still an open challenge. 
Yet another challenge is to characterize the joint PDF $W(x,p,t)$ and
in particular the correlations between position and momentum, which are
expected to be non-trivial. Again, some elegant ideas and tools
were recently developed \cite{Zabu} within the L\'evy walk framework,
however more work is needed for direct comparison 
with the physical picture under consideration. 


\subsection{Cold atom experiments}

 Experiments on  ultra-cold diffusing particles
in optical lattices have been  performed on a single 
$^{24}$Mg$^{+}$ ion
 \cite{Katori}
and an ensemble of Rb atoms \cite{Wickenbrock,Sagi,Hodapp}.
Experiments on the spreading of the density of atoms
yield ensemble averages like the mean square displacement and the density 
$P(x,t)$. Analysis of individual trajectories yields, in principle,
 deep insight on 
paths, for example information on the first passage time
in velocity space, i.e. $g(\tau)$, zero crossing times, statistics
of excursions and meanders etc. For a comprehensive understanding of the
dynamics of the particles, 
 both types of experiments are needed. The theory presented herein
 provides statistical
 information on the stochastic trajectories, e.g., zero crossing events,
which in turn are related to the spreading of the packet of particles.
Thus from single trajectories, one may estimate $g(\tau)$, $B(v_{3/2})$, and 
$\psi(\chi,\tau)$. From $g(\tau)$ one may
then obtain the exponent $\alpha$ which can be used to predict the qualitative
features of the spreading of the ensemble of particles.
 For example with an estimate of $\alpha$ we can 
determine the phase of motion, be it Richardson, L\'evy or Gauss. Of course,
a more quantitative investigation is now possible, since we have analytically
related $\alpha$,
the mean square displacement, and $K_\nu$ with microscopical parameters
like the optical potential depth $U_0$. In principle, statistics of Bessel
excursions, so
far of great interest mainly in the mathematical literature, could be detected
in single particle experiments.  
These single ion experiments are also ideal in the sense that 
collisions/interparticle interactions,
do not play a role and they also provide insights on ergodicity. 
Three specific unsolved issues are:
\begin{itemize}
\item[(i)] Can the Richardson phase be experimentally demonstrated?
 So far this phase
was obtained in our theory, and Monte Carlo simulations \cite{Wickenbrock}, while experiments
exhibit ballistic diffusion as the upper limit \cite{Katori,Wickenbrock,Sagi}.
 This might be related to the subtle
limit of taking time to infinity before the depth of optical lattice approaches
 zero, since clearly in the absence of an optical lattice the fastest particles
are ballistic. 
\item[(ii)] While our theory predicts correctly
 a L\'evy phase in agreement with experiments, in \cite{Sagi}
  two exponents were used  to fit the data.
In contrast the theory we developed suggests  a single L\'evy  exponent $\nu$. 
This could be due to the  
leakage of particles and also to the $t^{3/2}$ cutoff
of the   L\'evy density, which are due to the correlations between $\chi$ and
$\tau$ investigated in this manuscript. Avoiding the leakage of
particles is an experimental challenge, and once this is accomplished,
a more informed comparison of our theoretical prediction on the L\'evy phase
with a single characteristic exponent $\nu$ could be made. 
Success on this front
would provide an elegant demonstration of L\'evy's central limit theorem,
with the characteristic exponent controlled by the depth of 
the optical potential. 
\item[(iii)] As pointed out in \cite{Wickenbrock},
losses of atoms lead to an underestimate of the diffusion exponent,
as many super-diffusing atoms are lost. Characterisation of these losses
both theoretically and experimentally could advance the field,
since this yields insight on the underlying processes and leads as towards
better control of the particles.  Specifically, we do not know what is
the number of particles kept in the system, versus time,
for varying strength of the  optical potential. 
\item[(iv)] In the L\'evy phase, measurements of non-integer moments $\langle |x|^q \rangle$ 
 with $q<\nu<2$ will
according to theory exhibit L\'evy scaling 
$\langle x^2 \rangle  \sim t^{q/\nu}$. 
In contrast,
 the main focus of experiments so far was the second
moment $q=2$, which is difficult to determine statistically since  as we have shown
here it  depends 
on the tails of $P(x,t)$ and very fast particles.
Thus ideally a wide spectrum of moments $\langle |x|^q \rangle$ should be
recorded in experiment, the low order moments $q<\nu$ giving information on the center part of the density  
while higher order moments on the correlations and tails. 
\end{itemize}

In the case that  experiment and theory will not reconcile, we will
have  strong   indication that
the current semi-classical theory in not sufficient and then
we will be forced to investigate at least four other aspects of the
problem:
\begin{itemize}
\item[(a)] Effects of collisions on anomalous diffusion of the atoms.
\item[(b)] Effect of higher dimensions.
\item[(c)] Quantum effects beyond the semi-classical approach used here.
 In particular it would be very interesting to simulate this
system with full quantum Monte Carlo simulations \cite{Zoller},
 to  
compare the semiclassical theory with quantum dynamics.  
We note that the Richardson phase, which as mentioned
was  not observed in experiments,  is actually 
a heating 
phase and quantum simulations become difficult because the numerical
lattice introduces a cutoff on velocities which induces artificial ballistic
motion. 
\item[(d)] Other cutoff effects that modify the anomalous diffusion.
For example at high enough velocities, Doppler cooling is expected
to diminish the fast particles. In \cite{nonpub} we
 simulated the effect
of Doppler friction, and showed that the anomalous character of the 
diffusion is kept unchanged, at least for a certain reasonable
set of  parameters.  
However, a general rule on the influence of Doppler cooling
 is not yet established, and  an experimentalist  with  a specific
set of  parameters in mind might  
wish to test numerically the magnitude of this effect on the anomalous spreading.
\end{itemize}

 An interesting approach was recently suggested by Dechant and Lutz 
\cite{DechantPRL}. They
investigated the multi-fractal nature of the moments  $\langle |x|^{|q|} \rangle$ of the process
and consider initial conditions different than ours.
  They assume 
that in a first stage, of duration $t_c$, the particles are cooled in a confining
field (which inhibits the spreading). Then the momentum of particles
relaxes to a state described by the infinite covariant density 
\cite{KesslerPRL} which depends on $t_c$. The particles are then released
and their spreading is recorded for a duration $t$. We considered the
case $t\gg t_c$ while Ref. \cite{DechantPRL} considers the opposite case.
At-least the experiment in \cite{Sagi}
 is conducted under the conditions investigated
here, namely that the spreading time is
much longer than the preparation time.
As shown by Hirschberg et al.,  starting with power law distributions
will dramatically  influence
the spreading, both in momentum space \cite{Ori} and in space.
Indeed large $t_c$ implies power law initial conditions, Eq.
(\ref{eqequil}), with a $t_c$ dependent cutoff \cite{KesslerPRL}.
In that sense diffusivity is sensitive to the initial preparation
of the system, and an experimental verification of these effects
would indicate the fundamental difference between transport in
these systems and normal transport which does not depend
on initial conditions. This is clearly related to the strong sensitivity
we have found of the mean squared displacement on a single jump
event, described by the Bessel meander.  



{\bf Acknowledgement} 
This  work  was supported by the  Israel Science  Foundation. 
We thank Yoav Sagi and Nir Davidson for discussions on the
experiment \cite{Sagi} and Andreas Dechant and Eric Lutz
on collaborations on related problems.   
\appendix
\section{The waiting time PDF}

The waiting time is the time it takes the particle starting
with momentum $p_i>0$ to reach $p_f<p_i$ for the first
time. Here we investigate its PDF, $g(\tau)$. 
 
\subsection{First passage time for the Bessel process}

We first briefly investigate the first passage time
problem for the Bessel process following Ref. \cite{Lutz}. 
The Bessel process corresponds to the case $F(p) = -1/p$
so the the force diverges at the origin. In the next subsection we will
consider the  regularized force
$F(p) = -p /(1 + p^2)$. According to \cite{Gardiner},
the survival probability $S(\tau)$, namely the probability that a particle
initially at $p_i$ does not cross the boundary $p_f<p_i$ in the
time interval whose length is $\tau$, satisfies
\begin{equation}
\partial_\tau S = F(p_i) \partial_{p_i} S + D (\partial_{p_i})^2 S
\label{eqApA01}
\end{equation}
Here $S=1$ for $\tau=0$  since the particle's escape time cannot be zero.
Further $S \to 0$ 
 when $p_i\to p_f$, since in that case the particle starts out at
the boundary, and $S=1$ if one starts at $p_i=\infty$. 
The random time it takes a particle starting on $p_i$ to 
reach $p_f$ for the first time is $\tau$ and its PDF
is $g_B(\tau) = -\partial_\tau S(\tau)$. Here the subscript $B$ denotes the 
Bessel process.   Since $S(\tau)|_{\tau=0} = 1$ we
have in Laplace $\tau \to u$ space the following 
simple relation
\begin{equation}
 \hat{g}_B(u) = - u \hat{S}(u)  + 1 .
\label{eqApA02} 
\end{equation}
Using Eq. (\ref{eqApA02})
and the Laplace transform of Eq. (\ref{eqApA01}) we find
\begin{equation}
D (\partial_{p_i})^2 \hat{g}_B (u) - (1/p_i) \partial_{p_i} \hat{g}_B(u) - u \hat{g}_B (u) = 0 .
\label{eqApA03} 
\end{equation}
Note that  $\hat{g}_B(u)|_{p_i = p_f} = 1$ since $g_B(\tau)|_{p_i\to p_f} = \delta(\tau)$ . 
The solution of Eq.  (\ref{eqApA03}) 
is \cite{Abr}
\begin{equation}
\hat{g}_B(u) = { K_\alpha \left( p_i \sqrt{ { u \over D} } \right) \over
                 K_\alpha \left( p_f \sqrt{ { u \over D} } \right)}
             \left( {p_i  \over p_f} \right)^\alpha .
\label{eqApA04} 
\end{equation}
Here $K_\alpha(.)$ is the modified Bessel function of the
second kind and as before $\alpha = 1/2 + 1/(2 D)$.

 If $\alpha>1\!\! $ the small $u$ expansion of Eq. (\ref{eqApA04})
yields $\hat{g}_B (u) \sim 1 - u \langle \tau_B\rangle+ \cdots$
where $\langle \tau_B \rangle$ is the average first passage
time. Expanding Eq. (\ref{eqApA04}) we find
\begin{equation}
\langle \tau_B \rangle = {1 \over 4} { (p_i)^2 - (p_f)^2 \over \alpha - 1} {1 \over D} .
\label{eqApA05} 
\end{equation} 
Notice that $\langle \tau_B \rangle$ diverges when $\alpha\to 1$ corresponding
to $D\to 1$. Not surprisingly, the average time in 
Eq. (\ref{eqApA05}) is generally different from
the expression for the average waiting time for the regularized process
Eqs.
(\ref{eqAs04},
\ref{eqAs05}).

Expanding Eq. (\ref{eqApA04}) for small $u$ for the case $0<\alpha<1$ we get 
\begin{equation}
\hat{g}_B (u) \sim 1 - { \Gamma(1 -\alpha) \over \Gamma( 1 + \alpha)}
\left( {1 \over 2} \right)^{2 \alpha} \left( (p_i)^{2 \alpha} -(p_f)^{2 \alpha} \right) \left( { u \over D} \right)^\alpha  +\  \cdots.
\label{eqApA06u} 
\end{equation} 
Inverting to the time domain we find, using a Tauberian theorem,
 the long time behavior of 
the PDF, 
\begin{equation}
g_B(\tau) \sim \left( { 1 \over 2 \sqrt{D} }\right)^{2 \alpha} 
{ (p_i)^{2 \alpha} - (p_f)^{2 \alpha}  \over \Gamma(\alpha) } \tau^{-1 - \alpha}.
\label{eqApA06}
\end{equation}
One can show that this fat-tail behavior is valid also for the regime 
$\alpha>1$. The procedure involves expansion of Eq.
(\ref{eqApA04}) beyond the first two leading terms; for
example for $1<\alpha<2$ one finds $\hat{g}_B (u) = 1 - u \langle \tau_B \rangle + C u^{\alpha}  .... $ and the third term gives the tail of PDF. 
Note that the PDF $g_B(\tau)$ yields the same exponent
as in Eq. (\ref{eqSca01}). However to calculate
the amplitude $g^{*}$ 
we must consider the regularized process.

\subsection{First passage time for the regularized process}
\label{suSecFPTR}

 For the optical lattice problem we need to treat the regularized force
$F(p) = - p /(1 + p^2)$. Our aim is to find the asymptotic behavior
in Eq. (\ref{eqSca01}) while the average waiting time
is given already in Eqs.  
(\ref{eqAs04},\ref{eqAs05}). From these, we know
that for $\alpha<1$ the average first passage time
from $p_i$ to $p_f$ is infinite (the derivation can be easily generalized
for arbitrary initial and final states). Furthermore, for long time the large momentum behavior of $F(p)$ plays the crucial role and hence
we expect for small $u$
\begin{equation}
\hat{g}(u) \approx  1 - G u^{\alpha} ; \qquad 0<\alpha <1,
\label{eqApA07}
\end{equation}
where we must determine $G$ which then gives the amplitude $g^{*}$ introduced in Eq. (\ref{eqSca01}).
The  equation for $\hat{g}(u)$ is
the same as Eq. (\ref{eqApA03}) but with the regularized force: 
\begin{equation}
D (\partial_{p_i})^2 \hat{g}(u)  -{p_i \over 1 + (p_i)^2} \partial_{p_i} \hat{g}(u) - u \hat{g}(u) = 0. 
\label{eqApA08u}
\end{equation}
We need to solve for $\hat{g}(u)$ for small $u$, so to leading order we drop the last term  and find
\begin{equation}
D (\partial_{p_i})^2 \hat{g}(u) + F(p_i) \partial_{p_i} \hat{g}(u) = 0.
\label{eqApA08}
\end{equation}
Hence
\begin{equation}
\hat{g}(u) = C_1(u) \int_{p_f} ^{p_i} e^{ V(p') / D} {\rm d} p' + C_2(u)
\label{eqApA09}
\end{equation}
where $V(p) = \ln(1+p^2)^{1/2}$. 
Since $\hat{g}(u)=1$ when $p_i = p_f$, $C_2(u)=1$. 
 This approximation breaks down however when $p_i$ is too large, since then, 
$\hat{g}(u) \propto p_i^{1+1/D}$,
and the last term 
in Eq. (\ref{eqApA08u})
is comparable in size to the first two when $p_i \sim u^{-1/2}$.  In the large $p_i$ regime, however, $p_i \gg 1$, so we can approximate $F(p_i)$ by its Bessel 
form, $F(p_i) \approx- 1/p_i$, and the solution is, as above,
\begin{equation}
\hat{g}(u) \approx C_3(u) K_\alpha\left(p_i \sqrt{\frac{u}{D}}\right) p_i^\alpha
\end{equation}
where here we cannot use the $p_f\to p_i$ limit to normalize $\hat{g}(u)$, 
since $p_f$ is not necessarily large.  Nevertheless, $\hat{g}(u=0)=1$
and $K_\alpha(z) \sim (\Gamma(\alpha)/2) (z/2)^{-\alpha}$ for $z \to 0$,
implying that
\begin{equation}
C_3(u) \approx \frac{2}{\Gamma(\alpha)}\left(\frac{1}{2}\sqrt{\frac{u}{D}}\right)^\alpha
\end{equation}
to leading order in $u$.
For $1 \ll p_i \ll u^{-1/2}$, our two approximations must agree, and so we find
using the second order expansion of $K_\alpha(z)$ 
\begin{equation}
1 - \frac{\Gamma(1-\alpha)}{\Gamma(1+\alpha)}\left(p_i\sqrt{\frac{u}{D}}\right)^{2\alpha} \approx 1 + C_1(u) \frac{p_i^{1/D+1}}{1/D+1},
\end{equation}
and recognizing that $(1/D+1)=2\alpha$,
\begin{equation}
C_1(u) \approx -\frac{2\alpha\Gamma(1-\alpha)}{\Gamma(1+\alpha)}\left(\frac{u}{4D}\right)^\alpha .
\end{equation}
Thus, using Eq. (\ref{eqApA09}),
\begin{equation}
\hat{g}(u) \approx 1 - \left[2 { \Gamma\left( 1 - \alpha \right) \over \Gamma(\alpha)} \left( {1 \over 4D } \right)^{\alpha} \int_{p_f} ^{p_i} e^{ V(p') / D} {\rm d} p'\right] u^\alpha . 
\label{eqApA12}
\end{equation}
Using the Tauberian theorem, which implies 
$u^{\alpha} \to \tau^{-(1 + \alpha)} / \Gamma(-\alpha)$, we find
the large $\tau$ behavior
\begin{equation}
g(\tau) \sim { 2 \alpha \over \Gamma(\alpha) } { \int_{p_f} ^{p_i} e^{V(p')/D} {\rm d} p' \over \left(4D\right)^{\alpha} } \tau^{-(1+ \alpha)} .
\label{eqApA13}
\end{equation}
It can be shown, with additional work, that this result is also valid for $\alpha>1$. 
To conclude, we find for the final state $p_f=0$ and the initial state
 $p_i = \epsilon\ll 1$,
\begin{equation}
g^{*} = { 2 \alpha \over \left( 4D \right)^{\alpha}  \Gamma(\alpha) } 
\int_0 ^\epsilon e^{V(p') /D} {\rm d} p' \approx { 2 \alpha \epsilon \over \left( 4D \right)^{\alpha}  \Gamma(\alpha) } 
\label{eqApA13a}
\end{equation}
This of course vanishes as $\epsilon \to 0$, in this case linearly,
as opposed to the $p_i\to \epsilon$ and $p_i=0$ limit of Eq.
(\ref{eqApA06}), where the dependence is of higher order. 
 As shown in the manuscript,
for the purpose of calculation of the asymptotic behavior
of $P(x,t)$, all we need is $\partial_\epsilon g^{*}$ when $\epsilon \to 0$
which is a finite constant which depends on $D$ only.
The full shape of $V(p)$ is unimportant for the calculation
of $g^{*}$ indicating a degree of universality. In contrast 
$\langle \tau \rangle$ depends on ${\cal Z}$ and so is non-universal.
Since $g^{*}$ is a measure of the long time behavior, the detailed shape of
the potential
is not important, provided it is regularized.

\subsection{Moments of $g(\tau)$}

In the text we show that $\langle \tau \rangle$ and $\langle \tau^3 \rangle$
 are relevant when $D<1/5$ so that these averages do not diverge. 
The moments $\langle \tau \rangle$ and $\langle \tau^3 \rangle$
are found using
\begin{equation}
\hat{g}(u) = 1 - u \langle \tau\rangle + u^2 \langle \tau^2 \rangle/2
- u^3 \langle \tau^3 \rangle/6 + \cdots
\label{eqBD04}
\end{equation}
$\hat{g}(u)$ is the Laplace transform of the waiting time PDF $g(\tau)$.
From
\begin{equation}
\left[ D (\partial_p)^2 - { p \over 1+ p^2} \partial_p - u \right] \hat{g}(u) = 0
\label{eqBD05}
\end{equation}
we get
\begin{equation}
\left[ D (\partial_p)^2 - { p \over 1 + p^2} \partial_p \right] \langle \tau \rangle= -1 ,
\label{eqBD06}
\end{equation}
\begin{equation}
\left[ D (\partial_p)^2 - { p \over 1 + p^2} \partial_p \right] \langle \tau^2 \rangle= -2 \langle \tau \rangle
\label{eqBD07}
\end{equation}
\begin{equation}
\left[ D (\partial_p)^2 - { p \over 1 + p^2} \partial_p \right] \langle \tau^3 \rangle= -3 \langle \tau^2 \rangle .
\label{eqBD07a}
\end{equation}
Solving Eq.
(\ref{eqBD06})
\begin{equation}
\langle \tau \rangle = {1 \over D} \int_0 ^p {\rm d} y e^{ V(y)/D}
\int_y ^\infty e^{ - V(x)/D} {\rm d} x,
\label{eqBD08}
\end{equation}
so $\langle \tau \rangle =0$ when $p=0$ as expected for
the first passage from $p$ to the origin.
For the  second and third moments
\begin{equation}
\langle \tau^2  \rangle = {2 \over D} \int_0 ^p {\rm d} y e^{ V(y)/D}
\int_y ^\infty e^{ - V(x)/D} \langle \tau (x) \rangle  {\rm d} x,
\label{eqBD09}
\end{equation}
and
\begin{equation}
\langle \tau^3  \rangle = {3 \over D} \int_0 ^p {\rm d} y e^{ V(y)/D}
\int_y ^\infty e^{ - V(x)/D} \langle \tau^2 (x) \rangle  {\rm d} x.
\label{eqBD10}
\end{equation}
Thus, in the $p = \epsilon \to 0$ limit,
\begin{equation}
\langle \tau^3 \rangle \sim { 3 \epsilon \over D} \int_0 ^\infty e^{ - V(z)/D} \langle \tau^2 (z) \rangle {\rm d} z.
\label{eqBD11}
\end{equation}

\begin{widetext}

\section{The jump length  PDF $q(\chi)$}

The excursion length $\chi$  is the distance  the particle travels
 from its start with momentum 
$p_i>0$ until it reaches the  momentum origin $p_f=0$ for the first
time.  
 Here we investigate its PDF $q(\chi)$ 
which in  the limit of $p_i\to0$ and for the regularized force field gives
 the jump length PDF.
Previously this PDF was investigated
for the $-1/p$ force field in \cite{Zoller}.
Note that in what follows $p_i>0$ so $\chi>0$. As elsewhere  in this
paper,  from symmetry positive and negative
$\chi$ are equally probable so we may restrict our attention to $\chi>0$. 
 
\subsection{PDF of jump lengths for the Bessel process}

As mentioned, for the Bessel process $F(p) = - 1/p$. The PDF $q_B(\chi)$ which depends of course
on the start and end points $p_i$ and $p_f$ satisfies the following backward equation \cite{Gardiner}:
\begin{equation}
D (\partial_{p_i})^2 q_B - (1/p_i) \partial_{p_i} q_B - p_i \partial_\chi q_B =0,
\label{eqApB01}
\end{equation}
where the subscript $B$ again stands for Bessel.
Since $\chi>0$, we define the Laplace transform
\begin{equation}
\hat{q}_B(s) = \int_0 ^\infty \exp(- \chi s) q_B(\chi) {\rm d} \chi.
\label{eqApB02}
\end{equation}
When $p_f\to p_i$, we have $q_B(\chi) \to \delta(\chi)$ and $q_B(\chi)_{\chi=0}= 0$ if $p_i\neq p_f$ since it
takes time for the particle to reach the boundary and hence we cannot get an excursion
whose size is zero.  Of course, in the
opposite limit of large $\chi$, $\lim_{\chi\to \infty} q_B(\chi) = 0$. 
In Laplace space Eq. (\ref{eqApB01}) yields
\begin{equation}
(\partial_{p_i})^2 \hat{q}_B(s) -(1/ D p_i) \partial_{p_i} \hat{q}_B(s) - (p_i s /D) \hat{q}_B(s) = 0.
\label{eqApB03}
\end{equation}
It is easy to verify that the appropriate solution is 
\begin{equation}
\hat{q}_B(s) = { (p_i)^{3 \nu/2} K_\nu \left( { 2 \over 3} \sqrt{{s \over D}} (p_i)^{3/2} \right) \over
               (p_f)^{3 \nu/2} K_\nu \left( { 2 \over 3} \sqrt{{s \over D}} (p_f)^{3/2} \right) }.
\label{eqApB04}
\end{equation} 
with $\nu= (1 + D)/3D$ and as in the previous Appendix
 $K_\nu(\cdot)$ is the modified Bessel function of the second kind. 

 Our goal is to find the large $\chi$ behavior of $q_B(\chi)$,
 so we expand Eq.
(\ref{eqApB04}) in the small $s$ limit, finding, for $0<\nu<1$, 
\begin{equation}
\hat{q}_B(s) \sim 1 -
\left({1\over 3}\right)^{2 \nu}\left[ (p_i)^{3 \nu} - (p_f)^{3 \nu} \right]{ \Gamma(1 -\nu) \over \Gamma(1 + \nu) } \left( 3 \nu -1\right)^\nu s^\nu
\cdots .
\label{eqApB05}
\end{equation} 
The first term on the left hand side is simply the normalization, and the $s^\nu$ term indicates that the
average excursion length diverges since $\nu<1$. 
Passing from $s$ to $\chi$, we get for large $\chi$ 
\begin{equation}
q_B(\chi)\sim \left( {1 \over 3} \right)^{2 \nu} { (p_i)^{3 \nu} - (p_f)^{3 \nu} \over \Gamma(\nu) } 
\left( 3 \nu - 1 \right)^\nu \chi^{-1 - \nu}
\label{eqApB06}
\end{equation} 
With a similar method, one can show that Eq. (\ref{eqApB06}) holds also for
$1<\nu<2$.   There the expansion
Eq. (\ref{eqApB05})
 contains three terms; the additional term yields the average of $\chi$
which is now finite. 
In \cite{Zoller} 
a more complicated method was used to investigate the same problem and
a similar result was found  although with a typographical error. 
They report $(3 \nu + 1)^\nu$ in the prefactor whereas we find
 $(3 \nu-1)^\nu$, 
 a difference with some importance since  the pre-factor found here
goes to zero when $D\to \infty$, which 
is necessary to obtain reasonable physical results. 

\subsection{PDF of $\chi$  for regularized process}

To obtain the large $\chi$ behavior of $q(\chi)$ for the regularized force $F(p)= - p / (1 +p^2)$, we
follow the same steps carried out in Sec.  
\ref{suSecFPTR}. The equation to solve is 
\begin{equation}
D (\partial_{p_i})^2 q + F(p_i) \partial_{p_i} q - p_i \partial_{\chi} q=0.
\label{eqApB07}
\end{equation}
For $\nu<1$, we switch to Laplace space $\chi \to s$, 
i.e.
$D (\partial_{p_i})^2 \hat{q} + F(p_i) \partial_{p_i}
 \hat{q}-p_i s \hat{q}=0$, and 
drop the last term for small s, yielding
\begin{equation}
\hat{q}(s) \approx C_1(s) \int_{p_f}^{p_i} e^{V(p')/D} dp'
\end{equation}
after applying the boundary condition $\hat{q}|_{p_f \to p_i} = 1$.  
 Again, for $p_i \gg 1$, we can approximate $F(p_i)$ by its large $p_i$ approximation, yielding
\begin{equation}
\hat{q}(s) \approx C_3(s) p_i^{3\nu/2} K_\nu\left(\frac{2}{3}\sqrt{\frac{sp_i^3}{D}}\right).
\end{equation}
In the small $s$ limit, $q(s\to 0)=1$ implies
\begin{equation}
C_3(s) \approx \frac{2}{\Gamma(\nu)}\left(\frac{1}{3}\sqrt{\frac{s}{D}}\right)^\nu
\end{equation}
to leading order.  These two approximations must agree for $1 \ll p_i \ll s^{-1/3}$, yielding
\begin{equation}
C_1(s) \approx 3^{1-2\nu} (3\nu-1)^\nu \frac{\Gamma(1-\nu)}{\Gamma(\nu)} s^\nu.
\end{equation}
This implies that
\begin{equation}
q(\chi) \approx \left[\frac{1}{3^{2\nu-1}}\frac{(3\nu-1)^\nu}{\Gamma(\nu)}\nu\int_{p_f}^{p_i} e^{V(p')/D} dp' \right] \chi^{-1-\nu}.
\end{equation}
This calculation was done assuming $p_i$, $p_f$ positive.  Allowing also for negative momenta, and so negative $\chi$, we get
an additional prefactor of $1/2$:
\begin{equation}
q(\chi) \approx \left[\frac{1}{3^{2\nu-1}}\frac{(3\nu-1)^\nu}{2\Gamma(\nu)}\nu\int_{p_f}^{p_i} e^{V(p')/D} dp' \right] \left|\chi\right|^{-1-\nu}.
\end{equation}
One can show that this
result is valid in the whole domain of interest 
$1/3<\nu<2$, i.e. the domain where
the variance of $\chi$ is infinite.

\subsection{The variance of $\chi$}

We here obtain the finite $\langle \chi^2 \rangle$ for the case $\nu>2$ 
 for the regularized force $F(p) = - p/(1 + p^2)$.
Using the Laplace $\chi \to s$ transform of  Eq. (\ref{eqApB07})
we find the following backward equation for $\hat{q}(s)$:
\begin{equation}
D {\partial^2 \over \partial (p_i)^2} \hat{q}(s) + F(p_i) {\partial \over \partial p_i} \hat{q}(s) - p_i s \hat{q}(s)=0. 
\label{eqssvar01}
\end{equation}
Here we used $q(\chi)|_{\chi =0} =0$ since the particle starting with $p_i>0$
cannot reach zero momentum without traveling some finite distance. 
The Laplace transform $\hat{q}(s)$ is expanded in $s$:
\begin{equation}
\hat{q}(s) = \int_0 ^\infty e^{ - s \chi} q(\chi) {\rm d} \chi =
1 - s \langle \chi \rangle + s^2 \langle \chi^2 \rangle / 2 + {}\cdots .
\label{eqssvar02}
\end{equation}
Here $\langle \chi \rangle$ is the average jump size, for positive
excursions i.e. those that start with $p_i >0$ and $\langle \chi^2 \rangle$
is the second moment. Note that in the original model we have
both positive and negative excursions  and so 
we have $p_i=+\epsilon$ or 
$p_i = -\epsilon$ with probability $1/2$ hence for that case 
$-\infty < \chi<\infty$ and from symmetry $\langle \chi \rangle =0$. 
The variance of the original process is $\langle \chi^2 \rangle$
due to symmetry. As we restrict ourselves to  $p_i>0$,  $\chi>0$
and $\langle \chi \rangle$ in Eq. (\ref{eqssvar02}) is finite. 

Inserting Eq. (\ref{eqssvar02}) in Eq. (\ref{eqssvar01}) we find
\begin{equation}
D{\partial^2 \over \partial (p_i)^2} \left[ - s \langle \chi\rangle  + s^2 {\langle \chi^2 \rangle \over 2 }  +{} \cdots \right] - {p_i \over 1 + (p_i)^2 } 
{\partial  \over \partial p_i} \left[ - s \langle \chi \rangle + s^2{ \langle \chi^2 \rangle \over 2 }  + {}\cdots \right] -p_i s \left[ 1 - s \langle \chi \rangle +s^2  {\langle \chi^2 \rangle \over 2} +{} \cdots \right] = 0.
\label{eqssvar03}
\end{equation}
The $s^1$ terms give
\begin{equation}
- D {\partial^2 \over \partial (p_i)^2 } \langle \chi \rangle + { p_i \over 1 + (p_i)^2 } {\partial \over \partial p_i} \langle \chi \rangle - p_i = 0 ,
\label{eqssvar03a}
\end{equation}
while the $s^2$ terms are
\begin{equation}
 {D \over 2}  {\partial^2 \over \partial (p_i)^2 } \langle \chi^2 \rangle - {1\over 2} { p_i \over 1 + (p_i)^2 } {\partial \over \partial p_i} \langle \chi^2 \rangle + p_i\langle \chi \rangle = 0 .
\label{eqssvar04}
\end{equation}
This means that the first and then the  second moment can
be found by repeated integration.
The boundary conditions
are that both the first and second moments are zero if $p_i=0$
while they diverge if $p_i = \infty$. Using a reflecting boundary at
$p_i = \infty$ we find 
(see e.g. \cite{Gardiner}, chapter XX)
\begin{equation}
\langle \chi (p_i) \rangle = {1 \over D}  \int_0 ^{p_i} {\rm d} y \, e^{ V(y) /D} \int_{y} ^\infty 
  {\rm d} z  e^{ - V(z)/D} z,
\label{eqssvar05}
\end{equation}
where the lower bound is the momentum origin  
and the potential is given in Eq. (\ref{eqvp}).
Integrating Eq.
(\ref{eqssvar04})  we find
\begin{equation}
\langle \chi^2 \rangle = {2 \over D}  \int_0 ^{p_i} {\rm d} y\, e^{ V(y) /D} \int_{y} ^\infty {\rm d} z\, e^{ - V(z) / D} z \langle \chi(z) \rangle. 
\label{eqssvar06}
\end{equation}
As explained in the text we consider only the limit when $p_i= \epsilon$
is small so we have
\begin{equation}
\langle \chi^2 \rangle \sim {2 \epsilon \over D} \int_{0} ^\infty {\rm d} z \,e^{ - V(z) / D} z \langle \chi(z) \rangle,
\label{eqssvar07}
\end{equation}
since $V(0)=0$. 
Using Eq. (\ref{eqssvar05}) and integrating by parts
we get
\begin{equation}
\langle \chi^2 \rangle \sim {2 \epsilon \over D^2} \int_{0} ^\infty {\rm d} z\, e^{  V(z) / D} \left[ \int_{z} ^\infty {\rm d} z\, z e^{-V(z) /D}  \right]^2. 
\label{eqssvar08}
\end{equation}
Note that from
symmetry  $V(p) = V(-p)$, 
so we have $\int_z ^\infty {\rm d} z\, z \exp[ - V(z)/D]  = -
\int_{-\infty} ^{-z} {\rm d} z \, z \exp[- V(z)/D] $. 

\section{Expansion of Eq. (\ref{eqAs09}) }

Since $g(\tau)\sim g^{*} \tau^{ -3/2 - \gamma}$ for large $\tau$, with $\gamma=1/(2 D)$, we assume that
there exists some $t_0$ such that $g(\tau) = g^{*} \tau^{-(3/2+\gamma)}$ 
for $\tau>t_0$. Then 
using Eq. (\ref{eqAs07}) and  $\nu = 1/3 + 2 \gamma/3$,
\begin{equation}
\widehat{\psi}(k,u)\Big|_{u=0} = 1 - \int_0 ^{t_0} {\rm d} \tau \left[ 1 - \widehat{B} \left( k\sqrt{D\tau^3}  \right) \right] g\left( \tau \right) +
\int_{t_0} ^\infty {\rm d} \tau \left[ 1 - \tilde{B} \left( k \sqrt{D\tau^3} \right) \right] g_{*}  \tau^{-3/2 - \gamma } .
\label{eqAppC01} 
\end{equation}
The first term on the right hand side yields the normalization condition
$\widehat{\psi} (k,u)|_{k=u=0} = 1$, the second term is zero for $k=0$ and due 
to symmetry of the jumps $B(v_{3/2}) = B(-v_{3/2})$,
 this term when expanded in $k$ gives
a $k^2$ term. As we will show, the last term gives a $|k|^\nu$ term,
so as long as $\nu<2$ we can neglect the second term. Then changing
variables to $\tilde{k} = k \sqrt{D\tau^3}$ we find
\begin{equation}
\widehat{\psi}(k,u)\Big|_{u=0} = 1 - {2 \over 3} g_{*} (Dk^2)^{\nu/2} \int_0 ^\infty {1 - \tilde{B} (\tilde{k}) \over (\tilde{k})^{1 + \nu}} {\rm d} \tilde{k} . 
\label{eqAppC02} 
\end{equation} 
Here we have taken the small $k$ limit in such a way that 
$k (t_0)^{3/2}$ which appears in the lower limit of an integral is
negligible.

From Eq. (\ref{eqAppC02}) we must investigate
the integral
\begin{equation}
{\cal I}_\nu=  \int_0 ^\infty {1 - \tilde{B} (k) \over (\tilde{k})^{1 + \nu}} {\rm d} \tilde{k} . 
\label{eqAppC03} 
\end{equation} 
 From symmetry $B(v_{3/2}) = B(-v_{3/2})$
and hence  the Fourier transform 
$\tilde{B}(k)$ is a real function so
\begin{equation}
{\cal I}_\nu= Re\left[  \int_{k_c}  ^\infty {1 - \tilde{B} (k) \over (\tilde{k})^{1 + \nu}} {\rm d} \tilde{k} \right],
\label{eqAppC04} 
\end{equation} 
and $k_c$ will eventually be taken to  zero. 
Using the definition of the Fourier transform
\begin{equation}
{\cal I}_\nu = {{ k_c}^{-\nu} \over \nu} - 2\mbox{Re} \left[ \int_{0} ^\infty {\rm d} v_{3/2}  B(v_{3/2})  \int_{k_c} ^\infty { e^{ i k v_{3/2}} \over k^{1 + \nu} } {\rm d} k \right] .
\label{eqAppC05} 
\end{equation} 
Let $i v_{3/2} = - x$ and then using integration by parts,
\begin{equation}
\mbox{Re} \left[ \int_{k_c} ^\infty \,{e^{ i k v_{3/2}} \over k^{1 + \nu}} \right] =
\mbox{Re}\left[ e^{ - k_c x} {(k_c)^{-\nu} \over \nu} -{x}^\nu \Gamma(1 -\nu)/\nu \right]  . 
\label{eqAppC06} 
\end{equation} 
Inserting the last expression in Eq. (\ref{eqAppC05}), the
 diverging $(k_c)^{-\nu}$ terms cancel each other,
and we find 
\begin{equation}
{\cal I}_\nu= \frac{2\Gamma(1-\nu)}{\nu} \mbox{Re} \left[ \int_{0} ^\infty {\rm d} v_{3/2}  B(v_{3/2}) x^\nu  \right]. 
\label{eqAppC07} 
\end{equation} 
Using $x=-i v_{3/2}$ we get
\begin{equation}
{\cal I}_\nu = - \langle |v_{3/2}|^\nu \rangle \sin \left( { \pi(1 + \nu) \over 2} \right) \Gamma(-\nu),
\label{eqAppC07a} 
\end{equation} 
where $\langle \left|v_{3/2}\right|^\nu\rangle = \int_{-\infty} ^\infty |v_{3/2}|^\nu B(v_{3/2}) {\rm d} v_{3/2}$. 
Inserting Eq. (\ref{eqAppC07a})  
in (\ref{eqAppC02}), using  
 (\ref{eqAppC03})  we get  
Eq. (\ref{eqAs10}).

\end{widetext}

\section{On the simulations}
\label{SecAppDSim}

\subsection{The simulation}
In this appendix, we briefly discuss the simulations presented in 
the paper to test our predictions. We have two classes of simulations:
  one  generates
Bessel excursions, and the second solves the Langevin equations. 
The first is based  on a discrete random walk treatment of the excursion,
 wherein the particle takes a biased walk to the right or left 
in momentum space every time step.   
 The degree of bias varies with $p$ in accord with the force $F(p)$.  Our dimensionless control parameter $D$ is the ratio of the coefficient of the diffusion term to the coefficient of the $1/p$ large $p$ behavior of the force. Since for our simple random walk the diffusion constant is $1/2$, the parameter $D$ enters via the strength of the bias. The probability to move right or left is then given by
\begin{equation}
P_\pm(p) = \frac{1}{2} \left( 1 \pm \frac{1}{2D}F(p)\Delta_p\right)
\end{equation}
The continuum limit is approached, as usual, when $\Delta_p \to 0$.  Given that in our units $F(p)$ varies on the scale of unity, it is sufficient to take $\Delta_p \ll1$.  In practice, we took $\Delta_p=0.1$. We also monitor the position, adding $p$ to the position every time the particle takes a step starting at momentum $p$.

To simulate paths of the Langevin dynamics
we used  both a straightforward Euler-Maruyama integration of the 
Langevin equation  as well as one based on the random walk picture above.
Both methods give equivalent results. 

\subsection{The Bessel Excursion}
An efficient way to generate the set of Bessel excursions is 
to generate all discrete
Brownian excursions with equal probability and then to weight each excursion 
by its appropriate weight.  The task of generating the ensemble of 
Brownian excursions is at first sight a nontrivial problem, 
since the constraint that the random walk not cross the origin is
 nonlocal in the individual left/right steps constituting the walk.  However, 
it is easily accomplished using Callan's proof~\cite{Callan} of the 
famed Chung-Feller 
theorem~\cite{ChungFeller}.  Callan gives a explicit mapping of any 
$N$ left, $N$ right 
step random walk to a unique walk that does not cross the origin.
 This mapping maps exactly $N+1$ of the $2 N \choose N$ $N$ left, $N$ right 
 walks,
 to each of the non-zero crossing walks (of which there are ${2 N \choose N} / (N + 1)$).
  Since every Brownian excursion of $N+2$ steps consists 
of an initial right step, a non-zero-crossing $N$ left, $N$ right 
step walk and  a final left step back to the origin, Callan's mapping allows us
 to generate an $N+2$ step Brownian excursion by generating a random $N$ left,
 $N$ right step walk, apply the map, 
and pre- and post-pending the appropriate steps.  In this way every excursion is generated with exactly the same probability. 

Callan's  mapping is as follows. For a given $N$ left, $N$ right step walk,
 one accepts it as is if it does not cross the origin. Otherwise,
 one finds the leftmost point 
reached by the  walk
(i.e., the minimal $x$ of the path), 
 which we denote $x_n$, where $n$ is the number
of steps to reach this point.  
If this point is visited more than once, 
then one takes the first visit.  One then constructs a new walk based on the
original walk. We divide the original walk into two segments, 
the first $n-1$ steps and the remaining part. The first part
of the new walk consists of the second segment,  shifted
by $-x_n$ such that the  new walk  starts on the origin.
In the original walk the step before reaching $x_n$ is obviously a left
step. 
One  appends this  left step at the end  of the
walk under construction,  and then attaches the
first segment (of the original walk). This new walk clearly 
does not cross the origin, while preserving the number of left and right steps,
hence it starts and ends on the origin.  

To account for the bias, i.e., to switch from Brownian 
to Bessel excursions, 
it is sufficient to weight each excursion by the weight factor, $F$:
\begin{equation}
F= \prod_{i=1}^N f_i \ ; \quad f_i = \left\{\begin{array}{ll} 1 + { F(p_i) \Delta_p \over 2 D}  & \textrm{Step $i$ is a right step} \\ 1- { F(p_i)  \Delta_p\over 2D} \quad & \textrm{Step $i$ is a left step} \end{array}\right.
\end{equation}
This works well as long as $D$ is not too small. For very small $D$, the convergence is quite slow since small areas have anomalously large weight unless $N$ is very large.  As long as one is interested in the large $N$ behavior, better convergence is achieved by taking $F(p) = \theta(p-1)/p$.

\begin{figure}\begin{center}
\includegraphics[width=0.41\textwidth]{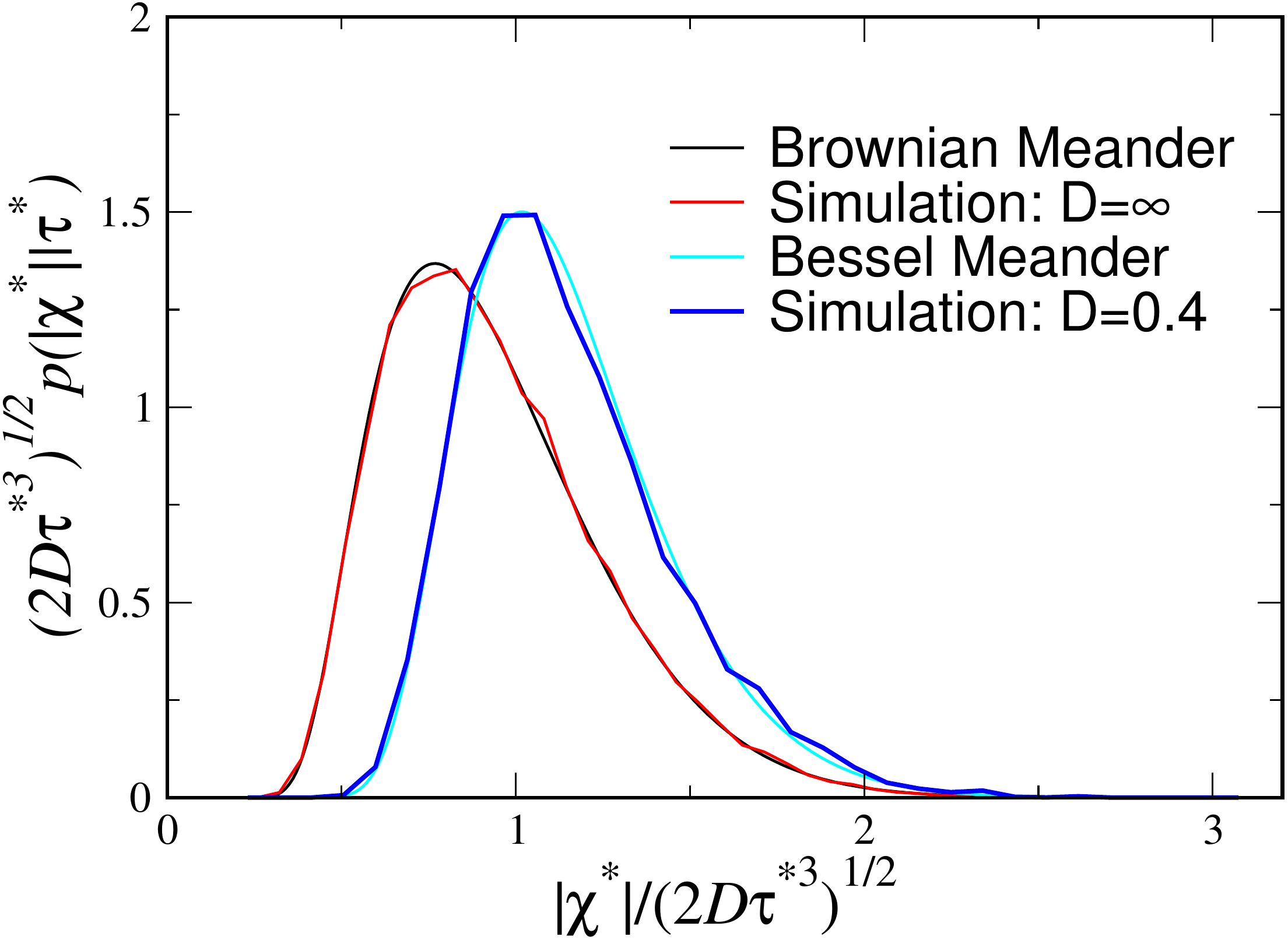}
\end{center}
\caption{
The conditional PDF 
of the area under the Bessel $(D=2/5)$  and Brownian $(D \rightarrow \infty)$
meanders,
Eq. (\ref{eqerez15}).
For $D=2/5$, we used  $5$  term summation 
in Eq. (\ref{eqerez15}), 
 with $a_1=0.849,a_2=-0.314,a_3=0.535,a_4=-0.3072,
a_5=0.441$
and values of $\tilde{d}_k$ in  Table \ref{table1}.
The simulation method is outlined in Appendix
\ref{SecAppDSim}
 and we averaged over $2\times 10^5$ samples with $\tau=10^5$.
Theoretical curve was plotted with Maple. 
}
\label{figE01}
\end{figure}

\section{Areal distribution of the Bessel Meander}

Our goal is to find the conditional PDF
 $p_M(\chi^*|\tau^*)$ where $\chi^*= \int_0 ^{\tau^*} p(t') {\rm d} t'$,
is the area under the Brownian meander.
The starting point for the calculation of $p_M(\chi^*|\tau^*)$
is a modification of Eq. 
(\ref{eqBessel02}).
As for the area under the Bessel excursion, 
we consider only the positive meander
where $0<\chi^*<\infty$, and later use the symmetry of the process
to find the areas under both positive and negative meanders.
 Since contrary to the condition on the excursion,
the meander is not bound to return to the origin, we now keep the
end point free and integrate the propagator $\hat{G}_{\tau^*}(s,p|p_{i})$
over all possible values of $p$. 
Hence
\begin{equation}
\hat{p}_{M}(s|\tau^*)=\lim_{\epsilon\rightarrow 0}\frac{\intop_{0}^{\infty}\hat{G}_{\tau^*}(s,p|\epsilon){\rm d} p}{\intop_{0}^{\infty}\hat{G}_{\tau^*}(s=0,p|\epsilon){\rm d} p};
\label{eqErez01}
\end{equation}
with
$\hat{p}_M(s|\tau^*) = \int_0 ^\infty p_M(\chi^*|\tau^*) \exp( - s D \chi^*) {\rm d} \chi^*$.

\begin{widetext}
We expand the propagator $\hat{G}_{\tau^*}(s,p|p_{i})$ in a complete orthonormal
basis, using the same approach as in the main text, while accounting for the
boundary conditions of the meander. We thus rewrite Eq.  
 (\ref{eqBessel02})
for the
case of the meander and integrate over $p$ 
\begin{equation}
\intop_{0}^{\infty}\hat{G}_{\tau^*}(s,p|\epsilon){\rm d} p=\epsilon^{1/2D}s^{1/3}\intop_{0}^{\infty}{\rm d}p  p^{-1/2D}\sum_{k}f_{k}(s^{1/3}\epsilon)f_{k}(s^{1/3}p)e^{-DE_{k}\tau^*},
\label{eqErez02}
\end{equation}
where the functions $f_{k}(.)$, as before, are the solutions of the
time independent equation
(\ref{eqBessel09}) 
and $E_{k}=s^{2/3}\lambda_{k}$. Changing variables to $\tilde{p}=s^{1/3}p$,
and using the small $p$ behavior  Eq. 
(\ref{eq28er}),
which gives 
$f_{k}(s^{1/3}\epsilon)\sim\tilde{d_{k}}s^{1/3+1/6D}\epsilon^{1+1/2D}$,
we find 
\begin{equation}
\intop_{0}^{\infty}\hat{G}_{\tau^*}(s,p|\epsilon){\rm d} p=\epsilon^{2\alpha}s^{\nu}\sum_{k}\tilde{d_{k}}e^{-Ds^{2/3}\lambda_{k}\tau^*}\intop_{0}^{\infty}\frac{f_{k}(\tilde{p})}{\tilde{p}^{1/2D}}d\tilde{p},
\label{eqErez03}
\end{equation}
for the numerator in Eq. (\ref{eqErez01}).
We now define a new k-dependent
coefficient 
\begin{equation}
a_{k}\equiv \intop_{0}^{\infty}\frac{f_{k}(\tilde{p})}{\tilde{p}^{1/2D}}d\tilde{p}.
\label{eqErez04}
\end{equation}
 The $a_{k}$'s, similar to the $\tilde{d_{k}}$'s,
are evaluated from a numerically exact
solution of Eq. 
(\ref{eqBessel09}).
 Rewriting in terms of the
new coefficients 
\begin{equation}
\intop_{0}^{\infty}\hat{G}_{\tau^*}(s,p|\epsilon){\rm d} p=\epsilon^{2\alpha}s^{\nu}\sum_{k}\tilde{d_{k}} a_{k}e^{-Ds^{2/3}\lambda_{k}\tau^*}.
\label{eqErez05}
\end{equation}
 For the denominator in Eq. (\ref{eqErez01}), we expand the propagator in the
orthonormal basis for the $s=0$ case,
\begin{equation}
\hat{G}_{\tau^*}(s=0,p|\epsilon)=\sum_{k}(\frac{\epsilon}{p})^{1/2D}\phi_{k}^{0}(\epsilon)\phi_{k}^{0}(p)e^{-DE_{k}\tau^*}.
\label{eqErez06}
\end{equation}
 Integrating over $p$, as required for the meander,
we get
\begin{equation}
\int_{0}^{\infty}{\rm d} p\hat{G}_{\tau^*}(s=0,p|\epsilon)=\int_{0}^{\infty}{\rm d} p\sum_{k}(\frac{\epsilon}{p})^{1/2D}\phi_{k}^{0}(\epsilon)\phi_{k}^{0}(p)e^{-DE_{k}\tau^*}.
\label{eqErez07}
\end{equation}
 Plugging in $\phi_{k}^{0}(\epsilon)$ and $\phi_{k}^{0}(p)$ using
Eq. 
(\ref{eqBessel16}),
with $B_{k}=\sqrt{\pi k/L}$, we find  
\begin{equation}
\int_{0}^{\infty}{\rm d} p\hat{G}_{\tau^*}(s=0,p|\epsilon)=\epsilon^{1/2+1/2D}\frac{\pi}{L}\int_{0}^{\infty}{\rm d} p\sum_{k}p^{-1/2D}kJ_{\alpha}(k\epsilon)\sqrt{p}J_{\alpha}(kp)e^{-DE_{k}\tau^*}.
\label{eqerez08}
\end{equation}
 Transforming the sum into an integral over $k$ : 
$\sum_{k}\rightarrow\frac{L}{\pi}\intop_{0}^{\infty}{\rm d} k$,
and using the small $p$ behavior of the Bessel function for $J_{\alpha}(k\epsilon)\sim\frac{(k\epsilon)^{\alpha}}{2^{\alpha}\Gamma(1+\alpha)},$
we get
\begin{equation}
\int_{0}^{\infty}{\rm d} p\hat{G}_{\tau^*}(s=0,p|\epsilon)=\frac{\epsilon^{2\alpha}}{2^{\alpha}\Gamma(1+\alpha)}\int_{0}^{\infty}{\rm d} p p^{{1\over 2}- {1\over 2D}}\int_{0}^{\infty}{\rm d} k k^{1+\alpha}J_{\alpha}(kp)e^{-DE_{k}\tau^*},
\label{eqerez09}
\end{equation}
with $E_{k}=k^{2}$ in this continuum limit. To solve Eq. (\ref{eqerez09}),
define $q=kp$, and so 
\begin{equation}
\int_{0}^{\infty}{\rm d} p\hat{G}_{\tau^*}(s=0,p|\epsilon)=\frac{\epsilon^{2\alpha}}{2^{\alpha}\Gamma(1+\alpha)}\int_{0}^{\infty}{\rm d}q\int_{0}^{\infty}{\rm d} k k^{1/D}q^{1/2-1/2D}J_{\alpha}(q)e^{-Dk^{2}\tau^*},
\label{eqerez10}
\end{equation}
then inserting $x=k^{2}$, we get 
\begin{equation}
\int_{0}^{\infty}\hat{G}_{\tau^*}(s=0,p|\epsilon){\rm d} p=\frac{\epsilon^{2\alpha}}{2^{\alpha+1}\Gamma(1+\alpha)}\intop_{0}^{\infty}{\rm d} x x^{\alpha-1}e^{-Dx\tau^*}\intop_{0}^{\infty}{\rm d} q q^{1/2-1/2D}J_{\alpha}(q).
\label{eqerez11}
\end{equation}
 The two integrals on the right-hand side of Eq. (\ref{eqerez11}) 
can be evaluated
analytically, and the final result for the denominator of Eq. (\ref{eqErez01})
is 
\begin{equation}
\int_{0}^{\infty}\hat{G}_{\tau^*}(s=0,p|\epsilon){\rm d} p=\frac{\epsilon^{2\alpha}}{2^{2\alpha}\Gamma(1+\alpha)}(D\tau^*)^{-\alpha}.
\label{eqerez12}
\end{equation}
 Plugging Eqs. (\ref{eqErez03},\ref{eqErez04}) and (\ref{eqerez12}) 
into Eq. (\ref{eqErez01}), we reach
our first main result for the area under the Bessel meander 
\begin{equation}
\hat{p}_{M}(s|\tau^*)=2^{2\alpha}\Gamma(1+\alpha)[s(D\tau^*)^{3/2}]^{\nu}\sum_{k}\tilde{d_{k}}a_{k}e^{-Ds^{2/3}\lambda_{k}\tau^*}.
\label{eqerez13}
\end{equation}
 The inverse of $\hat{p}_{M}(s|\tau^*)$, i.e., $s D \rightarrow \chi^*$, is 
\begin{equation}
p_{M}(\chi^*|\tau^*)=2^{2\alpha}\Gamma(1+\alpha)D^{\nu/2}[(\tau^*)^{3/2}]^{\nu}\sum_{k}\tilde{d_{k}}a_{k}\sum_{n=0}^{\infty}\frac{(-c_{k})^{n}(\chi^*)^{-(\nu+2n/3+1)}}{\Gamma(-\nu-2n/3)n!}.
\label{eqerez14}
\end{equation}
with $c_k = D^{1/3} \lambda_k \tau^*$. 
 And finally, summing over $n$ using Maple,
\begin{eqnarray}
p_{M}(\chi^*|\tau^*) & =\Gamma(1+\alpha)[\frac{4^{3/2} D^{1/2}(\tau^*)^{3/2}}{\chi^*}]^{\nu}(-\frac{1}{\pi\chi^*})\sum_{k}\tilde{d_{k}}a_{k}\times[\Gamma(1+\nu)\sin(\pi\nu)_{2}F_{2}(\frac{\nu}{2}+1,\frac{\nu}{2}+\frac{1}{2};\frac{1}{3},\frac{2}{3};\frac{-4(D^{1/3}\lambda_{k}\tau^*)^{3}}{27(\chi^*)^{2}})\nonumber \\
 & -(\frac{D^{1/3}\lambda_{k}\tau^*}{(\chi^*)^{2/3}})\Gamma(\frac{5}{3}+\nu)\sin(\pi\frac{2+3\nu}{3})_{2}F_{2}(\frac{\nu}{2}+\frac{4}{3},\frac{\nu}{2}+\frac{5}{6};\frac{2}{3},\frac{4}{3};\frac{-4(D^{1/3}\lambda_{k}\tau^*)^{3}}{27(\chi^*)^{2}})\nonumber \\
 & +\frac{1}{2}(\frac{D^{1/3}\lambda_{k}\tau^*}{(\chi^*)^{2/3}})^{2}\Gamma(\frac{7}{3}+\nu)\sin(\pi\frac{4+\nu}{3})_{2}F_{2}(\frac{\nu}{2}+\frac{7}{6},\frac{\nu}{2}+\frac{5}{3};\frac{4}{3},\frac{5}{3};\frac{-4(D^{1/3}\lambda_{k}\tau^*)^{3}}{27(\chi^*)^{2}})].
\label{eqerez15}
\end{eqnarray}
This is the main result of this Appendix. 
In Fig. \ref{figE01}) we plot this areal distribution,
comparing it with a histogram obtained from finite time
simulations.  
%

\end{widetext}
\section{The time interval straddling $t$}

 The velocity process $v(t')$ restricted to the zero-free interval
containing a fixed observation time $t$ is called the excursion process
straddling $t$, and the portion of it up to $t$ is the meandering
process ending at $t$. For Brownian and Bessel processes these random
paths were the subject of intense mathematical investigation,
e.g. \cite{Chung,Getoor,Bertoin} and references therein.
 The investigation of the duration
of the excursion straddling $t$,  the time interval
for the meander ending
at $t$, and statistical properties of the path e.g.,
its maximal height, have attracted mathematical
attention since these reveal deep
and beautiful properties of Brownian motion in dimension
$d$ (as is well known the Bessel process is constructed from
the radius of a $d$ dimensional
Brownian motion). One aspect of the problem is the quantification of the 
properties of the set of points
on the time axis on which the zero crossings take place. Clearly
the number of zero crossings for a Brownian path starting at the
origin within the time interval $(0,t)$ is infinite, due to the continuous
nature of the path (see more details below).
One might naively expect that the time between
points of zero crossing approaches zero,
 since a finite measurement time divided
by
infinity is zero. That, in the context of our paper, might be true
when $\langle \tau\rangle$ is finite. Then we can expect
to find a zero crossing in the close vicinity of the measurement time $t$
(i.e., at a distance of the order of $\langle \tau \rangle$ from $t$).
 However,
when $0<\alpha<1$ the situation is more subtle.
In this case the points on time axis will be clustered,
 and while
their number is infinite visualizing these dots with a simulation we will
observe a fractal dust.


 In this Appendix we investigate
the statistics of duration of the excursion straddling time $t$,
deriving some known results along the way.
Our approach is based on renewal theory, using methods given
by Godreche and Luck \cite{GL}.
The previous mathematical approaches are based
on direct analysis of Brownian and Bessel motion,
while we use the $\epsilon$ trick introduced in the main text.
As mentioned,
we  replace the continuous regularized Bessel process
 or Brownian path,
with a non-continuous one, with jumps of size $\epsilon$ after zero
crossing,
and  at the end $\epsilon$ is taken to zero.

 Consider the PDF of waiting times $g(\tau)$ with $0<\tau < \infty$
with the long time limiting behavior
\begin{equation}
g(\tau) \sim g^{*} \tau^{-(1 + \alpha)}.
\label{eqApz01}
\end{equation}
While $\alpha=(1 +  D)/2D$ for the problem in the main manuscript,
hence $\alpha\ge 1/2$, here we  will assume that  $0<\alpha<1$.
The case $\alpha=1/2$ is the Brownian case, since a particle starting on
$\epsilon$ will return for the first time to the origin according to the
well-known law
$g(\tau) \propto \tau^{-3/2}$.  In our physical problem,
the $\epsilon$ trick means that $g^{*}$ is $\epsilon$ dependent, Eq.
(\ref{eqAs07}).
As in the main text, we denote
$\{ t_1, t_2 , \cdots, t_k, \cdots\}$ as times of the renewal
events (i.e., in our problem
zero crossing events). This is a finite set of times if $\epsilon$
is finite. The waiting times $\{ \tau_1, \tau_2 , \cdots\}$ are independent
identically distributed random variables, due to the Markovian
property of the underlying paths,
and $t_1=\tau_1$, $t_k = \sum_{i=1} ^k \tau_i$.
 We call $t$ the observation time, and the number of renewals
in $(0,t)$ is denoted by $n$, so by definition $t_n < t<t_{n+1}$
(this is obviously true as long as $\epsilon$ is finite).
The time $B=t-t_n$ is called the backward recurrence time \cite{GL},
 the time $F=t_{n+1} - t$ is the forward recurrence time,
and $\Delta= t_{n+1} - t_n$ is the time interval straddling $t$.
In our problem, the backward recurrence time is the duration
of the
meandering process which starts at $t_n$ and  ends at $t$.
For Brownian motion and Bessel processes,
 where $t_n$ and $t_{n+1}$ are the zero hitting times
straddling $t$, the statistics of $F,B$ and $\Delta$
are non-trivial.

 We now obtain the PDF of $\Delta$, denoted $d_t(\Delta)$,
using the methods in \cite{GL}.
From the constraint $t_n<t<t_{n+1}$ we have
\begin{equation}
d_t (\Delta) = \sum_{n=0} ^\infty \langle
\delta\left[\Delta - (t_{n+1} - t_n)\right] I(t_n < t < t_{n+1}) \rangle
\label{eqApz02}
\end{equation}
where $\delta[\cdots]$ is the Dirac $\delta$ function, and
$I(t_n < t < t_{n+1}) =1$ if the condition in the parenthesis is
true, otherwise it is zero. We use the double Laplace
transform
\begin{equation}
d_s(u) = \int_0 ^\infty \int_0 ^\infty e^{ - s t}  e^{- u \Delta}
d_t(\Delta) {\rm d}t {\rm d}\Delta,
\label{eqApz03}
\end{equation}
and  compute
\begin{widetext}
\begin{equation}
 \left\langle 
\int_0 ^\infty  \int_0 ^\infty e^{ - s t} e^{ - u \Delta} \delta\left[ \Delta - ( t_{n+1} - t_{n}) \right] I (t_n < t< t_{n+1}) {\rm d} t {\rm d} \Delta \right\rangle = 
 {1 \over s}\left\langle \left( e^{ - s t_n}-e^{ - s t_{n+1}}\right) e^{- u(t_{n+1} - t_n) } \right \rangle.
\label{eqApz03a}
\end{equation}
\end{widetext}
Since the waiting times $\{\tau_i\}$
 are mutually independent, identically distributed
random variables we find using $t_n= \sum_{i=1} ^n \tau_i$
\begin{equation}
\langle e^{ - s t_n} \rangle= \hat{g}^n (s),
\label{eqApz04}
\end{equation}
where $\hat{g}(s)= \int_0 ^\infty e^{ - s \tau} g(\tau) {\rm d} \tau$ is the
Laplace transform of $g(\tau)$. A similar result holds  for the
 other expressions
in Eq. (\ref{eqApz03a}). It is then easy to find
\begin{equation}
d_s(u) = \sum_{n = 0} ^\infty \hat{g}^n (s) { \hat{g}(u) - \hat{g}(u + s) \over s},
\label{eqApz05}
\end{equation}
and summing the geometric series we get
\begin{equation}
d_{s} (u) = { \hat{g}(s) - \hat{g} (s + u) \over s \left[ 1 - \hat{g} (s) \right] } .
\label{eqApz06}
\end{equation}
When $g(\tau) = \exp( - \tau)$ we find in the limit of large $t$
$d_t(\Delta) \to  \Delta \exp( - \Delta)$ so the minimum of this
PDF is at
$\Delta=0$ and $\Delta=\infty$ which is expected.
For our more interesting case, $\hat{g}(u) \sim 1 - g^{*} |\Gamma(- \alpha)|
u^\alpha    $ for small $u$, hence we get in the small $u$ and $s$ limit,
with an arbitrary ratio between them
\begin{equation}
d_s(u) \sim { (u + s)^\alpha - u^{\alpha} \over s^{1 + \alpha} }.
\label{eqApz07}
\end{equation}
This is a  satisfying result implying that  the solution does not depend
on $g^{*}$ and  hence it does not depend on $\epsilon$. Thus the artificial
cutoff, $\epsilon$, which was introduced merely as a mathematical tool, does
not alter our final formulas.
Let $\tilde{\Delta} = \Delta/t$ and denote by $\tilde{d} (\tilde{\Delta})$
its PDF in the long time limit.
With  a useful inversion formula, given  in the Appendix of \cite{GL},
we invert Eq. (\ref{eqApz07}) to the
time domain and find
\begin{equation}
\tilde{d} \left( \tilde{\Delta} \right) = { \sin \pi \alpha \over \pi
(\tilde{\Delta})^{ 1 + \alpha} } \left[ 1 - \left( 1 - \tilde{\Delta}\right)^\alpha I\left(0<\tilde{\Delta} <1\right) \right],
\label{eqApz08}
\end{equation}
which exhibits a discontinuity of its derivative at $\tilde{\Delta} = 1$.
Thus the PDF of $0<\Delta<\infty$ has a cusp at $t$,
which allows for the identification of the measurement time $t$
from a histogram of $\Delta$.
For Brownian motion, $\alpha=1/2$
\begin{equation}
\tilde{d} \left( \tilde{\Delta} \right) = \left\{
\begin{array}{c  c}
{ 1 \over \pi} { 1 - (1 - \tilde{\Delta} )^{1/2} \over
\left( \tilde{\Delta}\right)^{3 /2} } & \tilde{\Delta} < 1 \\
{1 \over \left( \tilde{\Delta}\right)^{3 /2} } & \tilde{\Delta} > 1.
\end{array}
\right.
\label{eqApz09}
\end{equation}

 Instead of fixing $t$,  we may
draw $t_e>0$ from an exponential PDF, $R \exp( - R t_e)$,
 in such a way that the mean
of $t_e$, $ \langle t_e \rangle =1/R$,
 is very large,
 so that the  number of renewals in the time
interval $(0,1/R)$ is large. Similar to the previous case, we define
$\Delta_e = t_{n+1} - t_n$, which is called the duration of the interval
straddling an independent exponential time, and $ t_n<t_e <t_{n+1}$.
This case was treated rigorously by Bertoin et al. \cite{Bertoin},
 for the Bessel process.
Using the renewal theory approach, we  now obtain their main result
on  the statistics of $\Delta_e$, with a few hand waving arguments.
 From the definition of the Laplace
transform, $\hat{f}(s) = \int_0 ^\infty f(t_e)\exp( - s t_e) {\rm d} t_e$,
we see that for a function $f(t_e)$ depending on a random variable $t_e$,
the later being exponentially distributed with mean equal to $1/R$, the
averaged function is  the Laplace transform of $f(t_e)$
 evaluated at
$s=1/R$ followed by multiplication with $R$, namely
$\langle f(t_e) \rangle = R \hat{f}(R)$. Inserting $s=R$ in
Eq. (\ref{eqApz06}) followed by multiplication with $R$, we get
the Laplace transform of the PDF of $\Delta_e$
\begin{equation}
\langle \exp( - u \Delta_e) \rangle = [(u + R)^\alpha - u^\alpha]/R^\alpha,
\label{eqApzo9}
\end{equation}
where we used the small $R$ and $u$ limit and as usual
$\langle \exp(- u \Delta_e)\rangle = \int_0 ^\infty e^{- u \Delta_e }
d_{e} (\Delta_e) {\rm d} \Delta_e$ where $d_e (\Delta_e)$ is the PDF
of $\Delta_e$.
This  is the known result for the Bessel process when scaled properly,
i.e., $R=1$ in \cite{Bertoin} who also give the
inverse Laplace transform of Eq. (\ref{eqApzo9}),
thus providing an explicit formula for the
PDF  $d_e(\Delta_e)$.

  To see the connection between renewal theory,
and statistics of the duration of Bessel excursion
straddling an exponential time,
notice that in \cite{Bertoin}   a Bessel
process in dimension
$0<d<2$ is considered with the relation $d=2(1-\alpha)$.
 In our work we consider
motion in a logarithmic potential in one dimension, which is easily
 mapped onto a Bessel process in dimension $d=1-1/D$.
Thus, the exponent $\alpha$ in Ref.  \cite{Bertoin}
is the same as ours
since as we have shown $\alpha=(1+ D)/(2 D)$, Eq. 
(\ref{eqBessel15}).
Notice that for optical lattices $D>0$, hence this system
is a physical
example  for a regularized Bessel process in dimension $-\infty < d <1$.
Finally, note that the original Bessel process considers the distance
$|r|$ from the origin of a Brownian motion in $d$ dimension, this
being non-negative $|r|\ge 0$. Hence the Bessel process
does not exhibit zero crossings, so the points on the time axis
are zero hitting points, not zero crossing points. This is a minor technical
issue, due the symmetry of the binding effective potential discussed
in the manuscript (i.e., negative and positive excursions in the logarithmic
potential are statistically identical). 

 Similarly, we can obtain the limiting distributions of $B$ and $F$
using renewal theory \cite{GL}.
The PDF of $\tilde{B} = B /t$ (here $t$ is fixed) is
\begin{equation}
\mbox{PDF}(\tilde{B}) \sim { \sin \pi \alpha \over \pi} \left( \tilde{B}\right)^{ - \alpha} \left( 1 - \tilde{B}\right)^{ \alpha -1} \ \ \ \ 0<\tilde{B} < 1
\label{eqApz10}
\end{equation}
and the PDF of $\tilde{F}= F/t$ originally derived by Dynkin
\cite{Dynkin}
\begin{equation}
\mbox{PDF}(\tilde{F}) \sim { \sin \pi \alpha \over \pi} { 1 \over \left( \tilde{F}\right)^\alpha \left(1 + \tilde{F} \right)} \ \ \ 0<\tilde{F} < \infty.
\label{eqApz11}
\end{equation}
Inserting $\alpha=1/2$ we get the result for $\tilde{B}$,
 for zero crossing of  Brownian
motion, obtained by Chung (see Eq. $2.22$ there).
One can also quantify other aspects of zero crossings. For example,
it is well known that the averaged
number of renewals follows
$\langle n \rangle \sim t^\alpha / g^{*} |\Gamma(-\alpha)|$ and
hence using  Eq.
(\ref{eqAs07}) $\langle n \rangle \propto t^\alpha/ \epsilon$ which diverges
in the continuum limit $\epsilon \to 0$, as it should.
The  distribution of $n/\langle n \rangle$ is
the well known Mittag-Leffler distribution,
 with index $\alpha$ and unit mean. 

 To conclude we see that the backward and forward recurrence time
scale linearly with $t$, and they exhibit non-trivial behavior,
which can be obtained either from analysis of Brownian or
Bessel processes, or using renewal theory with the $\epsilon$ trick.
The latter is a very simple approach, which requires some
basic results in renewal theory. Importantly, given the tools
of renewal theory, the exponent $\alpha$
 of the first
passage time PDF $g(\tau) \sim \tau^{-1 -\alpha}$
which is investigated in  Appendix A, 
determines uniquely the statistics
of $F,B$ and $\Delta$, in the long measurement time limit
and when $\alpha<1$.
The fact that the backward recurrence
time is long in the sense that it scales with the observation time
$t$, explains why the mean square displacement
$\langle x^2 \rangle$,  we have found in the main
text,  depends on
the properties of the  meander.


\begin{thebibliography}{99}
\bibitem{remark2} 
For atoms in molasses 
with a $J_g=1/2 \rightarrow J_e =3/2$
Zeeman substructure in a lin $\bot$ lin laser configuration $c=12.3$
\cite{Zoller}.
Ref. \cite{CT,Lutz,Renzoni}
give  $c=22$. 
As pointed out  in \cite{Zoller} different notations are used in
the literature.
\end{thebibliography}
\end{document}